%% file: main.tex
\documentclass[letterpaper,journal]{IEEEtran}
\usepackage{amsmath,amsfonts}
\usepackage{algorithmic}
\usepackage{algorithm}
\usepackage{array}
\usepackage[caption=false,font=normalsize,labelfont=rm,textfont=rm]{subfig}
\usepackage{textcomp}
\usepackage{stfloats}
\usepackage{url}
\usepackage{verbatim}
\usepackage{graphicx}
\usepackage{cite}
\usepackage{balance} 
\IEEEoverridecommandlockouts
\usepackage{cite}
\usepackage{amsmath,amssymb,amsfonts,amsthm}
\usepackage{algorithmic}
\usepackage{graphicx}
\usepackage{textcomp}
\usepackage{algorithm,algorithmic}
\usepackage{xcolor}
\usepackage{mathtools}
\usepackage{bm}
\usepackage[
    colorlinks=true,
    linkcolor=blue,
    citecolor=blue,
    urlcolor=black
]{hyperref}

\usepackage{float}
\def\BibTeX{{\rm B\kern-.05em{\sc i\kern-.025em b}\kern-.08em
    T\kern-.1667em\lower.7ex\hbox{E}\kern-.125emX}}
    \newtheorem{theorem}{Theorem}

\newtheorem{lemma}{Lemma}
\newtheorem{remark}{Remark}

\newtheorem{proposition}{Proposition}

\newtheorem{assumption}{Assumption}

\newcommand{\sE}{\mathcal{E}}

\newcommand{\sL}{\mathcal{L}}      
      
\newcommand{\sN}{\mathcal{N}}

\newcommand{\sS}{\mathcal{S}}

\newcommand{\sX}{\mathcal{X}}      
      
\newcommand{\sZ}{\mathcal{Z}}

 \newcommand{\bP}{\mathbf{P}}

 \newcommand{\Vx}{\boldsymbol{x}}

\newcommand{\Vz}{\boldsymbol{z}}
\newcommand{\Vlambda}{\boldsymbol{\lambda}}

\newcommand{\R}{\mathbb{R}}                     

\newcommand{\bp}{\bm{p}}

\newcommand{\bx}{\bm{x}}
\newcommand{\by}{\bm{y}}
\newcommand{\bz}{\bm{z}}

\newcommand{\bla}{\bm{\lambda}}

\newcommand{\tr}{\mathrm{r}}   
\newcommand{\td}{\mathrm{d}}   
\newcommand{\tM}{\mathrm{M}}     
\newcommand{\tL}{\mathrm{L}}   
\usepackage{booktabs}
\usepackage{multirow}
\allowdisplaybreaks 
\begin{document}
\raggedbottom

\title{Distributed Coordination of Grid-Forming and Grid-Following Inverters for Optimal Frequency Control in Power Systems
}

\author{Xiaoyang Wang,~\IEEEmembership{Graduate Student Member,~IEEE,}
        Xin Chen,~\IEEEmembership{Member,~IEEE}
        \thanks{X. Wang and X. Chen are with the Department of Electrical and Computer Engineering, Texas A\&M University,  College Station, TX 77840 USA (emails: wangxy@tamu.edu, xin\_chen@tamu.edu).} 
        \thanks{This work was supported in part by NSF AMPS 2523934, in part by NSF CAREER 2541998, and in part by the Consortium on AI and Large Flexible Load (CALL) at Texas A\&M University. 
        \emph{(Corresponding author: Xin Chen)}.}
}
%



\maketitle 
\input{Abstract}

\input{Introduction}

\input{Problem}

\input{Algorithm}

\input{Proof}

\input{Simulation}

\input{Conclusion}
\input{Appendix}
\bibliographystyle{IEEEtran}
\bibliography{IEEEabrv,mybibfile}


\end{document}

%% file: Abstract.tex
\begin{abstract}
The large-scale integration of inverter-interfaced renewable energy sources presents significant challenges to maintaining power balance and nominal frequency in modern power systems. 
This paper studies grid-level coordinated control of grid-forming (GFM) and grid-following (GFL) inverter-based resources (IBRs) for scalable and optimal frequency control. 
We propose a fully distributed optimal frequency control algorithm based on the projected primal-dual gradient method and by leveraging the structure of the underlying physical system dynamics. 
The proposed algorithm i) restores the nominal system frequency while minimizing total control cost and enforcing time-varying IBR power capacity limits and line thermal constraints, and ii) operates in a distributed manner that only needs local measurements and neighbor-to-neighbor communication. In particular, when the line thermal constraints are disregarded, the proposed algorithm admits a fully local implementation that requires no communication, while still ensuring optimality and satisfying IBR power capacity limits. We establish the global asymptotic stability of the algorithm using Lyapunov stability analysis. The effectiveness and optimality of the proposed algorithms are validated through high-fidelity, 100\% inverter-based electromagnetic transient (EMT) simulations on the IEEE 39-bus system. The simulations further demonstrate the effective coordination between synchronous generators and IBRs under the proposed control algorithms, as well as the robustness to measurement noise, communication delays, and parameter uncertainties.
\end{abstract}

\begin{IEEEkeywords}
Secondary frequency control, grid-forming inverter, grid-following inverter, distributed optimal control, projected primal-dual gradient dynamics.
\end{IEEEkeywords}

%% file: Introduction.tex
\section{Introduction}

\IEEEPARstart{I}{n} recent decades, modern power systems have witnessed a widespread integration of inverter-interfaced energy resources, such as photovoltaic (PV) panels, wind turbines, and battery energy storage systems. Unlike conventional synchronous generators (SG), inverter-based resources (IBRs) exhibit lower system inertia and faster dynamic behaviors. As a result, renewable generation uncertainties \cite{chen2017robust}, frequent power fluctuations, and fast system dynamics present significant challenges to the classic three-level frequency control architecture \cite{Rebours2007Jan} in power grids. In particular, 
traditional secondary and tertiary frequency controls,  typically operating on timescales of minutes to hours, become inadequate for promptly restoring frequency and ensuring economic efficiency in IBR-rich systems. Nevertheless, large-scale IBRs also bring new opportunities, as their fast response and high power flexibility can substantially enhance grid operation when properly coordinated \cite{chen2019aggregate}.

However, coordinating a huge number of IBRs for frequency control is challenging in practice, due to the heavy communication and computational burdens inherent in centralized control schemes\cite{Lyu2024Dec,Shen2023Oct}. Moreover, the dynamic models of IBR units are typically unknown to grid operators, as their control mechanisms are proprietary and not disclosed by manufacturers. In addition, real-time system disturbance information is often unavailable, which further complicates the grid-level coordination of IBRs. These limitations highlight the urgent need for advanced control strategies that enable distributed coordination of IBRs to enhance scalability and reduce reliance on detailed system model information. 

The primary control mechanisms for IBRs can be mainly categorized as grid-forming (GFM) and grid-following (GFL) \cite{Du2020Aug}. Among others, droop control, virtual synchronous machine control, and dispatchable virtual oscillator control are widely used for GFM inverters; see \cite{Pogaku2007Mar,Johnson2022Jan} for detailed introductions. In \cite{9779512},
 a reinforcement learning-based neural network controller incorporating Lyapunov functions is designed for IBRs to optimize the primary frequency control while ensuring system stability. 

 Extensive and mature research efforts have been devoted to distributed frequency control based on synchronous generators and controllable loads. Reference \cite{Li2015Jul} proposes a distributed generator control scheme that integrates economic dispatch with automatic generation control (AGC) to achieve fast and economical frequency regulation. More recent studies have extended this framework to the distributed control of flexible loads for optimal primary \cite{Zhao2014Jan} and secondary frequency control \cite{Chen2020Sep, Mallada2017Jun, Wang2018Aug}. Further, the challenges posed by non-smooth control cost functions have been addressed in \cite{WANG2020104607,10488734} through the use of Clarke's generalized gradient and the proximal method. Collaborative frequency control methods that coordinate generators and loads have also been proposed in \cite{Wu2024Sep} and \cite{Cherukuri2016Dec}, employing a similar framework as in \cite{Chen2020Sep}. Conventional synchronous machine-based power systems typically rely on AGC for secondary frequency regulation, which is implemented in a centralized/distributed manner and involves only a small number of participating generator units. Constrained by physical mechanical inertia and governor response characteristics, AGC usually operates on relatively slow time scales, ranging from tens of seconds to minutes. In contrast to synchronous machine-based systems, IBRs-dominated systems exhibit two fundamental characteristics, as noted in \cite{8892668}: 1) \emph{fast dynamics and flexible responses with hard power capacity constraints} due to power electronic inverter control, and 
2) \emph{large-scale deployment of a huge number of IBRs}. These characteristics highlight the urgent need for distributed (or even fully local) secondary frequency control algorithms that are tailored to the physical properties and fast dynamics of IBRs, while enabling scalable coordination of massive IBRs for frequency control.
 
 \begin{table*}[ht]
\centering
\footnotesize
\setlength{\tabcolsep}{4pt}
\caption{Comparison of representative secondary frequency control methods for IBR-dominated power systems.
}
\label{tab:secondary_control_summary}
\renewcommand{\arraystretch}{1.15}
\begin{tabular}{ l p{6.5cm} c >{\centering\arraybackslash}p{1.6cm} >{\centering\arraybackslash}p{1.9cm} >{\centering\arraybackslash}p{2.2cm}}
\toprule
\textbf{Control Architecture} 
& \multicolumn{1}{c}{\textbf{Method}}
& \textbf{Optimality} 
& \textbf{Capacity Constraints} 
& \textbf{Line Thermal Constraints} 
& \textbf{Communication} \\
\hline

\multirow{2}{*}{Centralized}
& Data-driven virtual reference feedback tuning \cite{10955180} 
& No
& No 
& No 
& Centralized \\

& Data-driven online feedback optimization \cite{9442926} 
& No
& Yes 
& No
& Centralized \\
\hline

\multirow{4}{*}{Distributed}
& PI-based AGC with consensus \cite{Li2022Nov} 
& No 
& No 
& No 
& Neighbor \\

& Distributed averaging / integral control \cite{7112129,9869334} 
& Global\textsuperscript{$\dagger$}
& No 
& No 
& Neighbor \\

& Primal-dual distributed optimization \cite{Xu2021Jun,Yi2016Dec} 
& Global\textsuperscript{$\dagger$}
& Yes 
& No 
& Neighbor \\

& Learning-based distributed control \cite{10696981} 
& No
& No 
& No 
& Neighbor \\
\hline

\multirow{2}{*}{Decentralized}
& Leaky-integral control \cite{8556089} 
& Near-optimal 
& No 
& No 
& Fully local \\

& Analytical closed-form controllers  
\cite{8588380,10124785} 
& Global\textsuperscript{$\dagger$}
& No 
& No 
& Fully local \\
\hline

\textbf{Our proposed method}
& \textbf{Projected primal-dual gradient dynamics integrated with inverter physical dynamics}
& \textbf{Global}\textbf{\textsuperscript{$\dagger$}}
& \textbf{Yes}
& \textbf{Yes}
& \textbf{Neighbor / Fully local\textsuperscript{*}} \\
\bottomrule
\end{tabular}

\begin{minipage}{\linewidth}
\footnotesize
\noindent
* When line thermal constraints are not taken into account, the proposed method admits a fully local implementation without any communication, while still ensuring system-level control optimality and satisfying IBR capacity limits.\\
$\dagger$ ``Global" refers to the global optimality using an approximate convex optimization problem formulation.
\end{minipage}
\vspace{-5pt}
\end{table*}
 
For IBRs' secondary frequency control, the control methods can be classified into centralized, distributed, and decentralized methods \cite{8892668}. 
 Unlike synchronous generators, IBRs are
characterized by fast power-electronic dynamics, and
strict capacity constraints, which necessitates dedicated control designs that explicitly account
for inverter dynamics and constraints.
A single multi-task central controller is used to manage all DERs in the microgrid based on a data-driven method\cite{10955180,9442926}. A distributed AGC method based on area control error and proportional-integral (PI) control is introduced in \cite{Li2022Nov}. However, the optimal control cost is not considered, and the power regulation movements are regulated to be in proportion to IBRs'
available capacity based on distributed consensus.
References \cite{Cherukuri2016Dec,Xu2021Jun} employ the idea of Laplacian non-smooth gradient to achieve grid-level control optimality. A similar technique, known as the distributed averaging-based or consensus-based integral method \cite{7112129}, is applied in \cite{9869334} to enable distributed implementation.
References \cite{Xu2021Jun}  and  \cite{Yi2016Dec} follow a similar approach to \cite{Chen2020Sep} combining the projected primal-dual gradient method with the distributed-averaging method for control 
algorithm design, and the dual variables and projection are used for power capacity constraints. However, the line thermal constraints are not considered. {Reinforcement-learning-based secondary frequency control is investigated in \cite{10696981}, where consensus-based reward sharing is used to estimate globally averaged rewards, and a distributed observation-sharing scheme is developed to infer global system information. Consequently, a dedicated communication network is required to interconnect all agents.

To eliminate communication, a decentralized leaky-integral controller is proposed in \cite{8556089}. In this approach, a small steady-state frequency deviation is intentionally retained to guarantee identical marginal costs at optimality, revealing an inherent trade-off between frequency regulation accuracy and communication requirements.
Fully decentralized optimal secondary control methods are developed in \cite{8588380} and \cite{10124785} by analytically solving optimal dynamic control problems without constraints. In \cite{8588380}, linearized frequency dynamics with quadratic cost functions are solved explicitly using Riccati equations. In contrast, \cite{10124785} models nonlinear phase-coupling dynamics via Kuramoto oscillators, and the decentralized controller is constructed analytically using Lyapunov arguments. However, analytical optimal or passivity-based dynamic controllers generally cannot enforce hard operational constraints. Once explicit hard constraints, such as line thermal limits or capacity bounds, are imposed, closed-form dynamic controllers are no longer sufficient.

Unlike prior works, this paper first proposes an optimal secondary frequency control framework based on IBRs, incorporating both line thermal constraints and IBRs' power capacity limits, and only local communication with the neighbor bus is needed. Besides, we also propose a fully local version for optimal secondary frequency control. The comparison of our proposed method with the existing secondary frequency control method is shown in Table \ref{tab:secondary_control_summary}.

To enforce the power (or current) capacity limits of IBR units, dynamic projection methods, including the continuous global projection \cite{Xia2000Jul, Gao2003Mar} and discontinuous tangent cone projection \cite{Feijer2010Dec, Cherukuri2016Jan}, are widely used to project state variables onto their corresponding feasible sets. In \cite{8399490, Tang2019Mar}, an augmented Lagrangian method is adopted to avoid discontinuities associated with the tangent cone projection, but it needs careful tuning of a hyper-penalty parameter. In this work, we employ the global projection approach introduced in \cite{Gao2003Mar} to avoid computing the tangent cone at each iteration and to eliminate the need for tuning additional hyperparameters. Moreover, we extend the problem formulation and stability analysis in \cite{Gao2003Mar, Chen2022Jun, Chen2020Sep} to complex settings that involve both equality and inequality constraints. This generalization is motivated by the optimal frequency control problem, where power flow equations and line thermal limits are typically modeled as equality and inequality constraints, respectively.

In this paper, we study the grid-level coordinated control of a mix of GFM and GFL IBRs for power system frequency regulation. By leveraging the projected primal-dual gradient dynamics method and the intrinsic physical dynamics of inverter control, we develop a fully distributed optimal frequency control algorithm for coordinating IBRs. Specifically, this algorithm dynamically adjusts the power setpoints of IBR units to achieve optimal secondary frequency control. 
The key contributions of this work are summarized as follows:
  \begin{itemize}
      \item [1)] The proposed algorithm can achieve grid-level optimal coordination of IBRs for frequency control, which can restore the nominal system frequency, minimize the total control cost, and satisfy time-varying power capacity limits of IBR units and line thermal capacity constraints. Particularly, the algorithm relies solely on real-time system measurements of frequency and line power flow and does not require information of real-time disturbances.
      \item [2)] The proposed algorithm is fully distributed, requiring only local measurements and neighbor-to-neighbor communication. Moreover, when the line thermal constraints are not considered, we further design a fully local version of the algorithm that does not need any communication while still ensuring system-level control optimality and IBR power capacity limits. These distributed and local implementations significantly enhance control scalability, and preserve the privacy of individual IBR units. 
      \item [3)] We establish the asymptotic stability of the proposed algorithm to an optimal solution using Lyapunov stability analysis. To our knowledge, this is the first work that proves the global asymptotic stability of the globally projected primal-dual gradient dynamics for optimization problems with both equality and inequality constraints. 
  \end{itemize}

Moreover, we develop a high-fidelity, 100\% inverter-based, electromagnetic transient (EMT) model of the IEEE 39-bus system in MATLAB Simulink to validate the effectiveness of the proposed algorithms via extensive simulations. The simulation model is released as open-source on GitHub \cite{wang2025ieee39ibr}.

Compared to its conference version~\cite{Wang2025PESGM}, this paper introduces several substantial new contributions:
\begin{itemize}
    \item[1)] A fully local optimal secondary frequency control algorithm is proposed, which operates without communication while ensuring system-level optimality and respecting IBR capacity constraints. To the best of our knowledge, 
    this algorithm represents the first communication-free implementation of an optimal secondary control scheme. 
    It is also the first to realize a local integral-type controller that simultaneously enforces capacity limits and achieves control optimality, offering an alternative to conventional local integral controllers that lack optimality guarantees. 
    
    \item[2)] A Lyapunov-based analysis is presented to prove the global asymptotic stability of the projected primal-dual dynamics for solving constrained optimization problems with both equality and inequality constraints. It provides the first rigorous stability proof for this class of dynamical systems. 
    
    \item[3)] Extensive additional numerical experiments are conducted to validate both the fully local and distributed optimal secondary frequency control algorithms using high-fidelity EMT simulations, considering systems with multiple GFM and GFL IBRs, hybrid IBR configurations, and SG-integrated cases. 
\end{itemize}

%% file: Problem.tex
\section{Dynamic Models and Problem Formulation}
In this section, we first present the dynamic models of GFM and GFL inverters, along with the power network model. We then formulate the optimal frequency control problem.
\subsection{Dynamic Models of Inverters and Power Network} \label{Section:Dynaic model of GFM and GFL}

Consider a power network delineated by a graph $G({\mathcal{N}},\mathcal{E})$, where  ${\mathcal{N}}\!\coloneq\!\left\{1,2,\cdots,|\mathcal{N}|\right\}$  denotes the set of buses and $\mathcal{E}\subset {\mathcal{N}} \!\times\! {\mathcal{N}}$  denotes the set of lines connecting the buses. Let $\sN_\tM$ and $\sN_\tL$ be the set of buses connecting to a GFM or GFL inverter, respectively. 
Without loss of generality, let $\sN_\tM \cap \sN_\tL =\emptyset$ and $\sN_\tM \cup \sN_\tL = \sN$; see Remark \ref{remark:dyn} for more details.
We then present the dynamic models below.

\subsubsection{Grid-Forming Inverters} We consider \emph{droop control-based} GFM inverters \cite{Markovic2021Feb}, and the dynamic model is given by \eqref{eq:gfm:s} for bus $i\in\sN_\tM$:
  \begin{align} \label{eq:gfm:s}
   k_i^\tM {\omega _i} =  {f_i^\tM}(s)\big(P_{\tM,i}^{\tr} - {P_{\tM,i}}\big), \ \   {f_i^\tM}(s) =\frac{\beta_i}{\beta_i+s},
\end{align}  
where $\omega_i$ denotes the frequency deviation from the nominal value at bus $i$.  $P_{\tM,i}^\tr$ and $P_{\tM,i}$ are the active power reference setpoint and the actual active power output of the GFM IBR at bus $i$, respectively. $f^\tM_i(s)$ is the GFM low-pass filter,  $\beta_i>0$ is the cut-off frequency, and $s$ denotes the complex frequency variable. $k^\tM_i>0$ is the droop coefficient. Equation \eqref{eq:gfm:s} can be equivalently reformulated as \eqref{eq:gfm:d}:
\begin{align}\label{eq:gfm:d}
     \frac{k_i^\tM}{\beta_i}  \dot{\omega}_i = -k_i^\tM \omega_i + P_{\tM,i}^\tr - P_{\tM,i},\qquad   i\in\sN_\tM. 
\end{align}

\subsubsection{Grid-Following Inverters} We consider GFL inverters under the power-frequency droop control \eqref{eq:gfl}: 
\begin{align}\label{eq:gfl}
   \quad \quad P_{\tL,i}^\tr -P_{\tL,i}  = k^\tL_i \omega_i, \qquad  i\in\sN_\tL,
\end{align}
where $P_{\tL,i}^\tr$ and $P_{\tL,i}$ are the active power reference setpoint and the actual active power output of the GFL IBR at bus $i$, respectively.
$k^\tL_i >0$ is the droop control coefficient.

 Based on the inverter models \eqref{eq:gfm:d} and \eqref{eq:gfl}, the power network dynamic model is formulated as \eqref{eq:net}:
 \begin{subequations} \label{eq:net}
    \begin{align}
         \frac{k_i^\tM}{\beta_i}  \dot{\omega}_i & =  -{k_i^\tM} \omega_i + P_{\tM,i}^{\tr} -P_i^\td-\!\! \sum_{j:ij\in \mathcal{E}}P_{ij},  && i\in\sN_\tM \label{eq: Dynamic:GFM}\\
          0 & = -k_i^\tL\omega_i+ P_{\tL,i}^{\tr} - P_i^\td- \sum_{j:ij\in \mathcal{E}}P_{ij}, &&i\in\sN_\tL  \label{eq: Dynamic:GFL}\\
          \dot{P}_{ij}& =B_{ij}\left(\omega_i-\omega_j\right),  && ij\in \mathcal{E}  \label{eq: Dynamic:Line}
          \end{align}
    \end{subequations}
where $P_{ij}$ is the active power flow from bus $i$ to bus $j$, and $B_{ij}$ denotes the admittance parameter of line $ij$ \cite{Chen2020Sep}.  $P_i^\td$ denotes the real-time net load demand at bus $i$, including both uncontrollable generation and load, which 
captures the \emph{power disturbances} in the system. 
Equations \eqref{eq: Dynamic:GFM} and \eqref{eq: Dynamic:GFL} represent the nodal power balances at GFM and GFL buses, and \eqref{eq: Dynamic:Line} is the DC power flow dynamic model.

This paper studies the dynamic adjustment of power reference setpoints ($P_{\tM,i}^\tr$, $P_{\tL,i}^\tr$) of GFM and GFL IBRs for secondary frequency control, in response to a power disturbance. 
\begin{remark} \label{remark:dyn}
\normalfont
In model \eqref{eq:net},
the dynamics of synchronous generators can be captured by $P_i^\td$ due to their much slower response compared to IBRs. Alternatively, the bus dynamics with synchronous generators \cite{Zhao2014Jan} can be incorporated as \eqref{eq:SG}:
\begin{align}\label{eq:SG}
	M_i \dot{\omega}_i &=-  D_i\omega_i +P_{\mathrm{G},i}^{\tr} - P_i^{\td}-\sum_{j:ij\in \mathcal{E}}P_{ij},\quad  i\in \mathcal{N}_\mathrm{G}
\end{align}
where $\mathcal{N}_\mathrm{G}$ is the set of buses connecting to synchronous generators, $M_i$ denotes the inertia of generator, $D_i$ is the damping coefficient, and $P_{\mathrm{G},i}^{\tr}$ is the generator mechanical power reference setpoint. Notice that the synchronous generator model \eqref{eq:SG} exhibits a similar structure to the GFM IBR model \eqref{eq: Dynamic:GFM}. In addition, buses without any power injection can be eliminated from the power network, while buses without any controllable devices can be treated as GFL buses with zero control capacity and $k_i^\tL=0$ in \eqref{eq: Dynamic:GFL}. {\qed}
\end{remark}  
\subsection{Optimal Frequency Control (OFC) Problem}
To ensure the control optimality and safety at the grid level, we formulate the OFC problem as \eqref{eq:ofc}:
\begin{subequations} \label{eq:ofc}
\begin{align}
\mathrm{Obj.} \, &\mathop {\min }  \sum\limits_{i \in \mathcal{N}_\tM} c_i^\tM( {P_{\tM,i}^\tr}) +\sum\limits_{i \in \mathcal{N}_\tL} c_i^\tL ( {P_{\tL,i}^\tr})\hspace{-5pt}&\label{eq:ofc_obj} \\ 
 \text{s.t. }&P_{\tM,i}^\tr = P_i^\td + \sum\limits_{j:ij \in {\mathcal{E}}} {{B_{ij}(\theta_i-\theta_j)}} ,&&\hspace{-5pt} i\in \mathcal{N}_\tM&  \label{eq:ofc_GFM} \\
&P_{\tL,i}^\tr = P_i^\td + \sum\limits_{j:ij \in {\mathcal{E}}} {{B_{ij}(\theta_i-\theta_j)}} ,&&\hspace{-5pt} i\in \mathcal{N}_\tL&   \label{eq:ofc_GFL}\\
& \underline{P}_{\tM,i}^\tr \leq P_{\tM,i}^\tr \leq \overline P_{\tM,i}^\tr,&&\hspace{-5pt} i\in \mathcal{N}_\tM&  \label{eq:ofc_limits_GFM}\\
& \underline{P}_{\tL,i}^\tr \leq P_{\tL,i}^\tr \leq \overline P_{\tL,i}^\tr, &&\hspace{-5pt} i\in \mathcal{N}_\tL &  \label{eq:ofc_limits_GFL}\\
& \underline{P}_{ij} \leq B_{ij}(\theta_i-\theta_j)  \leq \overline P_{ij}, &&\hspace{-5pt} ij\in \mathcal{E}&  \label{eq:ofc_limits_line}\\
&  \theta_{\rm{ref}}=0,  && &
\end{align}
\end{subequations}
where $c_{i}^\tM(\cdot)$ and $c_{i}^\tL(\cdot)$ are the control cost functions. $\theta_i$ is the phase angle of bus $i$, and $\theta_{\rm{ref}}$ denotes the phase angle of the reference bus. $\underline{P}_{\tM,i}^\tr$, $\overline P_{\tM,i}^\tr$ and $\underline{P}_{\tL,i}^\tr$, $\overline P_{\tL,i}^\tr$ are the lower and upper power limits of IBRs, which capture the inverter current limits and available power capacity. $\underline{P}_{ij}$ and $\overline P_{ij}$ are the lower and upper thermal capacity limits of line $ij$.  

The objective in \eqref{eq:ofc_obj} is to minimize the total control cost associated with the power adjustments of IBRs. Equations \eqref{eq:ofc_GFM} and \eqref{eq:ofc_GFL} represent the nodal power balance constraints. Equations \eqref{eq:ofc_limits_GFM} and \eqref{eq:ofc_limits_GFL} impose the power capacity limits of the IBRs, while \eqref{eq:ofc_limits_line} captures the line thermal constraints.

\begin{remark}
    \normalfont
    All variables in the OFC problem \eqref{eq:ofc} represent deviations from their nominal values that are set by the most recent economic dispatch results. 
    Moreover, the simplified IBR dynamic models and linear power flow models used in \eqref{eq:net} and \eqref{eq:ofc} are primarily intended to facilitate algorithm design and theoretical analysis. 
    Nevertheless, the proposed algorithms are applicable to practical systems with accurate models, as demonstrated by our simulations in Section~\ref{sec:simu}, which employ high-fidelity EMT models. {\qed}
\end{remark}

\begin{remark}
    \normalfont
    The OFC problem \eqref{eq:ofc} can be extended to optimal frequency control across multiple balancing areas, where each area absorbs its own power disturbance \cite{Li2015Jul,Chen2020Sep}. 
    Moreover, the power exchange adjustments across tie-lines between two areas can be regulated to a fixed value, such as zero, by setting both the upper and lower bounds ($\overline{P}_{ij}, \underline{P}_{ij}$) of the corresponding tie line to that value. {\qed}
\end{remark}

%% file: Algorithm.tex
\section{Distributed Control Algorithm Design}
In this section, we first modify the OFC problem \eqref{eq:ofc} to incorporate the goal of restoring nominal frequency, and then solve it using the projected primal-dual gradient method to develop a distributed frequency control algorithm. Building on this, we further derive a fully local optimal IBR control algorithm that operates without any communication.
\subsection{Modified OFC Problem}
To further eliminate frequency deviations and achieve optimal secondary frequency control, we modify the OFC 
problem \eqref{eq:ofc} as \eqref{eq:ofc2}:
\begin{subequations} \label{eq:ofc2}
\allowdisplaybreaks
\begin{align}
 \mathrm{Obj.} \   &\min \, \sum_{i\in \mathcal{N}_\tM} c_{i}^\tM(P_{\tM,i}^\tr) + \sum_{i\in \mathcal{N}_\tL} c^\tL_i(P^{\tr}_{\tL,i}) \hspace{-5pt}& \nonumber \\
        & \quad \quad \quad \quad+ \frac{1}{2}\sum_{i\in \mathcal{N}_\tM} k_i^\tM \omega_i^2 + \frac{1}{2}\sum_{i\in \mathcal{N}_\tL} k_i^\tL \omega_i^2 \hspace{-15pt}& \label{eq:ofc2:obj} \\
    \mathrm{s.t.}\ &P_{\tM,i}^\tr = k_i^\tM \omega_i + P^{\td}_i + \sum_{j:ij\in \mathcal{E}}  P_{ij} , &&\hspace{-15pt} i \in \mathcal{N}_\tM \label{ofc:GFM_unstable} \\
  &  P_{\tL,i}^\tr = k_i^\tL \omega_i + P^{\td}_i + \sum_{j:ij\in \mathcal{E}}  P_{ij}, &&\hspace{-15pt} i \in \mathcal{N}_\tL \label{ofc:GFL_unstable} \\
   & P_{\tM,i}^\tr = P^{\td}_i + \sum_{j:ij\in \mathcal{E}} B_{ij} (\psi_i - \psi_j),&& \hspace{-15pt} i \in \mathcal{N}_\tM  \label{ofc:GFM:satble}\\
 &   P_{\tL,i}^\tr = P^{\td}_i + \sum_{j:ij\in \mathcal{E}} B_{ij} (\psi_i - \psi_j) ,&& \hspace{-15pt} i \in \mathcal{N}_\tL \label{ofc:GFL:satble} \\
   & \underline{P}_{\tM,i}^\tr \leq P_{\tM,i}^\tr \leq \overline{P}_{\tM,i}^\tr,  && \hspace{-15pt} i \in \mathcal{N}_\tM  \label{ofc:power_limits:M_w=0} \\
  &  \underline{P}_{\tL,i}^\tr \leq P_{\tL,i}^\tr \leq \overline{P}_{\tL,i}^\tr,  && \hspace{-15pt} i \in \mathcal{N}_\tL \label{ofc:power_limits:L_w=0} \\
   & \underline{P}_{ij} \leq B_{ij} (\psi_i - \psi_j) \leq \overline{P}_{ij}.  &&\hspace{-15pt}  ij \in \mathcal{E} \label{ofc:thermal_limits}
\end{align}
\end{subequations}
Here, the variable $\psi_i$ denotes the virtual phase angle of bus $i$, introduced to enforce the power flow equations and line thermal limits in steady states \cite{Chen2020Sep, Mallada2017Jun}, while it may differ from the actual phase angle $\theta_i$ during transients. The objective function \eqref{eq:ofc2:obj} and constraints \eqref{ofc:GFM:satble} and \eqref{ofc:GFL:satble} are modified to ensure that the frequency deviation $\omega_i$ is zero in steady states (or optimal solutions). This property is established in Lemma~\ref{lemma1}, which also shows the equivalence between the two OFC models \eqref{eq:ofc} and \eqref{eq:ofc2}. 

Define $\bm{\omega}\coloneqq (\omega_i)_{i\in\sN},  \bm{P}_\tM^\tr  \coloneqq (P_{\tM, i}^{\tr})_{i\in\sN_{\tM}},  \bm{P}_\tL^\tr  \coloneqq (P_{\tL, i}^{\tr})_{i\in\sN_{\tL}}, \bm{P}\coloneqq(P_{ij})_{ij\in \sE}$, and $\bm{\psi} \coloneqq (\psi_i)_{i\in\sN}$. 

\begin{lemma}
    Suppose $(\bm{\omega}^*,  {\bm{P}_\tM^{\tr*}}, {\bm{P}_\tL^{\tr*}},\bm{P}^*,\bm{\psi}^*)$ is an optimal solution to problem \eqref{eq:ofc2}. Then,  $\bm{\omega}^*=\bm{0}$, and $(  {\bm{P}_\tM^{\tr*}}, {\bm{P}_\tL^{\tr*}},\bm{\theta}^* \!=\! \bm{\psi}^*-\psi_{\rm{ref}})$ 
    is an optimal solution to problem \eqref{eq:ofc}. \label{lemma1} 
\end{lemma}

\textit{Proof:} We prove $\bm{\omega}^*\!=\!\bm{0}$ using contradiction arguments. If $\omega_i^* \neq 0$ for some $i \in \sN$, one can construct a better solution  $(\hat{\bm{\omega}}=\bm{0},  {\bm{P}_\tM^{\tr*}}, {\bm{P}_\tL^{\tr*}}, \hat{\bm{P}} \coloneqq(B_{ij}(\psi_i^*-\psi_j^*))_{ij \in \sE})$, which is feasible to \eqref{eq:ofc2} and yields a smaller objective value \eqref{eq:ofc2:obj} since $k_i^\tM, k_i^\tL > 0$. Thus, $\bm{\omega}^*\!=\!\bm{0}$. 
In addition, for any feasible solution $(\bm{\omega}, \bm{P}_{\tM}^\tr, \bm{P}_{\tL}^\tr, \bm{P}, \bm{\psi})$ to problem \eqref{eq:ofc2}, the construction $(\bm{P}_{\tM}^\tr, \bm{P}_{\tL}^\tr, \bm{\theta} = \bm{\psi} - \psi_{\rm{ref}})$ is a feasible solution to problem \eqref{eq:ofc}. Since the objectives \eqref{eq:ofc_obj} and \eqref{eq:ofc2:obj} are the same when $\bm{\omega}^*\!=\!\bm{0}$, $({\bm{P}_\tM^{\tr*}}, {\bm{P}_\tL^{\tr*}}, \bm{\theta}^* = \bm{\psi}^* - \psi_{\rm{ref}})$ must be an optimal solution to problem \eqref{eq:ofc}. \qed

Additionally, we make the following assumption on problem \eqref{eq:ofc2} for the purpose of theoretical analysis.

\begin{assumption} \label{assumption:convex}
The cost functions $c^\tM_i(\cdot), c_i^\tL(\cdot)$ are convex and twice continuously differentiable. The problem \eqref{eq:ofc2} is feasible and  
Slater’s conditions hold for it. 
\end{assumption}

\subsection{Projected Primal-Dual Gradient Dynamics Method}\label{Algorithm}
We design a projected primal-dual gradient method to solve the modified OFC problem \eqref{eq:ofc2}. 
The Lagrangian function of \eqref{eq:ofc2} is given by:
\begin{align}  \label{eq:Lagrangian}
\sL =& \sum_{i\in \mathcal{N}_\tM} c_{i}^\tM(P_{\tM,i}^\tr) + \sum_{i\in \mathcal{N}_\tL} c^\tL_i(P^{\tr}_{\tL,i}) \nonumber\\
& + \frac{1}{2}\sum_{i\in \mathcal{N}_\tM} k_i^\tM \omega_i^2 + \frac{1}{2}\sum_{i\in \mathcal{N}_\tL} k_i^\tL \omega_i^2  \nonumber\\
& + \sum_{i\in\sN_\tM} \lambda_i^\tM \Big( P_{\tM,i}^\tr \!\!- k_i^\tM \omega_i \!\! - P^{\td}_i - \sum_{j:ij\in \mathcal{E}} P_{ij}\Big)  \nonumber\\
& + \sum_{i\in\sN_\tL} \lambda_i^\tL \Big( P_{\tL,i}^\tr - k_i^\tL \omega_i - P^{\td}_i - \sum_{j:ij\in \mathcal{E}} P_{ij}  \Big)  \nonumber\\
& + \sum_{i\in\sN_\tM} \mu_i^\tM \Big( P_{\tM,i}^\tr - P^{\td}_i - \sum_{j:ij\in \mathcal{E}} B_{ij}(\psi_i - \psi_j) \Big)  \nonumber\\
& + \sum_{i\in\sN_\tL} \mu_i^\tL \Big( P_{\tL,i}^\tr - P^{\td}_i - \sum_{j:ij\in \mathcal{E}} B_{ij}(\psi_i - \psi_j) \Big)  \nonumber\\
& + \sum_{ij\in\sE} \sigma_{ij}^+ \Big( B_{ij}(\psi_i - \psi_j) - \overline{P}_{ij} \Big)  \nonumber\\
& + \sum_{ij\in\sE} \sigma_{ij}^- \Big( \underline{P}_{ij} - B_{ij}(\psi_i - \psi_j) \Big),
\end{align}
where $\lambda_i^\tM, \lambda_i^\tL,\mu_i^\tM,\mu_i^\tL$ are the dual variables associated with equality constraints \eqref{ofc:GFM_unstable}-\eqref{ofc:GFL:satble}. $\sigma_{ij}^+\!\geq\!0$ and $\sigma_{ij}^-\!\geq\!0$ are the dual variables associated with inequality constraints \eqref{ofc:thermal_limits}. For IBRs power limit constraints \eqref{ofc:power_limits:M_w=0} and \eqref{ofc:power_limits:L_w=0}, 
we will then employ the global projection method \cite{Gao2003Mar, Xia2000Jul} to ensure that these constraints are always satisfied during the transient process.

\begin{remark}
    \normalfont
    In \eqref{eq:Lagrangian}, the dual variables $\lambda_i$ associated with the DC power flow equality constraints \eqref{ofc:GFM_unstable} and \eqref{ofc:GFL_unstable} have the physical interpretation of penalizing the frequency variation term $\frac{k_i^\tM}{\beta_i} \dot{\omega}_i$. This can be obtained by substituting \eqref{ofc:GFM_unstable} and \eqref{ofc:GFL_unstable} into the GFM and GFL dynamics equations \eqref{eq: Dynamic:GFM} and \eqref{eq: Dynamic:GFL}, respectively. These equality constraints effectively enforce that system frequencies converge to steady-state values, i.e., $\dot{\omega}_i = 0$, rather than continue to vary over time.
The dual variables $\mu_i$, associated with the DC power flow equality constraints in \eqref{ofc:GFM:satble} and \eqref{ofc:GFL:satble}, penalize both the frequency variation and deviation terms $\{\tfrac{k_i^{\mathrm{M}}}{\beta_i}\dot{\omega}_i + k_i \omega_i\}$ as well as the line power flow mismatch terms $\{P_{ij} - B_{ij}(\psi_i - \psi_j)\}$. This interpretation follows directly from substituting \eqref{ofc:GFM:satble} and \eqref{ofc:GFL:satble} into the GFM and GFL dynamic equations \eqref{eq: Dynamic:GFM} and \eqref{eq: Dynamic:GFL}, respectively.
The variables $\mu_i$ act as penalty terms that enforce the DC power flow equalities and ensure that the virtual phase angles $\psi_i$ converge to the actual phase angles in the steady state. As a result, the line thermal constraints can be consistently enforced using the virtual phase variables $\psi_i$.
    The variables $\sigma_{ij}^+$ and $\sigma_{ij}^-$ penalize line thermal constraint violations in the positive and negative directions, respectively. 
    {\qed}
\end{remark}
Define $\bm{\sigma} \coloneqq (\sigma_{ij}^+,\sigma_{ij}^-)_{ij \in \sE}, \bm{\lambda} \coloneqq (\lambda_i)_{i \in \sN }, \bm{\mu} \coloneqq (\mu_i)_{i \in \sN }$. 
Let $\mathcal{P}_\tM\coloneqq \prod_{i\in\sN_\tM}[ \underline{P}_{\tM,i}^\tr, \overline{P}_{\tM,i}^\tr]$, $ \mathcal{P}_\tL\coloneqq \prod_{i\in\sN_\tL}[\underline{P}_{\tL,i}^\tr, \overline{P}_{\tL,i}^\tr]$ be the power capacity feasible sets of GFM and GFL IBR units. 
The saddle point problem of \eqref{eq:ofc2} is then given by \eqref{eq:maxminL}:
\begin{equation} \label{eq:maxminL}
 \hspace{-10pt}   \min_{\bm{P}_\tM^\tr\in \mathcal{P}_\tM, \bm{P}_\tL^\tr\in\mathcal{P}_\tL, \bm{\omega}, \bm{\psi}, \bm{P}}\,     \max_{ {\bm{\sigma}\geq \bm{0},\bm{\lambda}, \bm{\mu}}}  \,  \hspace{-5pt}     \sL({\bm{P}_\tM^\tr,\bm{P}_\tL^\tr, \bm{\omega}, \bm{\psi}, \bm{P},\bm{\sigma}, \bm{\lambda}, \bm{\mu}}).
\end{equation}

To solve the saddle point problem \eqref{eq:maxminL}, we first optimize it over $\bm{\omega}$ by taking $\frac{\partial \sL}{\partial \omega_i} = 0$, leading to \eqref{w=la}: 
\begin{align}
 \qquad\qquad \qquad  \omega_i = \lambda_i, \qquad \qquad\qquad i\in\sN \label{w=la}
\end{align}
which implies the equivalence between $\omega_i$ and $\lambda_i$. 
Mathematically, \eqref{w=la} implies that at the optimal solution, $\lambda_i = \omega_i$. 
This restricts the search for the optimal solution to the manifold defined by $\lambda_i = \omega_i$. From a physical standpoint, this condition indicates that the shadow price of frequency variation, represented by $\frac{k_i^\tM}{\beta_i} \dot{\omega}_i$, is equal to the frequency deviation $\omega_i$. Since frequency deviation directly reflects the severity of active power imbalance, this equality implies that the shadow price of frequency variation is precisely quantified by the frequency deviation. Importantly, $\lambda_i$ is not computed within the controller but is instead measured directly from the physical system as the frequency deviation $\omega_i$, as illustrated in Figure \ref{fig:localcontroller}. The shadow price $\lambda_i$ evolves dynamically with $\omega_i$ and continues to adjust until the system reaches steady state, i.e., $\dot{\omega}_i = 0$.

Then, we solve the saddle point problem \eqref{eq:maxminL} over the remaining variables using the projected primal-dual gradient dynamics \eqref{eq:pdg}:
\begin{subequations}\label{eq:pdg}
 \begin{align} 
   &\hspace{-5pt}\dot{\lambda}{_{i}^\tM} \!=\!\dot{\omega}_i\!=\!\epsilon_{\omega_i}\Big(P_{\tM,i}^\tr- \!\!k_i^\tM \omega_i- \!\!P^{\td}_i -\!\!\hspace{-5pt} \sum_{j:ij\in \mathcal{E}}P_{ij} \Big),&&\hspace{-5pt}i \in \mathcal{N}_\tM \label{eq:lambdaM}\\
   &\hspace{-5pt} \dot{\lambda}{_{i}^\tL }\!=\!\dot{\omega}_i\!=\! \epsilon_{\omega_i}\Big(P_{\tL,i}^\tr - k_i^\tL \omega_i - P^{\td}_i - \!\! \hspace{-5pt}\sum_{j:ij\in \mathcal{E}} P_{ij}\Big), &&\hspace{-5pt} i \in \mathcal{N}_\tL \label{eq:lambdaL}\\
   &\hspace{-5pt} \dot{P}_{ij} \!=\! \epsilon_{P_{ij}}(  \lambda_i - \lambda_j ) = \epsilon_{P_{ij}}(  \omega_i - \omega_j ), &&\hspace{-5pt}ij\in\sE \label{eq:Pij}\\
     &\hspace{-5pt}   \dot{P}_{\tM,i}^\tr \!=\!\epsilon_{{P}_{\tM,i}^\tr} \Big[\text{Proj}_{\mathcal{P}^\tM_i}\Big( {P}_{\tM,i}^\tr - \alpha \big( {c_{i}^\tM}'( P_{\tM,i}^\tr)  &&  \nonumber\\
     & \quad \quad \quad \quad \quad \quad+ \omega_i + \mu_i^\tM \big) \Big) - P_{\tM,i}^\tr \Big], &&\hspace{-5pt}  i \in \!\mathcal{N}_\tM \label{eq: PMs} \\
  & \hspace{-5pt}   \dot{P}_{\tL,i}^\tr \!=\! \epsilon_{{P}_{\tL,i}^\tr} \Big[\text{Proj}_{\mathcal{P}^\tL_i}\Big( {P}_{\tL,i}^\tr - \alpha \big( {c_{i}^\tL}'( P_{\tL,i}^\tr) &&  \nonumber\\
    &\quad  \quad \quad \quad \quad \quad \quad+ \omega_i + \mu_i^\tL \big) \Big) - P_{\tL,i}^\tr \Big], &&\hspace{-5pt}   i \in \mathcal{N}_\tL \label{eq: PLs} \\
 & \hspace{-5pt}     \dot{\mu}_i^\tM\!=\! \epsilon_{\mu_i}  \Big( P_{\tM,i}^\tr - P^{\td}_i - \! \! \sum_{j:ij\in \mathcal{E}} B_{ij}(\psi_i - \psi_j)  \Big), &&\hspace{-5pt}  i \in \mathcal{N}_\tM  \label{eq: muM}\\
  & \hspace{-5pt}     \dot{\mu}_i^\tL\!=\! \epsilon_{\mu_i}  \Big( P_{\tL,i}^\tr - P^{\td}_i - \! \! \sum_{j:ij\in \mathcal{E}} B_{ij}(\psi_i - \psi_j)  \Big), &&\hspace{-5pt}  i \in \mathcal{N}_\tL  \label{eq: muL}\\
  &  \hspace{-5pt}  \dot{\psi_i} = \epsilon_{\psi_i} \Big( \sum_{j:ij \in \mathcal{E}} (\mu_i  -   \mu_j  -   \sigma_{ij}^+  +   \sigma_{ij}^-) B_{ij} \Big),   &&\hspace{-5pt} i \in \mathcal{N} \label{eq:psi}\\
 &\hspace{-5pt}     \dot{\sigma}_{ij}^+ \!=\! \epsilon_{\sigma_{ij}^+} \Big[\text{Proj}_{\mathbb{R}_{\geq 0}} \Big( \sigma_{ij}^+ +\alpha\big(B_{ij} ( \psi_i - \psi_j ) &&\nonumber  \\
 &\qquad\qquad\qquad\qquad\qquad- \overline{P}_{ij}\big) \Big) - \sigma_{ij}^+\Big],&&\hspace{-5pt}  ij \in \mathcal{E}   \label{eq:sigma+}\\
  &\hspace{-5pt}     \dot{\sigma}_{ij}^- \!=\! \epsilon_{\sigma_{ij}^-} \Big[\text{Proj}_{\mathbb{R}_{\geq 0}} \Big( \sigma_{ij}^- +\alpha\big(-B_{ij} ( \psi_i - \psi_j ) &&\nonumber \\
 &\qquad\qquad\qquad\qquad\qquad+ \underline{P}_{ij}\big) \Big) - \sigma_{ij}^-\Big],&&\hspace{-5pt}  ij \in \mathcal{E}   \label{eq:sigma-}
\end{align}   
\end{subequations} 
where the notations with $\epsilon$ denote positive constant
step sizes. 
$\alpha\!>\!0$ is a small constant parameter, and $\text{Proj}_{\sX}(\cdot)$ is the Lipschitz projection operator defined as:
\begin{align} \label{eq:projection_definiation}
\text{Proj}_{ {\sX}}(\boldsymbol{x})=\text{arg} \min_{\boldsymbol{y} \in  {\sX}}||\boldsymbol{y}-\boldsymbol{x}||,
\end{align}
where $\sX$ is a convex feasible set. Here, we employ the \emph{global dynamics projection} method \cite{Gao2003Mar} to ensure that ${P}_{\tM,i}^\tr$,  ${P}_{\tL,i}^\tr$ always stay within their feasible sets  $\mathcal{P}_i^\tM (t) \coloneqq [ \underline{P}_{\tM,i}^\tr(t), \overline{P}_{\tM,i}^\tr(t)]$, $ \mathcal{P}_i^\tL (t) \coloneqq [\underline{P}_{\tL,i}^\tr (t), \overline{P}_{\tL,i}^\tr (t)]$  throughout the transient process. 
The upper and lower power capacity limits of the IBRs can be dynamically adjusted in real time by replacing static feasible sets with time-varying feasible sets $[\underline{P}_{\tL,i}^{\tr}(t), \overline{P}_{\tL,i}^{\tr}(t)]$ in our algorithm to reflect the time-varying inverter headroom. The projection operator immediately maps the power setpoints onto the feasible region, thereby ensuring that the power outputs remain within admissible limits throughout the transient process.
The effectiveness of the proposed approach under time-varying feasible sets is further validated through simulations in Section \ref{section:IBRandSG} and Section \ref{section:Timevaryinghupper}.
The same dynamics projection method is also used in \eqref{eq:sigma+} and \eqref{eq:sigma-} to enforce the non-negativity of ${\sigma}_{ij}^+$ and ${\sigma}_{ij}^-$, where $\mathbb{R}_{\geq 0}\!\coloneqq\![0,+\infty]$ denotes the set of non-negative real values.
In \eqref{eq:lambdaM}-\eqref{eq:Pij}, we use the equivalence \eqref{w=la} between $\omega_i$ and $\lambda_i$.

In the following subsection, we develop a distributed optimal IBR control algorithm for real-time frequency control, based on the projected primal-dual gradient dynamics \eqref{eq:pdg}.

\vspace{-4mm}
\subsection{Distributed Optimal IBR Control Algorithm}
Due to the special design of the modified OFC problem \eqref{eq:ofc2}, the dynamics \eqref{eq:lambdaM}-\eqref{eq:Pij} become the same as the physical power network dynamic model \eqref{eq: Dynamic:GFM}-\eqref{eq: Dynamic:Line}, when setting the parameters $\epsilon_{\omega_i} \!=\!\frac{\beta_i}{k_i^\tM} $ and $ \epsilon_{P_{ij}}\! =\! B_{ij}$. 
In terms of \eqref{eq:lambdaL}, since the solution dynamics \eqref{eq:pdg} is performed in real time, we have $\dot{\lambda}_i^\tL =\dot{w}_i= 0$ due to the physical system dynamics \eqref{eq: Dynamic:GFL} for all GFL buses. 
    As a result, a portion of the projected primal-dual gradient dynamics \eqref{eq:pdg}, i.e., \eqref{eq:lambdaM}-\eqref{eq:Pij}, are essentially the power network dynamics \eqref{eq:net}, which is automatically executed by the physical power system itself. Hence, the remaining dynamics \eqref{eq: PMs}-\eqref{eq:sigma-} are adopted as our proposed control algorithm to coordinate IBRs for frequency control.

Moreover, as previously noted, $P_i^\td$ in \eqref{eq: muM} and \eqref{eq: muL} represents the real-time nodal power disturbance, which is inherently time-varying, and its real-time value is often unavailable in practice. 
To circumvent the information requirement of $P_i^\td$, we introduce new variables $\nu_i^\tM$ and $\nu_i^\tL$, defined as \eqref{eq:r}, to substitute $\mu_i^\tM$ and $\mu_i^\tL$ in the dynamics \eqref{eq:pdg}:
\begin{subequations} \label{eq:r}
\begin{align}
\nu_i^\tM&=\epsilon_{\nu_i} \Big(\frac{\mu_i^\tM}{\epsilon_{\mu_i}}-\frac{k_i^\tM \omega_i}{\beta_i}\Big), 
 &&  i\in \mathcal{N}_\tM \label{eq:rM} \\
\nu_i^\tL&=\frac{\epsilon_{\nu_i} }{\epsilon_{\mu_i}}\mu_i^\tL, &&   i\in \mathcal{N}_\tL  \label{eq:rL}
\end{align}
\end{subequations}
where $\epsilon_{\nu_i}$ is a positive parameter. Through this substitution of variables, we can equivalently replace the dynamics \eqref{eq: muM}, \eqref{eq: muL} of $\mu_i^\tM,\mu_i^\tL$ with the dynamics \eqref{eq:dr} of $\nu_i^\tM,\nu_i^\tL$: 
    \begin{align} \label{eq:dr}
   \dot{\nu}_i^{\tM/\tL} =\! \epsilon_{\nu_i}  \Big( k_i^{\tM/\tL} \omega_i + \! \! \!\! \sum_{j:ij \in \mathcal{E}} \! \! P_{ij} - \! \!\sum_{j:ij \in \mathcal{E}} \! \!\! B_{ij} &\left( \psi_i - \psi_j \right)  \Big), 
\end{align} 
which is derived using \eqref{eq: muM}, \eqref{eq: muL}, \eqref{eq:r}, \eqref{eq: Dynamic:GFM}, and \eqref{eq: Dynamic:GFL}.
Based on $\nu_i$ got from dynamic \eqref{eq:dr}, $\mu_i$ used in \eqref{eq: PMs}, \eqref{eq: PLs}, and \eqref{eq:psi} can be recovered by inverting \eqref{eq:r}.

\begin{figure}[ht]
    \centering
    \includegraphics[width=0.65\linewidth]{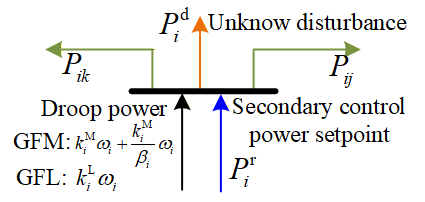}
    \caption{Bus power flow diagram.}
    \label{fig:buspowerdiagram}
\end{figure}

Through the algebraic equivalent substitution in \eqref{eq:r}, the requirement of real-time disturbance power $P_i^\td$ is eliminated, which is hard to obtain accurately in practice. Instead, such information is implicitly captured by the frequency deviation $\omega_i$ (and line power flows $P_{ij}$ when network constraints are considered). In the fully local algorithm introduced in the paper, this equivalence further allows the control law to rely only on local frequency measurements. Physically, this substitution reflects the fact that frequency deviation is a direct manifestation of droop power. Accordingly, the droop power admits two equivalent representations: 1):$P_{i}^\tr - P^{\td}_i - \sum_{j:ij\in \mathcal{E}}P_{ij}$, 2) $\frac{k_i^\tM}{\beta_i}  \dot{\omega}_i + {k_i^\tM} \omega_i$ for GFM, and ${k_i^\tL} \omega_i$ for GFL, as illustrated in Figure \ref{fig:buspowerdiagram}. Therefore, algebraic equivalent substitution \eqref{eq:r} can be interpreted as an equivalent reformulation that replaces real-time disturbance power $P_i^\text{d}$ with frequency measurements, thereby making it much easier for practical implementation while preserving control optimality.

\begin{figure}
    \centering
    \includegraphics[width=1\linewidth]{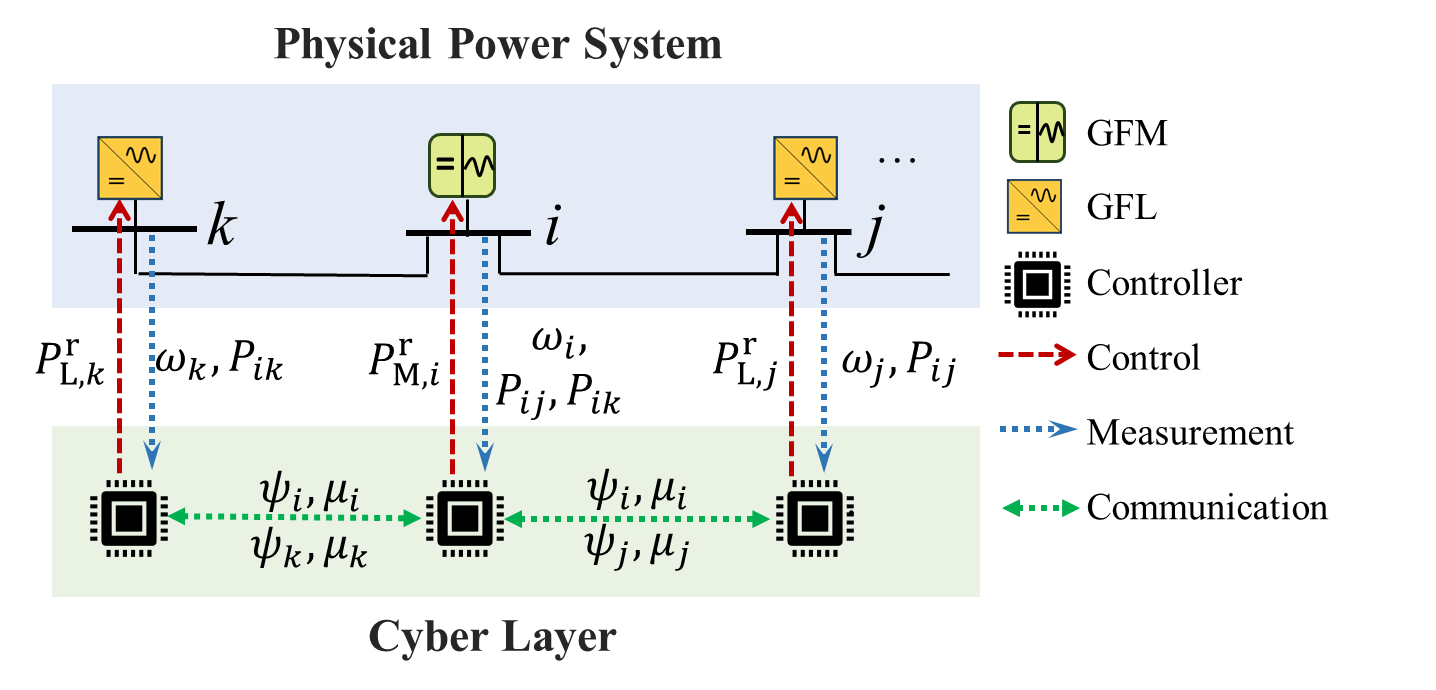}
    \caption{The fully distributed IBRs control algorithm (Algorithm \ref{alg:distributed_optimal_control}). 
 }
    \label{fig:2}
\end{figure}

\begin{algorithm}[b]
\caption{Distributed Optimal IBRs Control Algorithm}
\label{alg:distributed_optimal_control}
\begin{algorithmic}[1]
\STATE \textbf{Input:} Initial values of variables; parameters $(B_{ij})_{j:ij\in\sE}$ of the connected lines for each bus $i\!\in\!\sN$.
\FOR{each bus $i \in \mathcal{N}$ 
\textbf{in parallel}}
    \STATE Measure local frequency deviation $\omega_i$ and power flows $P_{ij}$ of connected lines $ij\in\sE$; exchange $(\psi_i,\mu_i)$ with neighbor buses $j: ij \in \mathcal{E}$ via local communication.\\
     \STATE \textbf{GFM IBR}:  update $P_{\tM,i}^\tr$ according to \eqref{eq: PMs} and execute control; update $(\mu^\tM_i,\nu^\tM_i)$ according to \eqref{eq:rM}, \eqref{eq:dr}. \\
       \textbf{GFL IBR}: update $P_{\tL,i}^\tr$ according to \eqref{eq: PLs} and execute control; update $(\mu^\tL_i,\nu^\tL_i)$ according to \eqref{eq:rL}, \eqref{eq:dr}. 
    \STATE Update $(\psi_i,\sigma_{ij}^+,\sigma_{ij}^-)$ according to \eqref{eq:psi}-\eqref{eq:sigma-}.  
\ENDFOR
\end{algorithmic}
\end{algorithm}
Consequently, based on the projected primal-dual gradient method \eqref{eq:pdg}, we develop the distributed optimal IBR control algorithm as Algorithm \ref{alg:distributed_optimal_control}. 
The implementation of Algorithm \ref{alg:distributed_optimal_control} is illustrated in Figure \ref{fig:2}. 
Each bus only needs to measure the local frequency and the power flows on its connected lines, along with local communication with its neighboring buses. Throughout the control process, the power capacity constraints of both GFM and GFL IBRs are satisfied at all times due to the use of projection. Since the combination of Algorithm \ref{alg:distributed_optimal_control} and the physical power network dynamics \eqref{eq:net} behaves as the projected primal-dual gradient method for solving the modified OFC problem \eqref{eq:ofc2}, the closed-loop system automatically steers the system states to an optimal solution of \eqref{eq:ofc2}. This indicates that the system frequency can be restored to the nominal value with $\omega^*_i=0$, while the total control cost is minimized and the line thermal capacity limits are respected in the steady state.

The global asymptotic stability of Algorithm \ref{alg:distributed_optimal_control},  equivalent to the projected primal-dual gradient dynamics \eqref{eq:pdg}, is established in Theorem \ref{theorem: convergence}; see Section \ref{Convergence proof} for its proof.

\begin{theorem} \label{theorem: convergence}
Under Assumption \ref{assumption:convex}, the projected primal-dual gradient dynamics \eqref{eq:pdg} (or equivalently Algorithm \ref{alg:distributed_optimal_control}) globally asymptotically converges to a fixed point 
($\bm{P}_\tM^{\tr*},\bm{P}_\tL^{\tr*}, \bm{\omega}^*\!=\!\bm{0}, \bm{\psi}^*, \bm{P}^*,\bm{\sigma}^*, \bm{\lambda}^*, \bm{\mu}^*$) in the equilibrium set of dynamics \eqref{eq:pdg}, 
which is an optimal solution of the saddle point problem \eqref{eq:maxminL}, and $ {(\bm{P}_\tM^{\tr*},\bm{P}_\tL^{\tr*}, \bm{\omega}^*\!=\!\bm{0}, \bm{\psi}^*, \bm{P}^*)}$ is an optimal solution of the modified OFC problem \eqref{eq:ofc2}. 
\end{theorem}

 \subsection{Fully Local Optimal IBR Control Algorithm} \label{section:fullylocal}

When the line thermal constraints \eqref{eq:ofc_limits_line} or \eqref{ofc:thermal_limits} are not considered, the distributed control algorithm above reduces to a \emph{fully local optimal IBR control algorithm} that operates without any communication. In this case, it is no longer necessary to introduce virtual phase angles $\psi_i$ to enforce line thermal limits. Instead, we simply replace the virtual phase angle with the actual bus phase angle by setting $\psi_i = \theta_i$, which implies $\dot{\psi}_i = \dot{\theta}_i = \omega_i$.

Under this condition, the line thermal constraints are absent, and the modified OFC problem \eqref{eq:ofc2} excludes \eqref{ofc:thermal_limits}. The constraints \eqref{ofc:GFM:satble} and \eqref{ofc:GFL:satble} can then be written as:
\begin{align} \label{ofc:GFMandGFLdistributesatble}
    P_{\tM/\tL,i}^\tr = P^{\td}_i + \sum_{j:ij\in \mathcal{E}} P_{ij}, \qquad i \in \mathcal{N},
\end{align}
which corresponds to the steady-state condition of the GFM and GFL dynamics \eqref{eq: Dynamic:GFM}-\eqref{eq: Dynamic:GFL} when $P_{ij}$ equals the measured real-time line power flow. Hence, \eqref{ofc:GFMandGFLdistributesatble} can be equivalently rewritten as:
\begin{subequations} \label{eq:wis0constrains}
 \begin{align}
        \frac{k_i^\tM}{\beta_i}\dot{\omega}_i + k_i^\tM \omega_i &= 0, && i \in \mathcal{N}_\tM, \label{eq: Mw+w=0GFM}\\
        k_i^\tL \omega_i &= 0, && i \in \mathcal{N}_\tL. \label{eq: Dw=0GFL}
 \end{align}
\end{subequations}
These conditions directly imply that any optimal solution of the OFC problem must satisfy $\omega_i = 0$ for all $i \in \mathcal{N}$.

By substituting \eqref{eq: Dynamic:GFM} and \eqref{eq: Dynamic:GFL} into \eqref{eq: muM} and \eqref{eq: muL}, with $P_{ij}=B_{ij}(\theta_i-\theta_j)$, we obtain the following fully local control dynamics:
\begin{subequations} \label{eq:mu_local}
\begin{align}
   \dot{\mu}_i^\tM &= \epsilon_{\mu_i} \Big( \frac{k_i^\tM}{\beta_i} \dot{\omega}_i + k_i^\tM \omega_i \Big), && i \in \mathcal{N}_\tM, \label{eq: muM_new}\\
   \dot{\mu}_i^\tL &= \epsilon_{\mu_i} \big( k_i^\tL \omega_i \big), && i \in \mathcal{N}_\tL. \label{eq: muL_new}
\end{align}
\end{subequations}

In the fully local setting, $\mu_i$ also serves as the dual variable associated with the constraints \eqref{eq:wis0constrains}. The control dynamics \eqref{eq:mu_local} can be derived directly by applying the projected primal-dual method to the associated saddle-point problem \eqref{eq:maxminL}.

Using the integral form of \eqref{eq:mu_local} and substituting it into \eqref{eq: PMs}, \eqref{eq: PLs}, we obtain the fully local control algorithm \eqref{eq:ps_local}: 
\begin{subequations} \label{eq:ps_local}
    \begin{align}
             &  \dot{P}_{\tM,i}^\tr \!=\!\epsilon_{{P}_{\tM,i}^\tr} \Big[\text{Proj}_{\mathcal{P}^\tM_i}\Big( {P}_{\tM,i}^\tr\! -\! \alpha \big( {c_{i}^\tM}'( P_{\tM,i}^\tr) \!+\!(1\!+\! \epsilon_{\mu_i} \frac{k_i^\tM}{\beta_i}) \omega_i   &&  \nonumber\\
     & \hspace{60pt}  + \epsilon_{\mu_i}{k_i^\tM} \int {{\omega _i}}  \big) \Big) - P_{\tM,i}^\tr \Big], &&\hspace{-70pt}  i \in \!\mathcal{N}_\tM \label{eq: PMs_local} \\
  &   \dot{P}_{\tL,i}^\tr \!=\! \epsilon_{{P}_{\tL,i}^\tr} \Big[\text{Proj}_{\mathcal{P}^\tL_i}\Big( {P}_{\tL,i}^\tr - \alpha \big( {c_{i}^\tL}'( P_{\tL,i}^\tr) + \omega_i &&  \nonumber\\
    &\hspace{60pt} + \epsilon_{\mu_i}{k_i^\tL} \int {{\omega _i}} \big) \Big) - P_{\tL,i}^\tr \Big], &&\hspace{-70pt}   i \in \mathcal{N}_\tL. \label{eq: PLs_local} 
    \end{align}
\end{subequations} 
\begin{figure}[t]
     \centering
     \includegraphics[width=1\linewidth]{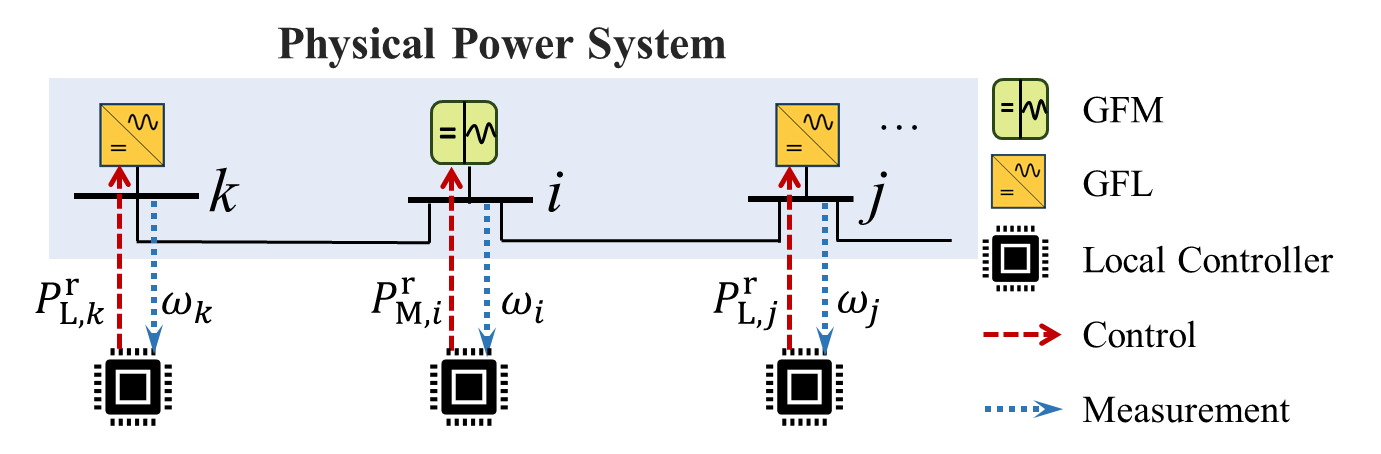}
     \caption{The fully local IBRs control algorithm.
     }
     \label{fig:fullylocal}
 \end{figure}
\begin{figure}[ht]
    \centering
    \begin{minipage}{0.95\linewidth}
        \centering
        \includegraphics[width=\linewidth]{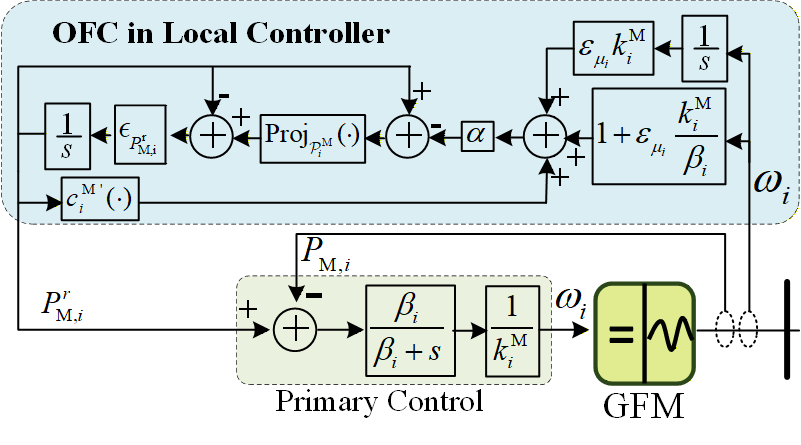}
    \end{minipage}

    \vspace{6pt}

    \begin{minipage}{0.95\linewidth}
        \centering
        \includegraphics[width=\linewidth]{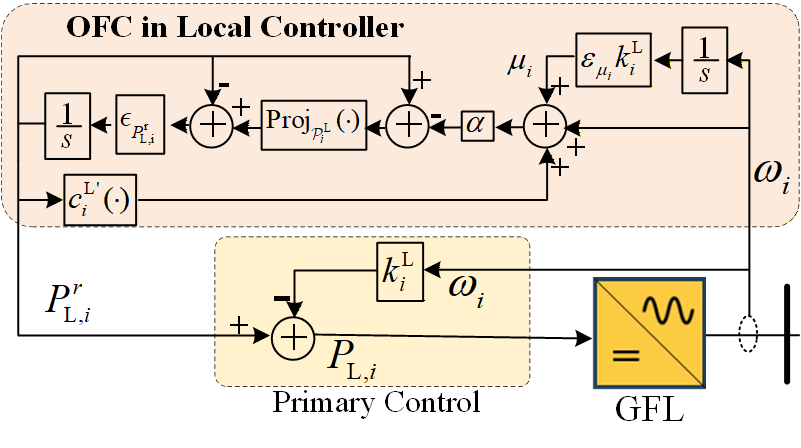}
    \end{minipage}

    \caption{The proposed local controllers for GFM and GFL IBRs.}
    \label{fig:localcontroller}
\end{figure}

As illustrated in Figure \ref{fig:localcontroller}, each IBR unit only needs to measure the local real-time frequency $\omega_i $ at its terminal bus and execute the control algorithm \eqref{eq:ps_local} to dynamically adjust the power reference setpoints ${P}_{\tM,i}^\tr, {P}_{\tL,i}^\tr$ for secondary frequency regulation. The local controllers for GFM and GFL, which implement \eqref{eq:ps_local}, are shown in Figure \ref{fig:fullylocal}. This forms a fully local control law for IBRs that operates without requiring any communication or real-time disturbance information. 
Essentially, the local control algorithm \eqref{eq:ps_local} can be viewed as a reduced version of the distributed control algorithm (i.e., Algorithm \ref{alg:distributed_optimal_control}) that omits line thermal constraints, while this local control algorithm can still restore the nominal frequency with optimal control cost and satisfy the IBR power capacity limits all the time. 
 
\begin{remark} \label{Remark:4}
\normalfont
In the fully local method, note that the constraints in \eqref{ofc:GFMandGFLdistributesatble} do not rely on any DC power-flow approximation. Summing the left and right sides of \eqref{ofc:GFMandGFLdistributesatble} over all buses yields
\begin{align}
  \sum_{i \in \mathcal{N}} P_{\tM/\tL,i}^\tr
  = \sum_{i \in \mathcal{N}} P^{\td}_i 
  + \sum_{ij \in \mathcal{E}} P^{\text{Loss}}_{ij}, \label{eq:lossprove}
\end{align}
where $P^{\text{Loss}}_{ij}=P_{ij}+P_{ji}$ denotes the real-power loss on line $ij$. This shows that the fully local method inherently accounts for line losses when achieving optimal frequency control.
\qed
\end{remark}

%% file: Proof.tex
\section{Theoretical Stability Analysis} 
\label{Convergence proof}
In this section, we analyze the Stability of the proposed control algorithm \eqref{eq:pdg} and prove Theorem \ref{theorem: convergence} using Lyapunov theory. To our knowledge, this is the first work to rigorously prove the asymptotic stability of the globally projected primal-dual dynamics algorithm for optimization problems involving both equality and inequality constraints and a feasible set. 
\subsection{General Optimization Problem Formulation}
Define $\bm{x}\coloneqq[{\bm{P}_\tM^\tr;\bm{P}_\tL^\tr; \bm{\omega}; \bm{\psi}; \bm{P}}]$ as the decision variable. The modified OFC problem \eqref{eq:ofc2} can be equivalently reformulated as a general constrained optimization problem \eqref{general problem}:
\begin{subequations} \label{general problem}
\begin{align} 
  \text{Obj.} \ & \min_{ \boldsymbol{x} \in  \sX} c(\boldsymbol{x}) \\
  \text{s.t.} \ &\bm{h}(\boldsymbol{x}) = \bm{0},\label{eq}\\
&   \bm{g}(\boldsymbol{x}) \label{ineq}\le  \bm{0},
\end{align}  
\end{subequations}
where ${\sX}\! \in\! \R^n$ is the feasible set that captures the IBR power capacity limits \eqref{ofc:power_limits:M_w=0} and \eqref{ofc:power_limits:L_w=0}. $c(\cdot): \R^n \to \R$ denotes the cost function \eqref{eq:ofc2:obj}. $\bm{h}(\cdot): \R^n \to \R^l$ is a vector of affine functions that represent the equality constraints \eqref{ofc:GFM_unstable}-\eqref{ofc:GFL:satble}. $\bm{g}(\cdot): \R^n \to \R^m$ represents inequality constrains \eqref{ofc:thermal_limits}.

In the following analysis, we focus on the general optimization problem \eqref{general problem} under Assumption \ref{ass:general}, and establish the global asymptotical stability of the projected primal-dual gradient dynamics for solving \eqref{general problem}. Then, the proof of Theorem \ref{theorem: convergence} follows as a special case of this general result.

\begin{assumption}\label{ass:general}
 The feasible set ${\sX}$ is nonempty, closed, and convex. The functions $c(\bx)$ and $\bm{g}(\bx)$ are convex, continuously twice differentiable, and have locally Lipschitz gradients on $\sX$. The problem \eqref{general problem} is feasible, and the  
Slater’s conditions hold.
\end{assumption}

\subsection{Dynamics Algorithm and Key Stability Result}

The Lagrangian function of \eqref{general problem} is given by:
\begin{align} \label{eq:Lag}
\sL(\boldsymbol{x, \lambda, \sigma}) = c(\boldsymbol{x}) +\boldsymbol{\lambda}^\top\bm{h}(\boldsymbol{x})+\boldsymbol{\sigma}^\top \bm{g}(\boldsymbol{x}),
\end{align}
where $\boldsymbol \lambda \in \R^l$ denotes the dual variables for the equality constrains \eqref{eq} and $\boldsymbol \sigma \in \R_{\geq 0}^m$ denotes  the dual variables for the inequality constrains \eqref{ineq}. Then, we formulate the saddle point problem as \eqref{saddle point problem}:
\begin{align}\label{saddle point problem}
    \max_{\boldsymbol \sigma \geq \bm{0}, \boldsymbol\lambda}\  \min_{\boldsymbol{x} \in\sX} \ \sL(\boldsymbol{x, \lambda, \sigma}).
\end{align}

Due to the strong duality under Assumption \ref{ass:general}, 
$\sL(\bx, \bla,\bm{\sigma})$ is convex in $\bx$  and concave in $\bla$ and $\bm{\sigma}$. Moreover, the $\boldsymbol{x}$-component of any saddle point of \eqref{saddle point problem} is an optimal solution to problem \eqref{general problem}. 
To solve the saddle point problem \eqref{saddle point problem}, we use the projected primal-dual gradient dynamics \eqref{dynamic}:
\begin{subequations}\label{dynamic}
\begin{align}
   \dot{ \boldsymbol{x}}&=\epsilon_x[\text{Proj}_{\sX}(\boldsymbol{x}-\alpha_{x} \nabla_{\boldsymbol{x}}\sL)-\boldsymbol{x}],\\
      \dot{ \boldsymbol{\lambda}}&=\epsilon_{\lambda}[\alpha_{\lambda} \nabla_{\boldsymbol{\lambda}}\sL],  \\  
    \dot{\boldsymbol{\sigma}}&=\epsilon_{\sigma}[\text{Proj}_{\R_{\geq 0}^m}(\boldsymbol{\sigma}+\alpha_{\sigma} \nabla_{\boldsymbol{\sigma}}\sL)-\boldsymbol{\sigma}],
\end{align}
\end{subequations}
where $\nabla_{\boldsymbol{x}}\sL=\nabla c(\boldsymbol{x})\!+\! \nabla \bm{h}(\boldsymbol{x})^\top \boldsymbol{\lambda}\!+\! \nabla \bm{g}(\boldsymbol{x})^\top \boldsymbol{\sigma}$, $ \nabla_{\boldsymbol{\lambda}}\sL=\bm{h}(\boldsymbol{x})$, and $ \nabla_{\boldsymbol{\sigma}}\sL=\bm{g}(\boldsymbol{x})$.
Denote $\boldsymbol{z}:=[\boldsymbol{x}; \bm{\lambda}; \bm{\sigma}]$, $\sZ :=\sX\times\R^l\times \R_{\geq 0}^m$, $\boldsymbol{\epsilon_z}:=\text{diag}[\boldsymbol{\epsilon}_x, \boldsymbol{\epsilon}_\lambda, \boldsymbol{\epsilon}_\sigma]$, $ \boldsymbol{\alpha_z}:=\text{diag}[\boldsymbol{\alpha}_x, \boldsymbol{\alpha}_\lambda\boldsymbol{\alpha}_\sigma]$. 

The dynamics \eqref{dynamic} can be equivalently rewritten as \eqref{dynamicz}:
\begin{align} \label{dynamicz}
\dot{\boldsymbol{z}}=\boldsymbol{\epsilon_z}\bm{f}(\boldsymbol{z})=\boldsymbol{\epsilon_z}\Big[\text{Proj}_{\sZ}\big(\boldsymbol{z}-\boldsymbol{\alpha_z} {\bm{\Omega}}(\boldsymbol{z})\big)-\boldsymbol{z}\Big],
\end{align}
where ${\bm{\Omega}}(\boldsymbol{z}):=[\nabla_{\boldsymbol{x}}\sL; - \nabla_{\boldsymbol{\lambda}}\sL;- \nabla_{\boldsymbol{\sigma}}\sL]$.
Without loss of generality, we let $\boldsymbol{\epsilon_z}$ be the identity matrix for simplicity. Proposition \ref{proposition:eq=saddle} below shows the equivalence between the equilibrium point set of \eqref{dynamic} and the saddle point set of \eqref{saddle point problem}, which is denoted as ${\sZ}^* \subseteq \sZ$. The proof of Proposition \ref{proposition:eq=saddle} is provided in Appendix \ref{proof:proposition:eq=saddle}.

\begin{proposition} \label{proposition:eq=saddle}
    Any equilibrium point of the globally projected primal-dual gradient dynamics \eqref{dynamic} is an optimal solution of the saddle point problem \eqref{saddle point problem}, and vice versa. 
\end{proposition}

We establish the stability results of the dynamics \eqref{dynamicz} in Theorem \ref{theorem:general dynamic}, which is a general form of Theorem \ref{theorem: convergence}. In the next subsection, we present the proof of Theorem \ref{theorem:general dynamic}.

\begin{theorem} \label{theorem:general dynamic}
    The globally projected primal-dual gradient dynamics \eqref{dynamic} or \eqref{dynamicz} with initial condition $\bm{z}(t_0) \in \sZ $ has a unique continuously differentiable solution $\bm{z}(t) \in \sZ $ for $ t \in [t_0, +\infty) $, which globally asymptotically converges to an optimal solution $\bm{z}^*$ of the saddle point problem \eqref{saddle point problem}. 
\end{theorem}

\subsection{Proof of Theorem \ref{theorem:general dynamic}}

Since $c(\boldsymbol{x})$ and $\bm{g}(\boldsymbol{x})$ have locally Lipschitz gradients and $\bm{h}(\boldsymbol{x})$ is an affine function, ${\bm{\Omega}}(\boldsymbol{z})$ is locally Lipschitz continuous. 
Since the projection function $\text{Proj}_{\sZ}(\cdot)$ is a singleton and globally Lipschitz with the Lipschitz constant $L=1$ \cite[Proposition~2.4.1]{clarke1990optimization}, the vector field  $\bm{f(z)}$ in \eqref{dynamicz} is locally Lipschitz on $\sZ$. Thus, there exists a unique continuous solution $\bm{z}(t) $ of \eqref{dynamicz} \cite[Corollary 1]{Cortes2008May}. Moreover, by \cite[Lemma 3]{Gao2003Mar}, we have that $\bm{z}(t) \in \sZ$ for all time $t \ge t_0$ when $\bm{z}(t_0) \in \sZ$. Below, Proposition \ref{proposition: projection_property} and Proposition \ref{proposition: gradientofprojection} present two key properties of the projection $\text{Proj}_{\sZ}(\cdot)$ for analysis. 

\begin{proposition}\cite{doi:10.1137/1023111} \label{proposition: projection_property} 
    Let ${\sZ}$ be a nonempty, closed, and convex subset of $\R^n$.
For every vector $\by \in \R^n$, a vector $\bm{q}\in {\sZ}$ is its projection if and only if :
\begin{align} \label{eq:propertyofproj}
    \langle \by - \bm{q}, \bp -  \bm{q} \rangle \leq 0,\qquad \forall \bp \in {\sZ}.
\end{align}
\end{proposition}
\begin{proposition} \cite[Theorem~10.18]{tisp_projection} \label{proposition: gradientofprojection}
Let ${\sX}$ be a nonempty, closed, and convex set.
The gradient of the squared distance function $\varphi_{{\sX}}(\bm{x})= \frac{1}{2}||\bx-{\rm{Proj}}_{\sX}(\bx)||^2$ 
at any point $\bx $ is given by:
\begin{align}
\nabla \varphi_{\sX}(\bx) = \bx - {\rm{Proj}}_{\sX}(\bx).
\end{align}
\end{proposition}

Let $\Vz^*=[\Vx^*; \Vlambda^*;\bm{\sigma}^*] \in {\sZ}^*  \subseteq {\sZ}$ be a saddle point of \eqref{saddle point problem} in the saddle point set $ {\sZ}^*$.
    Denote $\by:=\bz-\alpha_z{\bm{\Omega}}(\bz)$, $\bP_{{\sZ}}(\cdot): ={\text{Proj}}_{\sZ}(\cdot)$.
    Define the Lyapunov function $V$ \eqref{Lyapunov}:
    \begin{align}\label{Lyapunov}
        V(\bz)=&\frac{1}{2}||\bz-\bz^*||^2 + \alpha_z(\bz-\bP_{{\sZ}}(\by))^T{\bm{\Omega}}(\bz)  \nonumber \\
        &\hspace{10pt}-\frac{1}{2}||\bz-\bP_{{\sZ}}(\by)||^2 .
    \end{align}
    Then, we can obtain the following two key properties for the Lyapunov function \eqref{Lyapunov} shown in Lemma \ref{Property_Ly}.
    \begin{lemma}  \label{Property_Ly}
    For the Lyapunov function \eqref{Lyapunov}, we have:\\
(A) (Non-negativity): 
 \begin{align}  \label{eq: Non-negative}
      V(\bz) \ge \frac{1}{2}(||\bz-\bP_{{\sZ}}(\by)||^2+||\bz-\bz^*||^2)\ge0.
    \end{align}
(B) (Non-increasing): The Lyapunov function \eqref{Lyapunov} is non-increasing along the trajectory $\bm{z}(t)$, i.e., the Lie derivative of $V$ along the dynamics \eqref{dynamic} is non-positive:
\begin{align}
    \frac{dV}{dt} &= \nabla_{\bz} V(\bz)^\top \bm{f}(\bz) =  \nabla_{\bz} V(\bz)^\top(\bP_{{\sZ}}(\by)-\bz)   \nonumber\\
    & \le -\alpha_z(\bz-\bz^*)^\top{\bm{\Omega}}(\bz)  \le 0 .
\end{align} 
\end{lemma}
\noindent
\textit{Proof of Lemma \ref{Property_Ly}}: 
1) According to Proposition \ref{proposition: projection_property}, by letting  $\by=\bz-\alpha_z{\bm{\Omega}}(\bz)$, $\bp=\bz$ and $\bm{q}=\bP_{{\sZ}}(\by)$, we have:
    \begin{align} \label{eq:nonnegative2}
         \alpha_z(\bz-\bP_{{\sZ}}(\by))^\top{\bm{\Omega}}(\bz)\ge ||\bz-\bP_{{\sZ}}(\by)||^2, \  \  \forall\bz \in \sZ. 
    \end{align}
 By substituting \eqref{eq:nonnegative2} into \eqref{Lyapunov}, we obtain \eqref{eq: Non-negative}.

2) Let 
  $ 
  U(\by)=\frac{1}{2}||\bz-\alpha_z {\bm{\Omega}}(\bz)-\bP_{{\sZ}}(\by)||^2=\frac{1}{2}||\by-\bP_{{\sZ}}(\by)||^2  
 $.
 Since ${\bm{\Omega}}(\bz)$ is continuously differentiable on ${\sZ}$, according to Proposition \ref{proposition: gradientofprojection}, $  U(\by)$ is continuously differentiable, and its gradient is:
  \begin{align}
      \nabla_{\bz} U(\by)^\top&= \nabla_{\by} U(\by)^\top \nabla_{\bz}\by \nonumber \\
   &=\big(\by-\bP_{{\sZ}}(\by))^\top(\boldsymbol{I}-\alpha_z \nabla_{\bz} {\bm{\Omega}}(\bz)\big) \nonumber \\
     &=\big(\bz-\bP_{{\sZ}}(\by)\big)^\top \big(\boldsymbol{I}-\alpha_z \nabla_{\bz} {\bm{\Omega}}(\bz)\big) \nonumber\\
 & \hspace{15pt}-\alpha_z {\bm{\Omega}}(\bz)^\top+\alpha_z ^2  {\bm{\Omega}}(\bz)^\top\nabla_{\bz} {\bm{\Omega}}(\bz),
 \end{align}
 where $\bm{I}$ denotes the identity matrix.
 
Then, by substituting $U(\by)$ into \eqref{Lyapunov}, we obtain:
\begin{align}
  V(\bz)=&\frac{1}{2}||\bz-\bz^*||^2 -U(\by) +\frac{1}{2}||\alpha_z {\bm{\Omega}}(\bz)||^2.
  \end{align}
 $V(\bz)$ is continuously differentiable on ${\sZ}$, and its gradient is:
\begin{align}
\nabla_{\bz} V(\bz)^\top       &=(\bz-\bz^*)^\top -(\bz-\bP_{{\sZ}}(\by))^\top(\boldsymbol{I}-\alpha_z \nabla_{\bz} {\bm{\Omega}}(\bz)) \nonumber\\
&\hspace{10pt} +\alpha_z  {\bm{\Omega}}(\bz)^\top.
\end{align}
Hence, the Lie derivative of $V$ along the dynamics \eqref{dynamic} is:
\begin{align} \label{LieGradient_1}
\frac{dV}{dt}                   &= \nabla_{\bz} V(\bz)^\top \bm{f}(\bz)= \nabla_{\bz} V(\bz)^\top(\bP_{{\sZ}}(\by)-\bz) \nonumber\\
                                     &=(\bz-\bz^*+\alpha_z  {\bm{\Omega}}(\bz))^\top(\bP_{{\sZ}}(\by)-\bz) +||\bP_{{\sZ}}(\by)-\bz||^2 \nonumber \\
                                     &\hspace{15pt}-\alpha_z (\bP_{{\sZ}}(\by)-\bz)^\top \nabla_{\bz} {\bm{\Omega}}(\bz) (\bP_{{\sZ}}(\by)-\bz).
  \end{align}
  
 According to Proposition \ref{proposition: projection_property},
by letting $\by =\bz-\alpha_z{\bm{\Omega}}(\bz)$, $\bm{p}=\bz^*$ and $\bm{q}=\bP_{{\sZ}}(\by)$ in \eqref{eq:propertyofproj}, we obtain:
\begin{align}  \label{LieGradient_2}
    &(\bz-\alpha_z{\bm{\Omega}}(\bz)-\bP_{{\sZ}}(\by))^\top(\bz^*-\bP_{{\sZ}}(\by)) \le 0 \nonumber\\
    \hspace{-5pt} \Longleftrightarrow &(\bz-\alpha_z{\bm{\Omega}}(\bz)-\bP_{{\sZ}}(\by))^\top(\bz^*-\bz+\bz-\bP_{{\sZ}}(\by)) \le 0 \nonumber\\
   \Longleftrightarrow   & (\bz-\bz^*+\alpha_z{\bm{\Omega}}(\bz))^\top(\bP_{{\sZ}}(\by)-\bz) \le  \nonumber\\
     &\hspace{10pt} -||\bP_{{\sZ}}(\by)-\bz||^2-\alpha_z(\bz-\bz^*)^\top{\bm{\Omega}}(\bz).
\end{align}
Substituting \eqref{LieGradient_2} into \eqref{LieGradient_1}, we obtain:
\begin{align} 
    \frac{dV}{dt}     &\le -\alpha_z(\bz-\bz^*)^\top{\bm{\Omega}}(\bz) \nonumber\\
   &\hspace{15pt} -\alpha_z (\bP_{{\sZ}}(\by)-\bz)^\top \nabla_{\bz} {\bm{\Omega}}(\bz) (\bP_{{\sZ}}(\by)-\bz).
\end{align}

Since $\sL(\bx, \bla,\bm{\sigma})$ \eqref{eq:Lag} is convex in $\bx$ and concave in $\bla$ and $\bm{\sigma}$,
 ${\bm{\Omega}} (\boldsymbol{z})$  is monotone on $\sZ$.  Thus, its Jacobian matrix $\nabla_{\bz}{\bm{\Omega}}({\bz})$ is positive semi-definite, and we have
\begin{align}  \label{eq:dvdtle0}
    \frac{dV}{dt} &\le  -\alpha_z(\bz-\bz^*)^\top{\bm{\Omega}}(\bz)\nonumber\\
    & =  -\alpha_z \big[(\bx  -\bx^*)^\top \nabla_{\bx}\sL(\bx, \bla ,\bm{\sigma}) \nonumber\\
        &\quad -(\boldsymbol{\lambda}  \!-\!\boldsymbol{\lambda} ^*)^\top \nabla_{\boldsymbol{\lambda} }\sL(\bx, \bla ,\bm{\sigma})   \!   -\!(\boldsymbol{\sigma}\!-\!\boldsymbol{\sigma}^*)^\top \nabla_{\boldsymbol{\sigma} }\sL(\bx, \bla ,\bm{\sigma})   \big] \nonumber  \\
                        & \le\alpha_z  \big[ \sL(\bx^*, \bla ,\bm{\sigma})-\sL(\bx, \bla ,\bm{\sigma}) + \sL(\bx, \bla ,\bm{\sigma})  \nonumber\\
                        &\hspace{15pt}-\sL(\bx, \bla^* ,\bm{\sigma}) + \sL(\bx, \bla ,\bm{\sigma})-\sL(\bx, \bla ,\bm{\sigma}^*) \big] \nonumber\\
                        &  =\alpha_z  \big[\sL(\bx^*, \bla ,\bm{\sigma})-\sL(\bx, \bla ,\bm{\sigma}) + \sL(\bx, \bla ,\bm{\sigma}^*) \nonumber\\
                        & \hspace{15pt} -\sL(\bx, \bla^* ,\bm{\sigma}^*) + \sL(\bx, \bla ,\bm{\sigma})-\sL(\bx, \bla ,\bm{\sigma}^*) \big] \nonumber\\
                        &=\alpha_z  \big[  \sL(\bx^*, \bla ,\bm{\sigma})-\sL(\bx^*, \bla^* ,\bm{\sigma}^*) \nonumber \\
                        &\hspace{15pt}+ \sL(\bx^*, \bla^* ,\bm{\sigma}^*)-\sL(\bx, \bla^* ,\bm{\sigma}^*)   \big] \le 0,
\end{align}
where the second inequality is because $\sL(\bx, \bla,\bm{\sigma})$ is convex in $\bx$, and concave in $\bla$ and $\bm{\sigma}$. The second equality is because 
$\sL(\bx, \bla ,\bm{\sigma})-\sL(\bx, \bla^* ,\bm{\sigma})=(\bla-\bla^* )^\top \bm{h}(\bx)=\sL(\bx, \bla ,\bm{\sigma}^*)-\sL(\bx, \bla^* ,\bm{\sigma}^*)$. The third equality is because $(\bx^*, \bla^*,\bm{\sigma}^*)$ is one  saddle point of \eqref{saddle point problem}. Thus,
Lemma \ref{Property_Ly}-(B) is proved. \qed

\vspace{2pt}
From Lemma \ref{Property_Ly}-(B), $ \frac{dV}{dt} \le 0$ indicates that $\forall t \geq t_0$,
\begin{align}\label{eq:bound}
   \bz(t) \in \sS(\bz(t_0)) \coloneqq\big\{\bz \in {\sZ}| V(\bz) \le V(\bz(t_0)) \big\}.
\end{align}
From Lemma \ref{Property_Ly}, $V(\bz)$ is non-negative for all $\bz \in {\sZ}$ and non-increasing along the trajectory $\bz(t)$ of dynamics \eqref{dynamicz}, thus dynamics \eqref{dynamicz} is Lyapunov stable \cite[Theorem 3.2]{Slotine1991AppliedNC}.
From the LaSalle invariant set theorem \cite[Theorem~4.4]{khalil2002nonlinear}, the solution $\bz(t)$ of \eqref{dynamicz} converges to the largest invariant set $M$ contained in the set $B$ \eqref{eq:closure}:
\begin{align} \label{eq:closure}
    B\coloneqq \Big\{\bz \in \sS(\bz(t_0))| \frac{dV(\bz)}{dt}=0 \Big\}.
\end{align} 
Lemma \ref{asconvergence_final} below further shows that the solution \( \bm{z}(t) \) converges to a fixed equilibrium point that depends on the initial condition \( \bm{z}(t_0) \).
\begin{lemma} \label{asconvergence_final}
    The set \( B \) \eqref{eq:closure} is contained within the equilibrium point set of dynamics \eqref{dynamicz}, i.e., \(M \subseteq B \subseteq \sZ^* \). Moreover, the solution \( \bm{z}(t) \)  of dynamics \eqref{dynamicz} globally asymptotically converges to a fixed point in \( \sZ^* \). 
\end{lemma} 
The proof of Lemma \ref{asconvergence_final} is provided in Appendix \ref{Proofdvdt0eqequil}. From Proposition \ref{proposition:eq=saddle}, the solution \( \bm{z}(t) \)  of dynamics \eqref{dynamicz}  with $\bz(t_0)\in\sZ$  globally asymptotically converges to an optimal solution of the saddle point problem \eqref{saddle point problem}. Thus, Theorem \ref{theorem:general dynamic} is proved.

%% file: Simulation.tex
\section{Simulation Results}\label{sec:simu} 
In this section, high-fidelity EMT simulations are performed on a modified IEEE 39-bus system with 100\% inverter-based generation and a hybrid IBRs and SG-integrated case. The proposed distributed and fully local control algorithms are demonstrated under a step power disturbance and continuous power disturbances. The impacts of measurement noise, communication delay, parameter uncertainties, and time-varying IBRs' power capacity limits are also tested.
\subsection{Simulation Setup}
\begin{figure}
    \centering
    \includegraphics[scale=0.09]{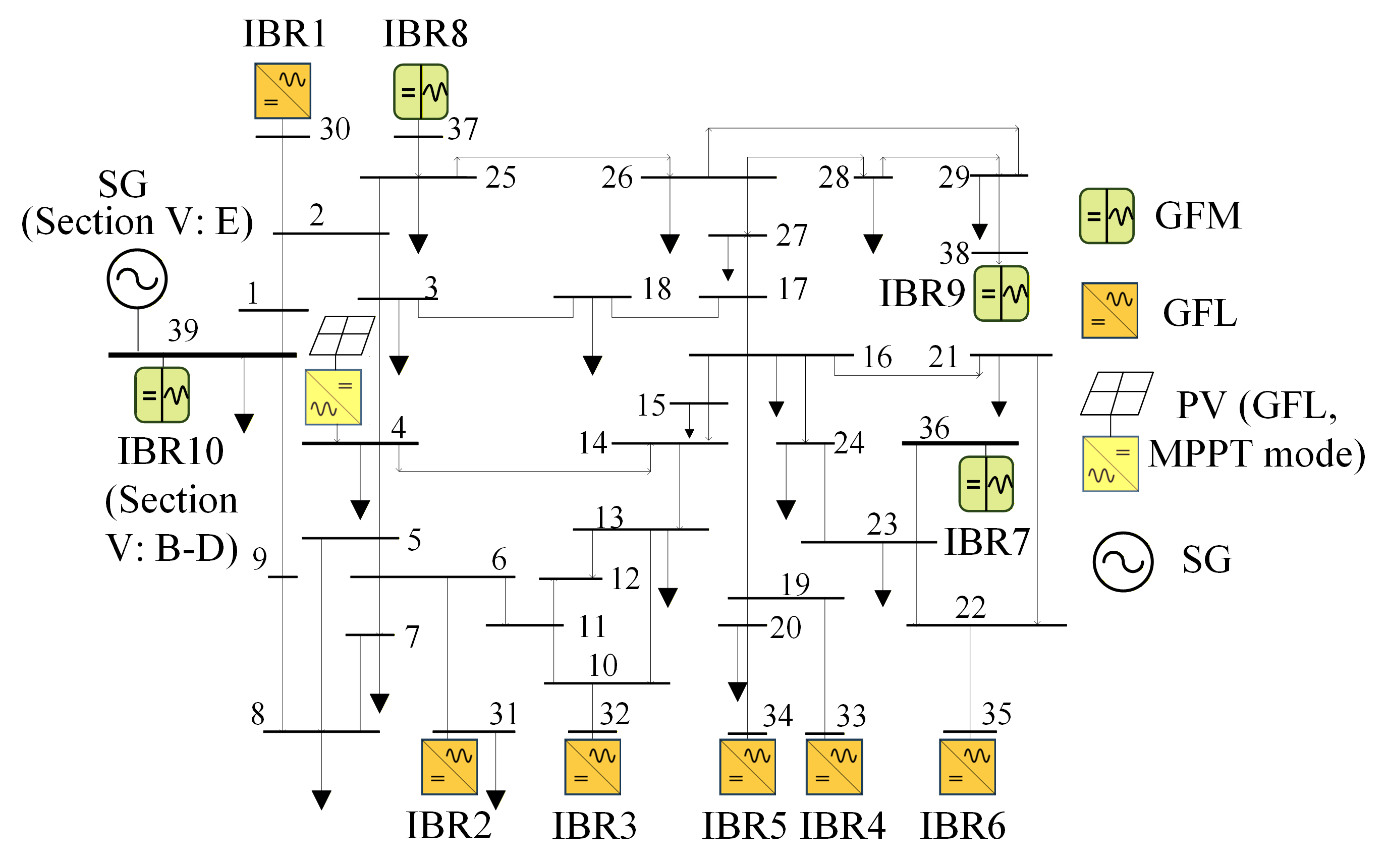}
    \caption{A modified IEEE 39-bus system with IBRs and synchronous generator.}
    \label{fig: IEEE39}
\end{figure}

We built a high-fidelity EMT model in MATLAB/Simulink for simulation studies, which has been open-sourced in~\cite{wang2025ieee39ibr}. The modified IEEE 39-bus system with 100\% IBRs, as shown in Figure \ref{fig: IEEE39}, is used as the test system for simulation Section \ref{section:fulltloct_sim} to  Section \ref{section:constiuouschangefullylocal}. 
All synchronous machines are replaced by IBR units at the same buses, including 4 GFM IBRs and 6 GFL IBRs. In Section \ref{section:IBRandSG}, IBR 10 is replaced by a synchronous generator to show the coordination of the proposed OFC with AGC. The primary frequency control of GFM and GFL IBRs is set according to Section \ref{Section:Dynaic model of GFM and GFL}. For the primary voltage control, the constant reactive power control is used for GFL IBRs (IBR 1-6), and the constant voltage control is used for GFM IBRs (IBR 7-10). Additionally, a PV unit with a GFL inverter operating under the maximum power point tracking (MPPT) mode is integrated at Bus 4 to emulate power disturbances. Unlike IBRs 1–10, this PV inverter is modeled as an uncontrollable device, with its real-time power output determined by solar irradiance. We consider the control cost functions ${c_i}(P_{i}^\tr) = {C_i}\cdot{P_i^\tr}^2$, which is quadratic on the power reference adjustment $P_i^\tr$. The cost coefficients $C_i$ for IBRs 1-4, 5-9, and 10 are set to 1, 2, and 0.5 (p.u.), respectively. The initial generation power of IBRs 1-6 and 7-10 are set to 2 MW and 2.6 MW, respectively. The power output limits for the GFL IBRs 1–6 are defined as 0 MW (lower) and 3 MW (upper), while those for the GFM IBRs 7–10 are set to 1.6 MW (lower) and 4 MW (upper), respectively.

\subsection{Fully Local Algorithm under Step Power Change} \label{section:fulltloct_sim}

\begin{figure}[ht]
    \includegraphics[width=1\linewidth]{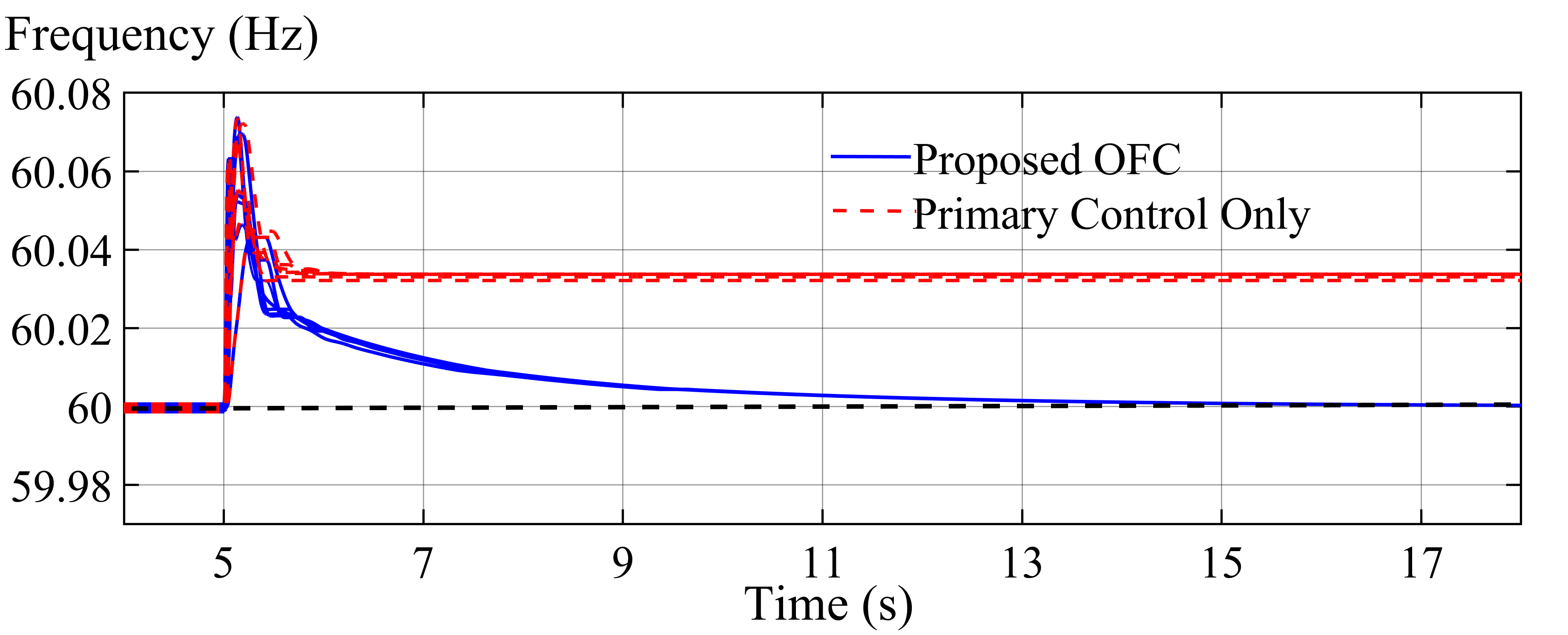}
    \caption{Frequency dynamics at IBRs terminal buses under step power change.}
    \label{fig:7}
\end{figure}

We implement the fully local algorithm described in section \ref{section:fullylocal} as Simulink modules. At $t\!=\!5$ s, the PV generation at bus 4 has a 5 MW  step power change (increase from 0  MW to 5 MW). The frequency dynamics with proposed OFC (solid line) and with primary droop control only (dotted line)  are shown in Figure \ref{fig:7}. Different lines represent the measured frequencies at different buses to which IBRs are connected. It is seen that the case with only primary control leads to a higher frequency level (60.034 Hz) due to increased generation. In contrast, the proposed OFC algorithm can effectively restore the system frequency back to the nominal value within a short time due to the fast dynamic response of IBRs. 

\begin{figure}[ht]
    \centering
 \vspace{-15pt}
\includegraphics[width=1\linewidth]{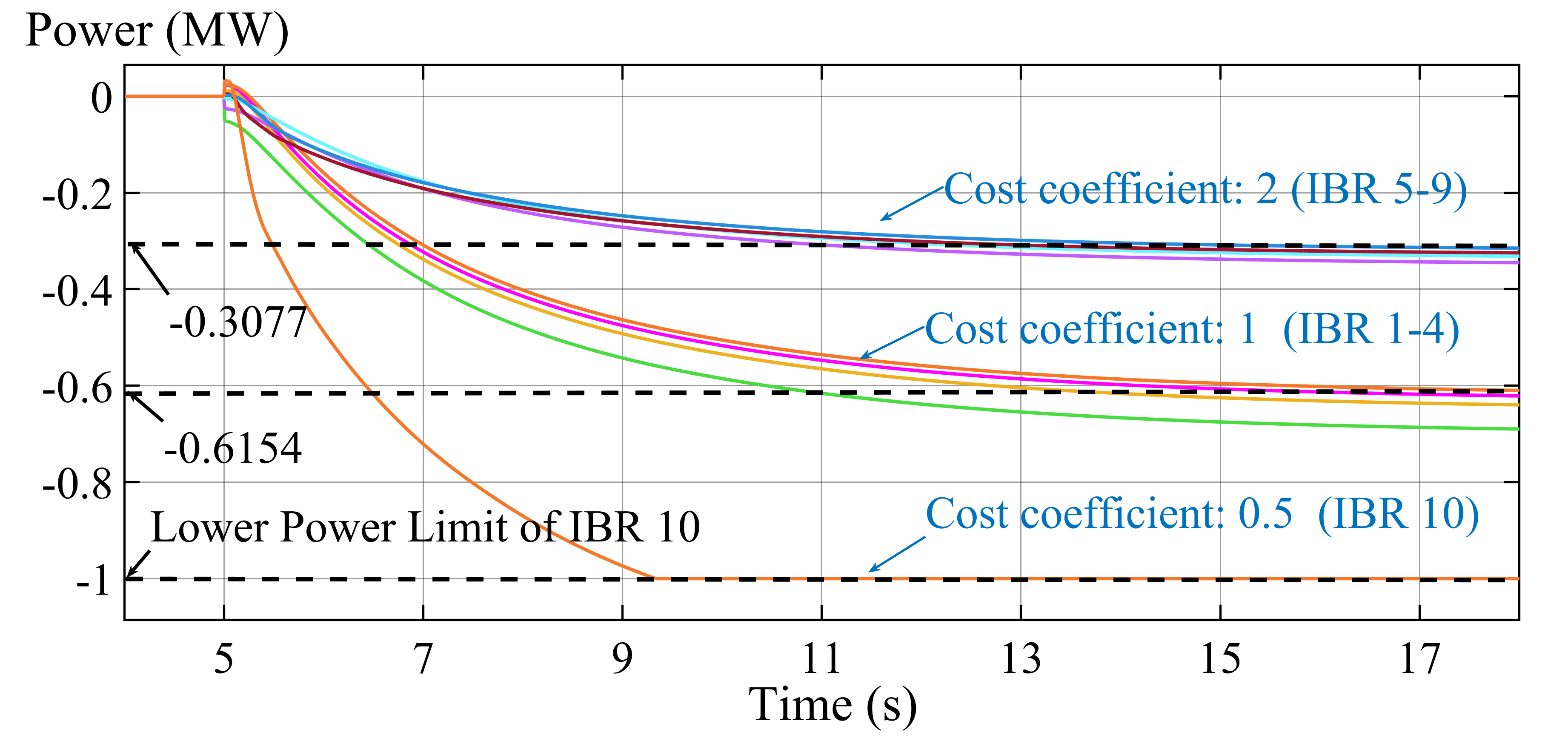}
    \vspace{-15pt}
    \caption{IBRs power setpoint adjustments with OFC. }
    \label{fig:8}
\end{figure}

\begin{figure}[ht]
    \vspace{-15pt}
    \centering
    \includegraphics[width=1\linewidth]{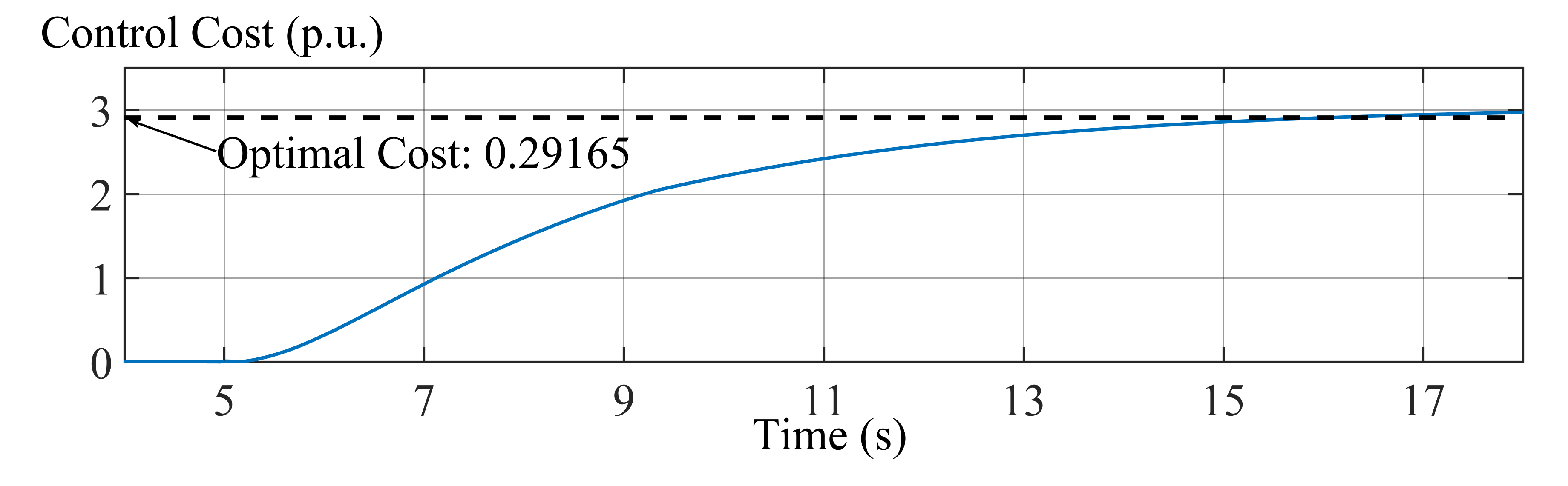}
     \vspace{-15pt}
    \caption{Total cost for IBRs power setpoint adjustment.  }
\label{fig:totalcontrol_costofPset}
\end{figure}

Figure \ref{fig:8} illustrates the dynamics of power adjustments of IBR units, which gradually converge to the optimal control decision with a different color line. It is observed that the IBR units with higher cost coefficients tend to make smaller power adjustments. 
IBR 10, which has the lowest cost coefficient, performs the largest adjustment until reaching its lower bound (the power adjustment lower bound is -1 MW, which equals the power capacity lower bound of 1.6 MW minus the initial value of 2.6 MW). The dynamics projection function in our control algorithm ensures that the IBR power capacity constraints are satisfied throughout the transient process.    
We solve the optimal frequency control problem \eqref{eq:ofc} to obtain the optimal control decision and cost
using MATLAB Optimization Toolbox. The solution results are depicted by the black dotted lines in  Figures \ref{fig:8} and \ref{fig:totalcontrol_costofPset}. 

Figure \ref{fig:totalcontrol_costofPset} shows the total cost during the control transient process calculated using \eqref{eq:ofc_obj}. 
The curves in different colors in Figure \ref{fig:8} represent the power adjustments of different IBRs under step change disturbance. 
Since IBRs 1–4 have identical cost coefficients, they admit the same optimal solution.
The small discrepancies observed in Figure \ref{fig:8} are attributed to power losses. These arise because our control algorithm is based on a linearized DC power flow model, whereas the high-fidelity simulations use the full AC power flow model. The magnitude of these discrepancies reflects how power losses are distributed, which depends on the power network configuration and the electrical distance between each IBR and the disturbance location.
These results demonstrate the control optimality of the proposed algorithm.

 \subsection{Distributed Algorithm under Step Power Change} \label{section:communication}
To satisfy the line thermal constraints, we introduce a local communication network as outlined in Algorithm \ref{alg:distributed_optimal_control}, where each bus communicates with its neighboring buses. The same step power change as described in Section \ref{section:fulltloct_sim} is applied. 
The distributed algorithm is then executed. 

\begin{figure}[hb]
     \centering
     \includegraphics[width=1\linewidth]{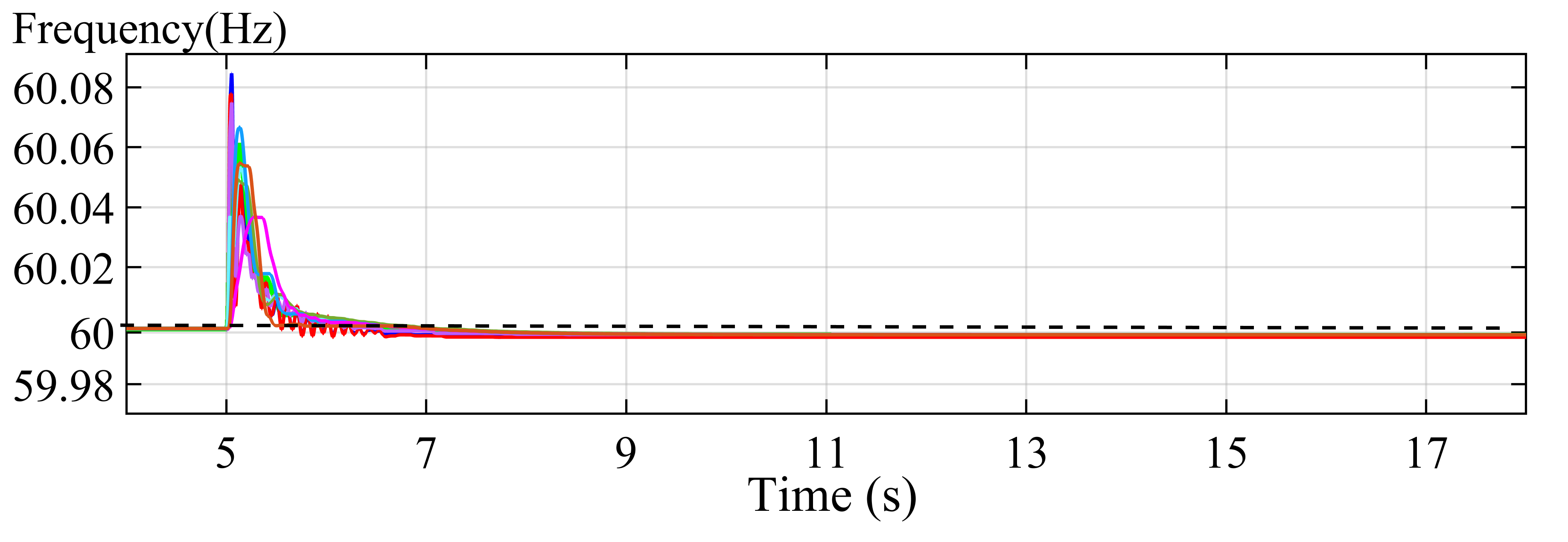}
     \caption{Frequency dynamics of IBRs' terminal Buses under distributed OFC.}
     \label{fig:frequencycommunication}
 \end{figure}
 \begin{figure}[hb]
     \centering
      \vspace{-15pt}
     \includegraphics[width=1\linewidth]{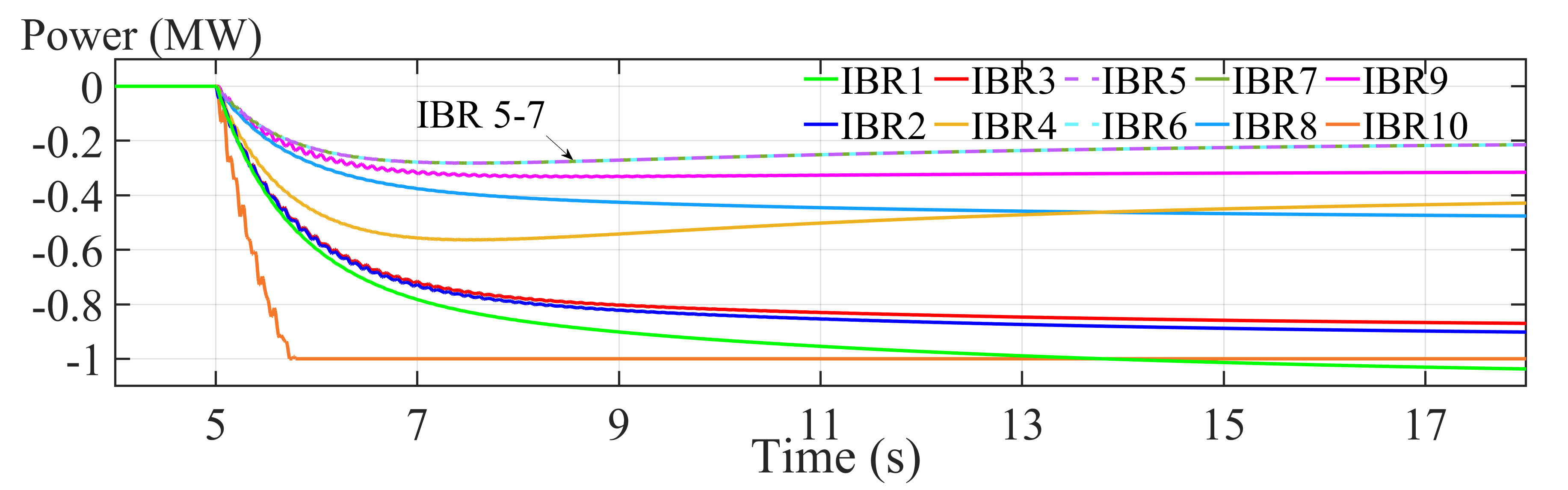}
     \caption{IBRs power setpoint adjustments with distributed OFC under line thermal constraints.}
     \label{fig:DeltaPIBRcommunication}
 \end{figure}
 
 \begin{figure}[hb]
    \centering
    \vspace{-15pt}
    \includegraphics[width=1\linewidth]{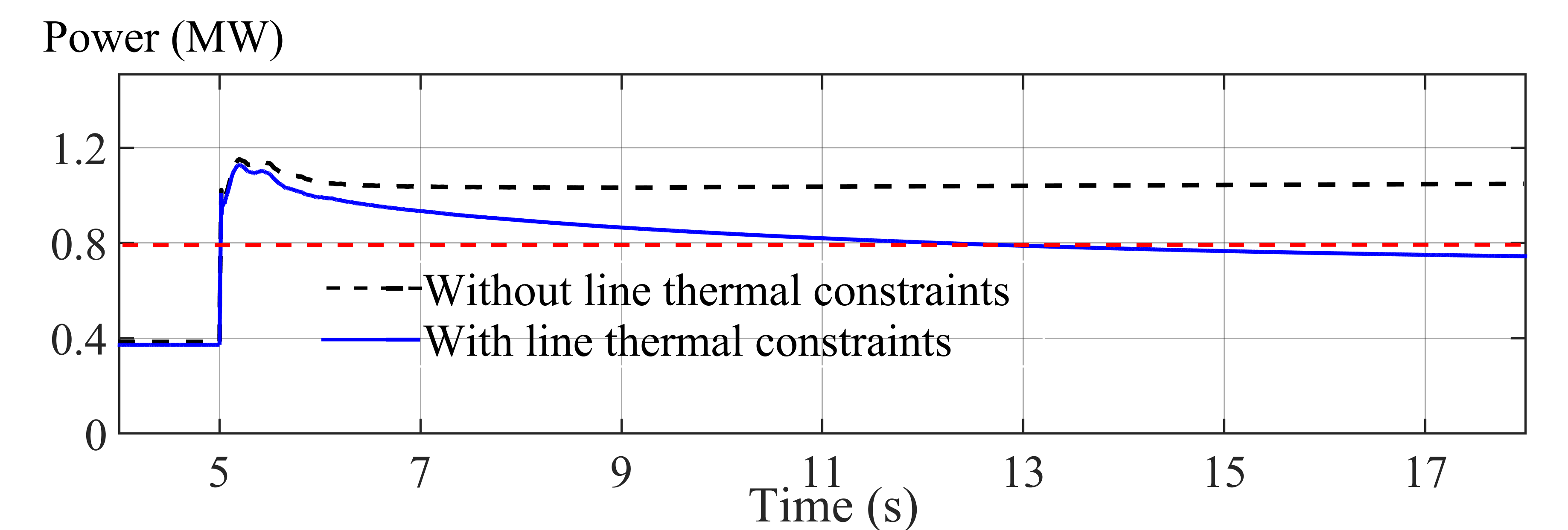}
    \caption{Power flow in line 3-18 under step power change. (Blue solid line: distributed control algorithm with line thermal constraints; black dashed line: local control algorithm without line thermal constraints; red dashed line: upper bound of line power flow.)}
    \label{fig:pij318}
\end{figure}
\begin{figure}[hb]
    \centering
     \vspace{-15pt}
    \includegraphics[width=1\linewidth]{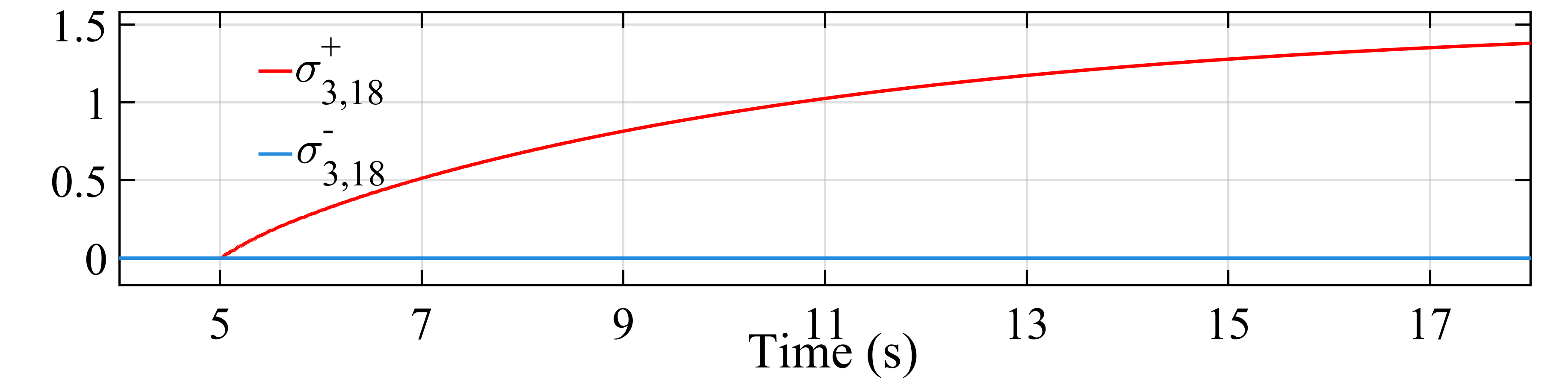}
     \vspace{-10pt}
    \caption{Dual variables $\sigma_{3,18}^+$ and $\sigma_{3,18}^-$ for line thermal constraints.}
    \label{fig:dualpij318}
\end{figure}

The frequency dynamics of IBRs buses are shown in Figure \ref{fig:frequencycommunication}. The power stepoints adjustments of IBRs are shown in Figure \ref{fig:DeltaPIBRcommunication}, and the power flow results of line 3-18 are shown in Figure \ref{fig:pij318}.
The upper and lower bounds for the power flow of line 3-18 are set to $\pm 0.8$ MW. Figure \ref{fig:pij318} compares the performance of the distributed algorithm, which incorporates line thermal constraints, with that of the local algorithm, which does not consider these constraints. It is observed that although the frequency can be restored to 60 Hz in a short time, as shown in Figure \ref{fig:frequencycommunication}, the line thermal constraint is temporarily violated during the transient process, while the distributed algorithm effectively regulates the line power flow and enforces the line thermal constraint in the steady state.

As shown in Figure \ref{fig:DeltaPIBRcommunication}, to regulate the power flow on line 3–18 within its feasible limits, IBRs 1–3 and IBR 8, which are located in the same area as the disturbance at bus 4, contribute more to imbalance compensation. In contrast, IBRs 4–7 and IBR 9, which are located in the opposite area and have a larger electrical distance from the disturbance, reduce their adjustments. As a result, the imbalanced power is redistributed to respect the line thermal constraints.

The corresponding dual variables $\sigma_{3,18}^+$ and $\sigma_{3,18}^-$ are shown in Figure \ref{fig:dualpij318}. When the upper transmission power limit of line 3–18 is violated, the dual variable $\sigma_{3,18}^+$ increases accordingly and influences the allocation of imbalance power until the line thermal constraints are satisfied. This behavior directly reflects the complementary slackness condition in the Karush–Kuhn–Tucker (KKT) optimality conditions, $B_{ij}(\psi_i^*-\psi_j^*)=P_{ij}^*$, and 
$ \sigma_{ij}^{+*} \bigl(B_{ij}(\psi_i^*-\psi_j^*) - \bar{P}_{ij}\bigr) = 0.$

\subsection{Fully Local Algorithm under Continuous Power Change} \label{section:constiuouschangefullylocal}
\begin{figure}[ht]
    \centering
    \vspace{-10pt}
    \includegraphics[width=1\linewidth]{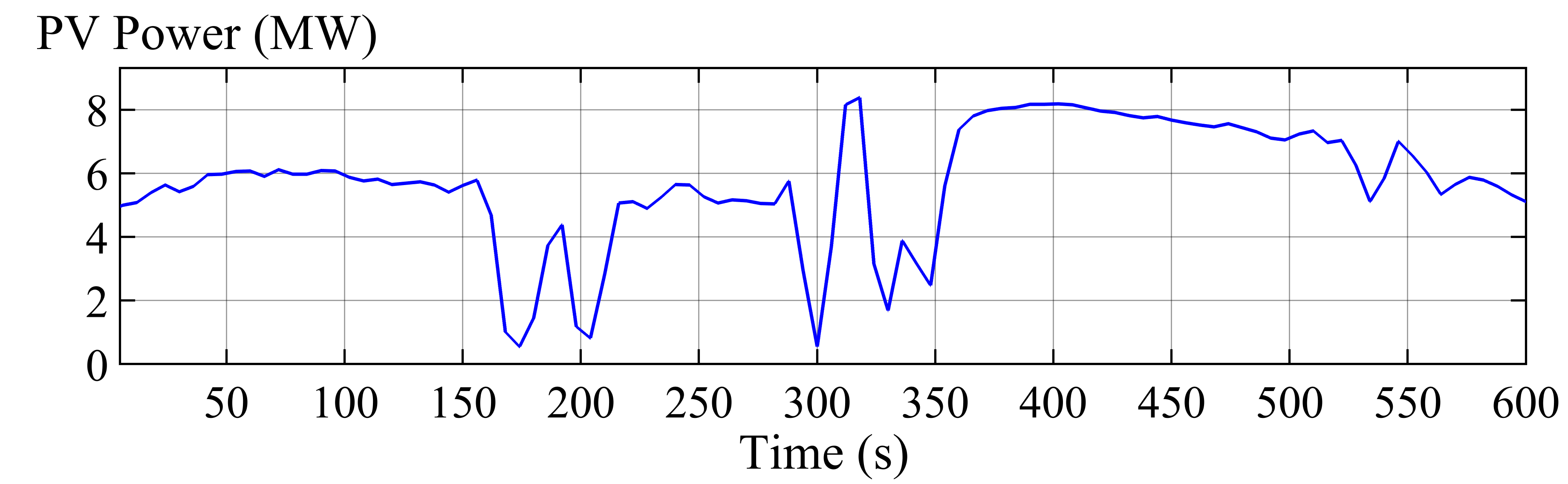}
    \caption{Time-varying power generation from the PV unit at Bus 4.}
    \label{fig:9}
\end{figure}

\begin{figure}[ht]
    \centering
    \vspace{-15pt}
    \includegraphics[width=1\linewidth]{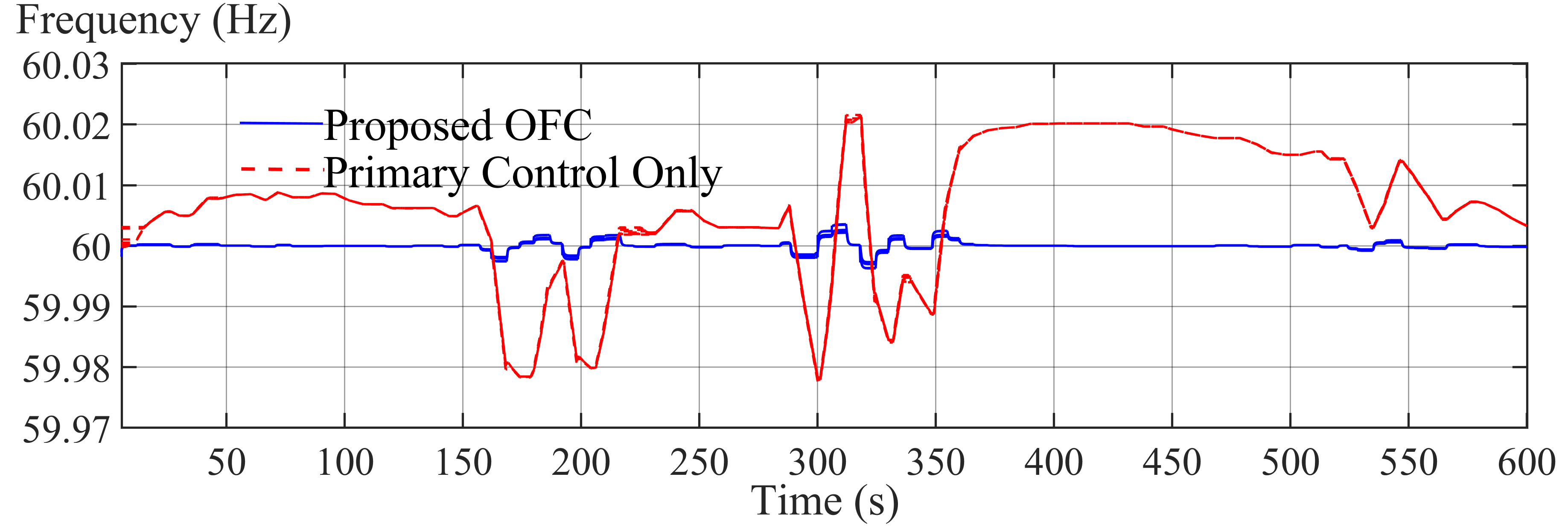}
    \caption{Frequency dynamics at IBRs' buses under continuous disturbance.}
      \label{fig:Frequency_Continous}
\end{figure}

We evaluate the performance of the proposed algorithm under continuously fluctuating power disturbances, which are simulated by the time-varying PV generation at bus 4. As shown in Figure~\ref{fig:9}, the PV generation profile is based on real-world data collected at a solar farm located in Texas, USA, and interpolated to match the simulation time resolution of 10~$\mu$s. Figure~\ref{fig:Frequency_Continous} presents a 10-minute simulation comparison of frequency dynamics under the proposed OFC algorithm and under primary control only.
The results demonstrate that the proposed algorithm effectively maintains the system frequency close to 60~Hz, despite the presence of substantial power fluctuations, in contrast to the case with primary control only.

\subsection{Control Coordination with Synchronous Generators} \label{section:IBRandSG}

In this section, we replace GFM-type IBR10 with a SG. The model of synchronous generator is the full-order IEEE standard model \cite{institute2003ieee}. The P-f droop control is used for the governor, and constant voltage control is used for the exciter. The proposed fully-local frequency control algorithm is implemented only for the IBRs (IBR 1-9). As for the secondary frequency control of SG, the area control error (ACE) is monitored. When the secondary frequency control capacity is insufficient, the SG will adjust the power setpoint of the governor based on the ACE. The PI controller is used for the AGC to eliminate the ACE. A ramping constraint is added to limit the mechanical power ramping (0.06 MW/s).

\begin{figure}[ht]
    \centering
    \includegraphics[width=1\linewidth]{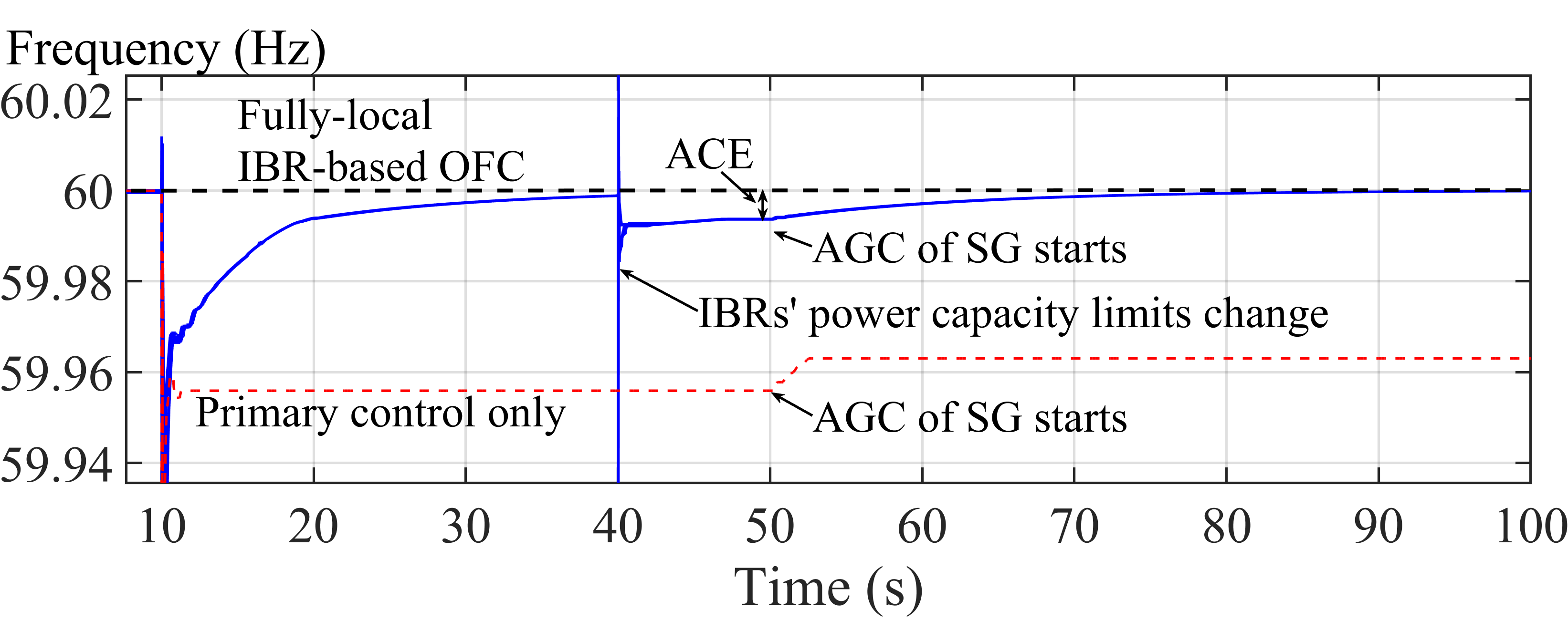}
    \caption{Frequency dynamics of buses connected with IBRs and SG.}
    \label{fig:frequencyIBRandSG}
\end{figure}

\begin{figure}[ht]
    \vspace{-17pt}
    \centering \includegraphics[width=1\linewidth]{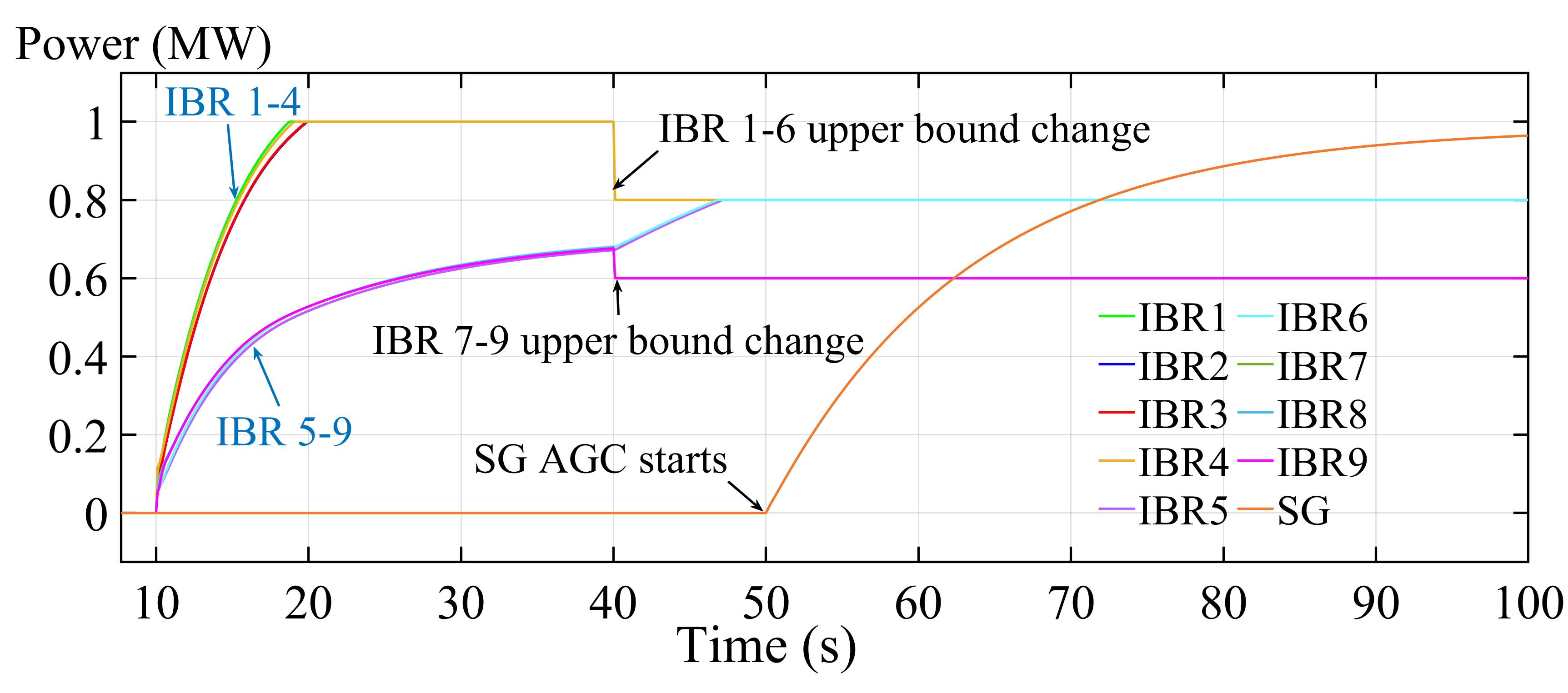}
    \caption{Power adjustments of IBRs and SG.}
    \label{fig:poweradjustmentIBRandSG}
\end{figure}

\begin{figure}[ht]
    \centering
    \vspace{-15pt}
    \includegraphics[width=1\linewidth]{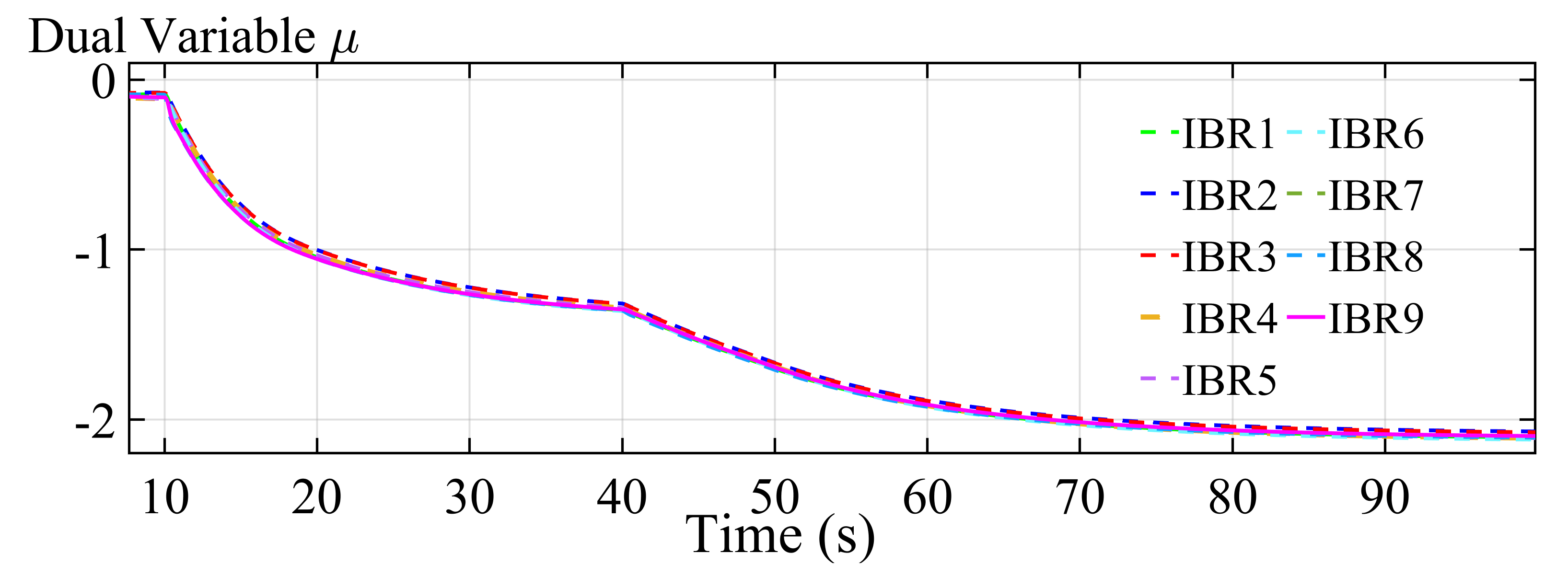}
    \caption{Dynamics of dual variables $\mu$. (Physical interpretation: the penalty coefficient on the frequency deviation and variation).}
    \label{fig:dual_mu}
\end{figure}

At $t=10s$, the PV at bus 4 experiences an 8 MW step decrease in power output. The frequency of IBRs' and SG's buses are shown in Figure \ref{fig:frequencyIBRandSG}. The power setpoint adjustments of IBR and SG are shown in Figure \ref{fig:poweradjustmentIBRandSG}. IBRs 1–9 implement the proposed secondary optimal frequency control algorithm.  At the beginning, the primary and secondary control of IBRs 1–9 and the primary control of the synchronous generator act together. The AGC of the synchronous generator has not yet been activated. IBR 1-4 with a lower cost coefficient will contribute more until they reach their limits of 3 MW (which means the power adjustment limits are 1MW). When $t=40s$, the frequency is regulated to near 60 Hz.

At $t=40s$, the upper bound of IBR 1-6 (GFL) are changed from 3 MW to be 2.8 MW (The secondary frequency control margin is reduced from 1 MW to 0.8 MW relative to the initial value). The upper bound of IBR 7-9 (GFM) is changed from 4 MW to 3.2 MW (The secondary frequency control margin is reduced from 1.4 MW to 0.6 MW relative to the initial value). The projection function in the proposed algorithm immediately adjusts the power setpoints into the feasible region. At this time, the capacity of all IBRs participating in secondary frequency control is no longer sufficient, and the frequency cannot be restored to 60 Hz, and there is a 0.01 Hz ACE.

At $t=50s$, the SG starts its AGC, and the ACE is input to its PI-based AGC. The mechanical power setpoint of the synchronous generator’s governor will gradually increase according to the signal from AGC until the frequency is restored to 60 Hz. 
 Besides, the dotted red curve in Figure \ref{fig:frequencyIBRandSG} shows the frequency dynamics without the proposed OFC in IBRs. Before $t=50s$, only primary control of IBRs and SG act, and there is a large ACE 0.044Hz. After $t=50s$, the AGC of the synchronous generator acts to restore the frequency; however, even after reaching the generator’s power limit, the frequency cannot be fully restored. 

\begin{figure}[ht]
    \centering
    \includegraphics[width=1\linewidth]{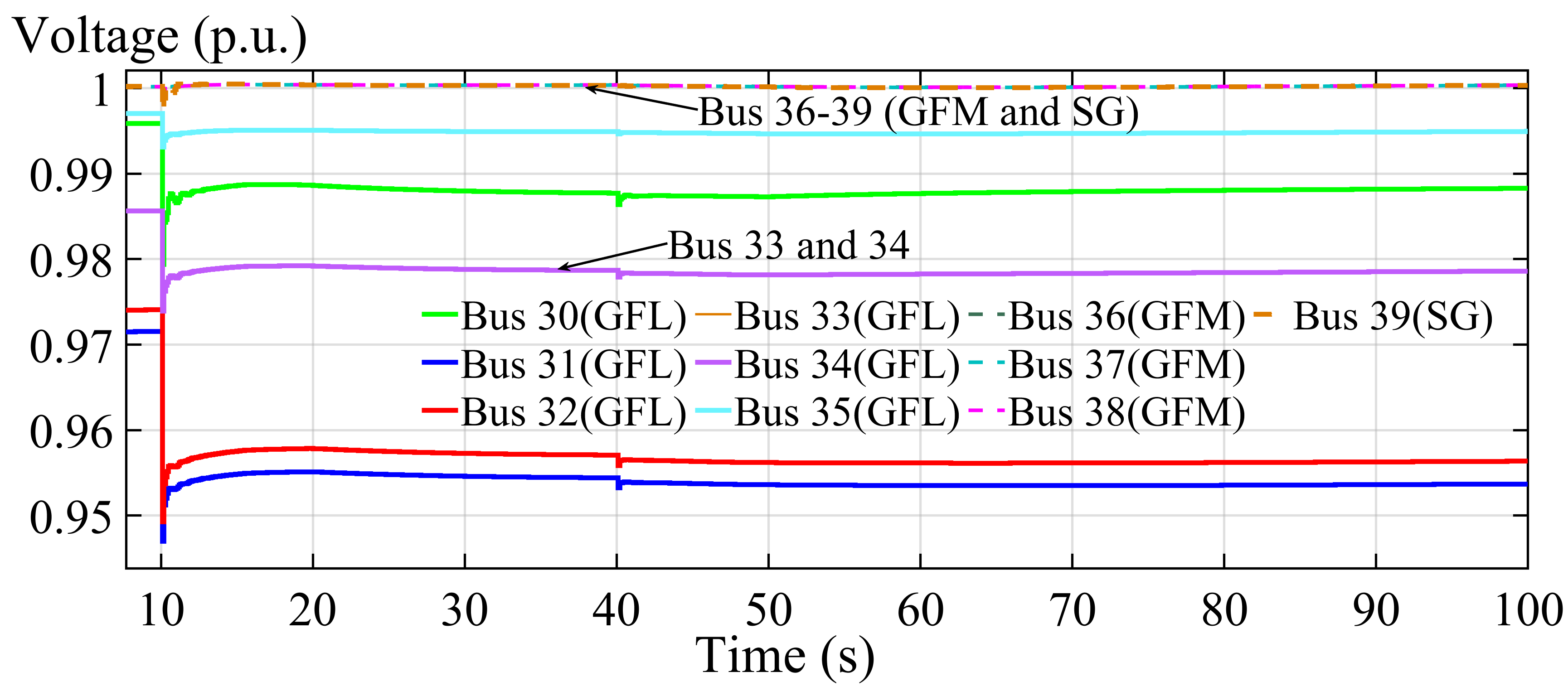}
    \caption{Bus voltages of GFL, GFM, and synchronous generator buses.}
    \label{fig:busvoltage}
\end{figure}

The dual variables $\mu_i$ as mentioned in \eqref{eq:mu_local} of each controller are shown in Figure \ref{fig:dual_mu}. The dual variable $\mu$ represents the penalty on the frequency deviation $k_i \omega_i$ and the frequency  variation $\frac{k_i}{\beta_i}\dot{\omega}_i$, as shown in~\eqref{eq:Lagrangian} and \eqref{eq:mu_local}. Accordingly, $\mu_i$ continues to increase in response to frequency deviation until frequency restoration is achieved, at which point the complementary slackness conditions \eqref{eq:kktmu} are satisfied:
\begin{subequations} \label{eq:kktmu}
  \begin{align}
\mu_i^{\tM *}\!\left(\frac{k_i^{\tM}}{\beta_i}\dot{\omega}_i^* + k_i^{\tM} \omega_i^*\right) &= 0, 
&& i \in \mathcal{N}_{\tM}, \label{eq:Mw+w=0GFM}\\
\mu_i^{\tL*}\!\left(k_i^{\tL} \omega_i^*\right) &= 0, 
&& i \in \mathcal{N}_{\tL}.
\end{align}  
\end{subequations}
The bus voltage of IBRs and SG are shown in Figure \ref{fig:busvoltage}. Because GFMs and the synchronous generator employ constant-voltage control, while GFLs adopt constant reactive-power control, the bus voltages connected to GFMs and SG (Bus 36-39) exhibit almost no fluctuation and are rapidly regulated to the reference value of 1~p.u. In contrast, the voltages at GFL buses (Bus 30-35) decrease due to the increase in active-power transfer. After the IBR-based secondary frequency control is activated at $t = 10\,\text{s}$, the power imbalance is redistributed among the IBRs, which facilitates voltage recovery. 

Through this case study, we demonstrate the coordination between the proposed IBR-based OFC and the synchronous generator’s AGC. The proposed IBR-based optimal secondary frequency control can rapidly restore the frequency in a cost-optimal manner, and once the IBR capacities reach their limits, the synchronous generator’s AGC restores the frequency based on the ACE.
\subsection{Impact of Measurement Noise}

\begin{figure}[hb]
    \centering
    \begin{minipage}{0.48\linewidth}
        \centering
        \includegraphics[width=\linewidth]{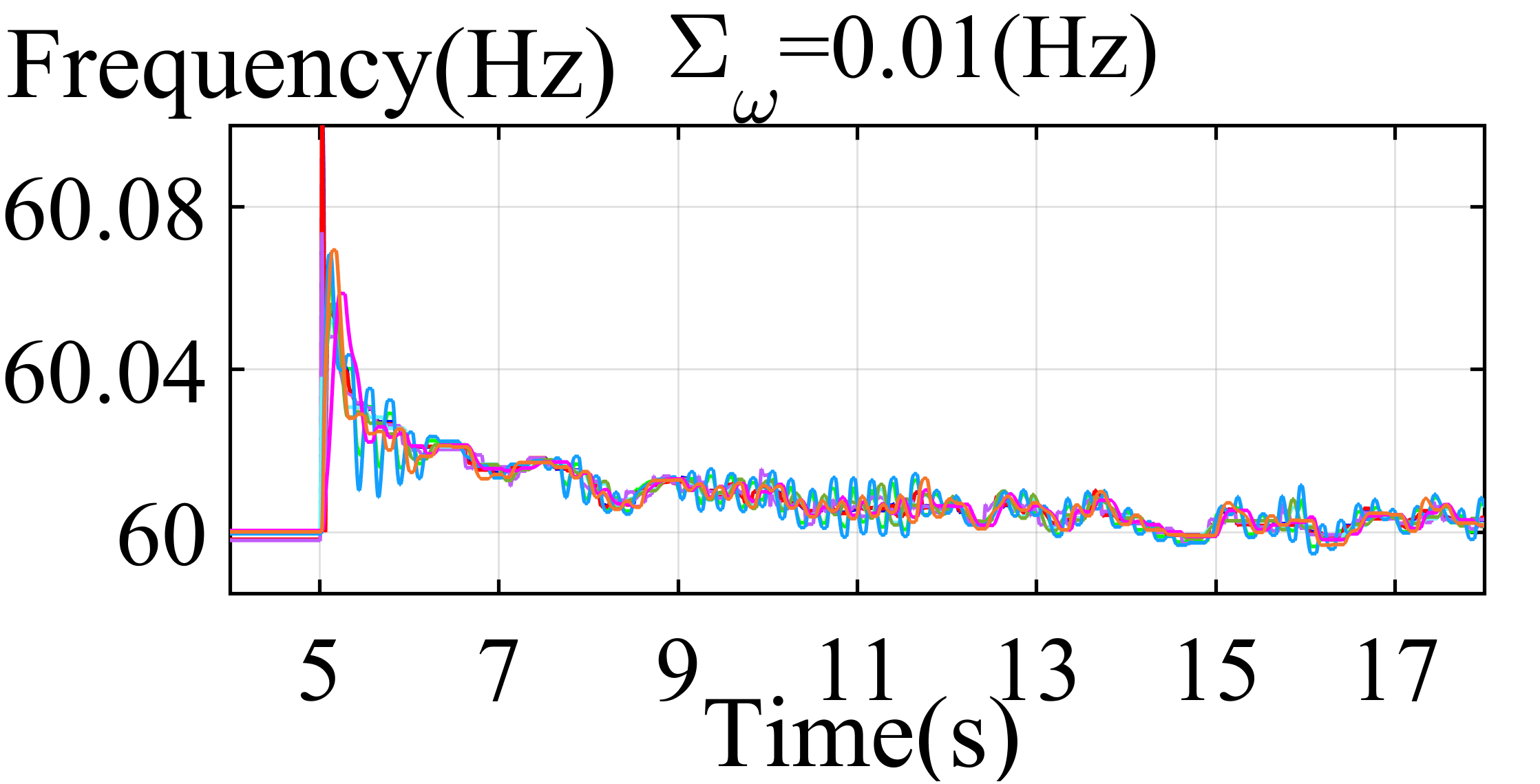}
    \end{minipage}
    \hfill
    \begin{minipage}{0.48\linewidth}
        \centering
        \includegraphics[width=\linewidth]{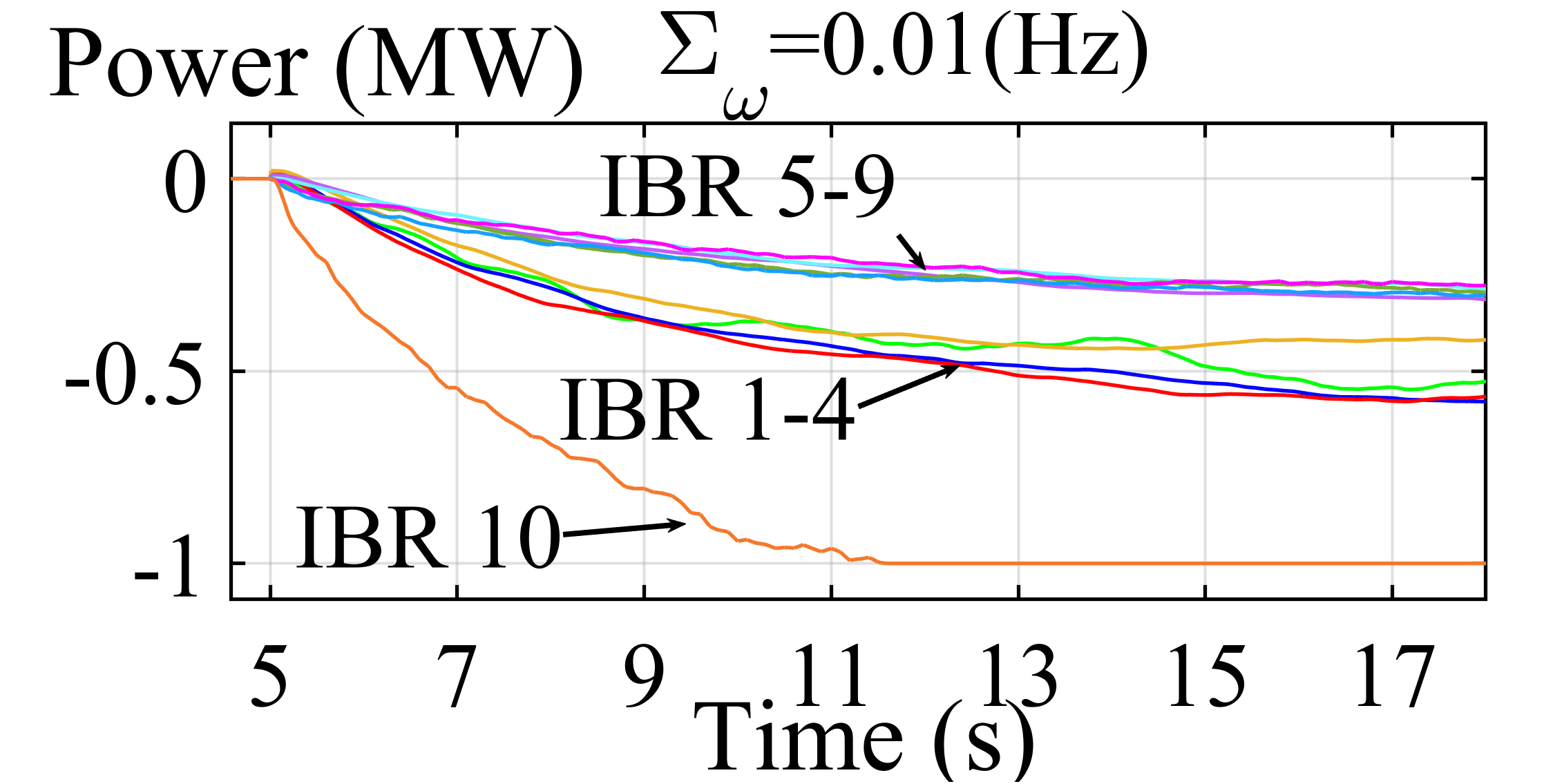}
    \end{minipage}

    \vspace{10pt}

    \begin{minipage}{0.48\linewidth}
        \centering
        \includegraphics[width=\linewidth]{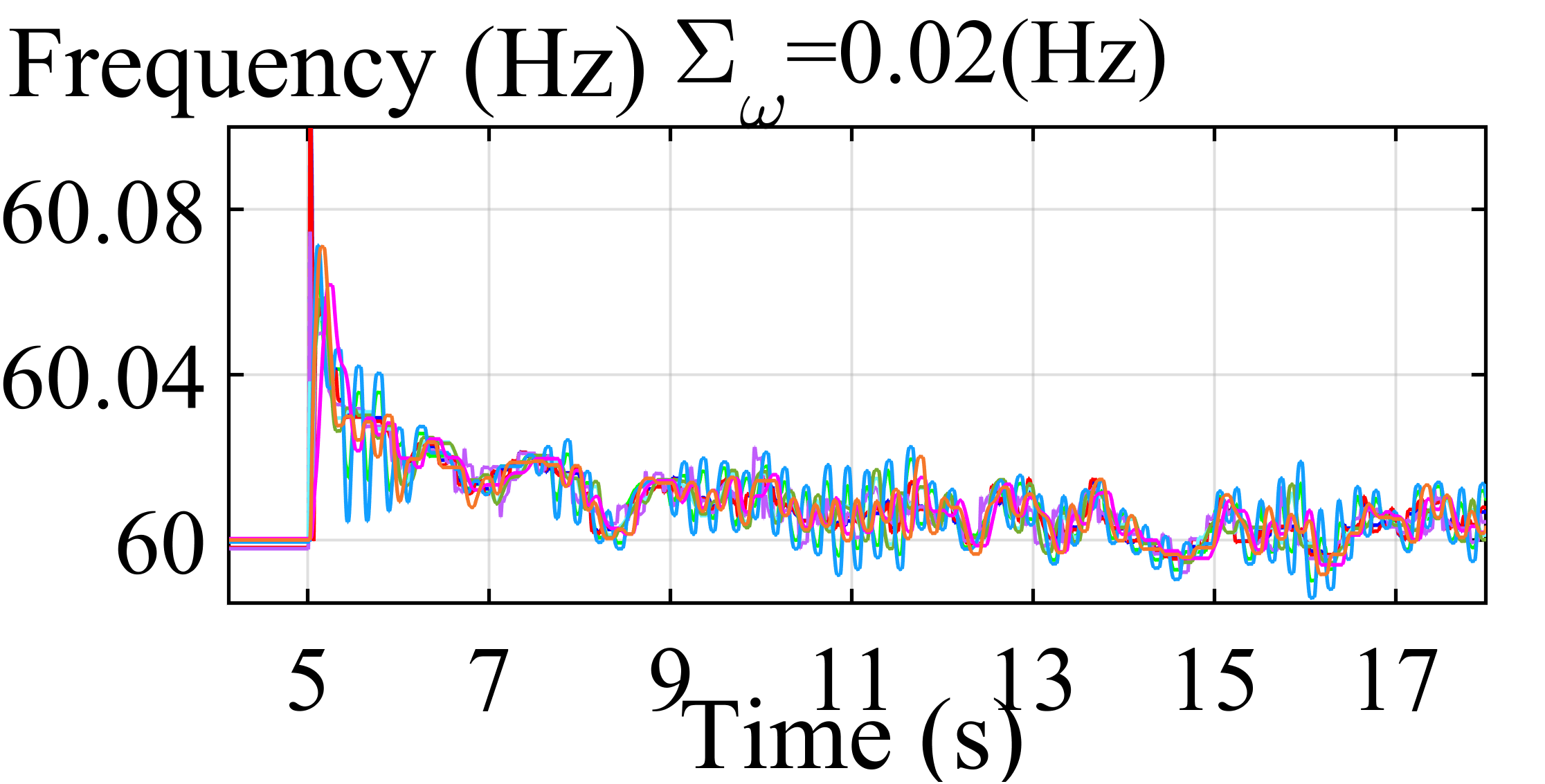}
    \end{minipage}
    \hfill
    \begin{minipage}{0.48\linewidth}
        \centering
        \includegraphics[width=\linewidth]{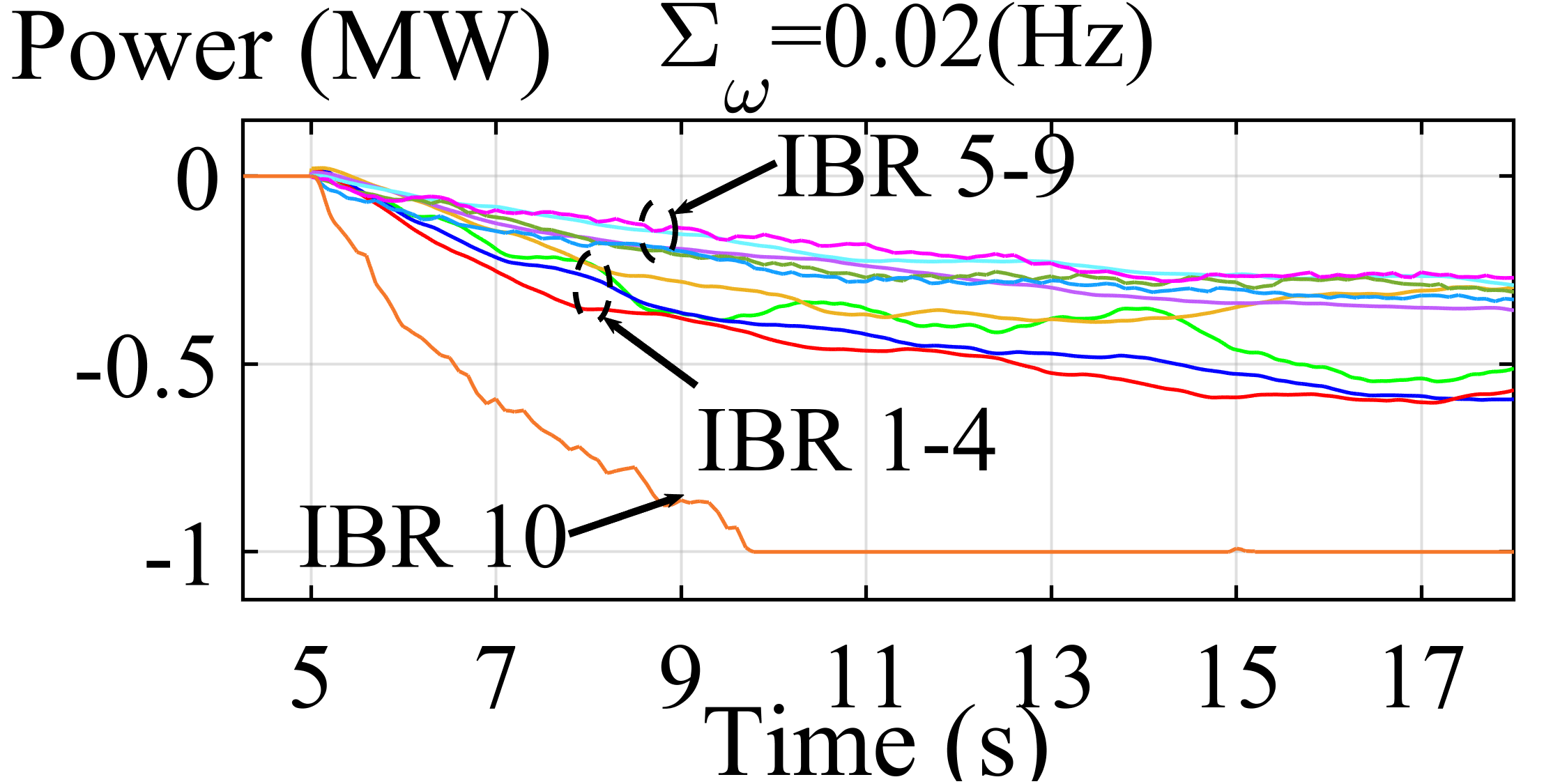}
    \end{minipage}
   \vspace{5pt}
    \caption{System response under frequency measurement noise:
    $\Sigma_\omega = 0.01$~Hz (top row) and $\Sigma_\omega = 0.02$~Hz (bottom row).}
    \label{fig:noise_response}
\end{figure}

 We study the influence of measurement noise on the proposed optimal secondary frequency control method. 
For the proposed fully local frequency control in Section \ref{section:fullylocal}, the control method only requires the measurements of frequency deviation $\omega_i$ as shown in Figure \ref{fig:localcontroller}. Consider the noise $\xi_{\omega_i}$ in the measurement $\omega_i$  
distribution $\mathcal{N}(0, \Sigma_\omega^2)$. With the same step-increasing PV power disturbance considered in Section \ref{section:fulltloct_sim}, the standard deviation of the frequency measurement noise is set to $\Sigma_\omega = 0.01$ Hz and $0.02$ Hz, respectively. The resulting bus frequency responses and IBR power setpoint adjustments under measurement noise are shown in Figure \ref{fig:noise_response}. With larger frequency measurement noise, the system frequency can still be restored close to 60 Hz, albeit with increased oscillations. The IBR power outputs are continuously modulated in response to the measurement noise while following the optimal cost allocation. As the noise magnitude increases, larger deviations from the optimal cost are observed. However, the measurement noise does not affect the hard power capacity constraints of the IBRs. For example, IBR 10 in Figure \ref{fig:noise_response} always satisfies its capacity limits, as the projection operator guarantees that the power setpoints remain within the feasible region.

\subsection{Impact of Communication Delay}
For the proposed distributed OFC algorithm, the dual variables $\nu$ and the virtual phase $\psi$ are exchanged among IBRs through communication links. In the presence of communication delays, the regulation of line thermal constraints requires a longer time to converge to feasibility. To model this effect, a delayed low-pass filter is applied to the communicated variables, with time constants of $\tau = 0.2 \mathrm{s}$ and $1 \mathrm{s}$. All other settings are identical to those of the test case for the proposed distributed OFC method in Section~\ref{section:communication}.

\begin{figure}[ht]
    \centering
    \includegraphics[width=1\linewidth]{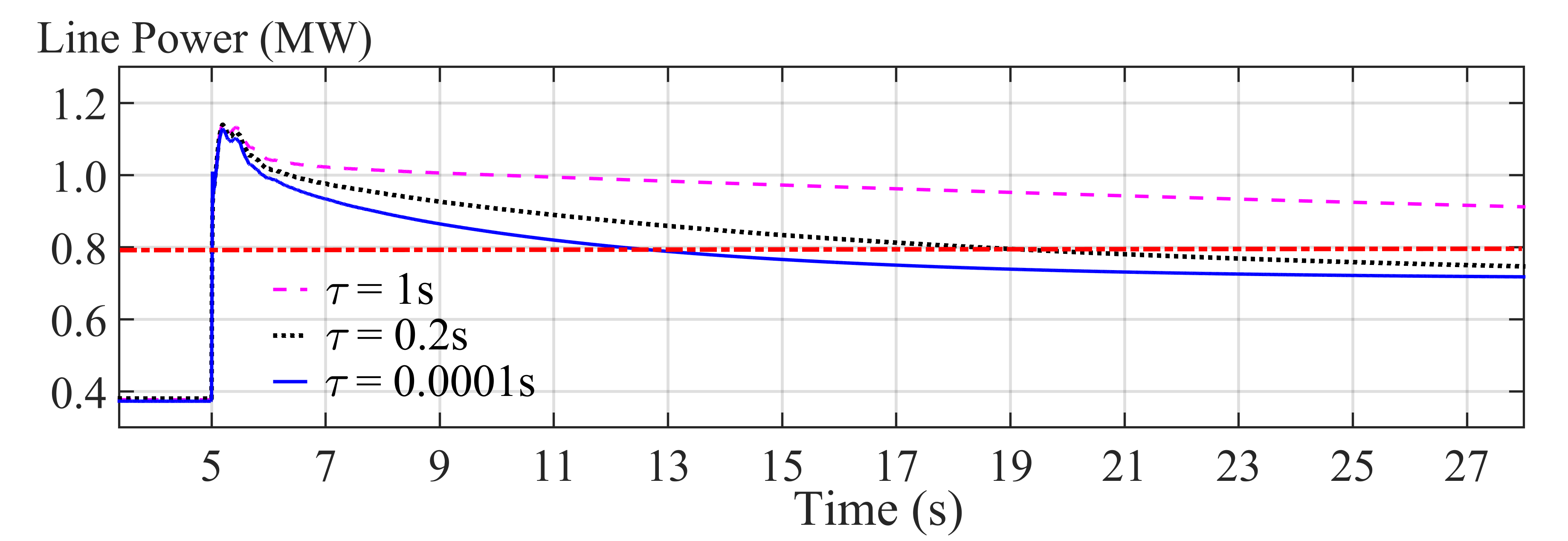}
    \caption{Line power flow control affected by communication delay.} 
    \label{fig:communicationdelay}
\end{figure}

The line power flow $P_{3,18}$ on line 3–18 is shown in Figure~\ref{fig:communicationdelay}. The blue curve corresponds to the baseline case in Section~\ref{section:communication}, where $\tau = 0.0001 \mathrm{s}$, equal to the simulation time step. It is observed that as the communication delay increases, the transient period during which the line power flow exceeds its thermal limit becomes longer, and consequently, more time is required to regulate the line power flow back within its upper and lower bounds. Since line thermal constraints do not require instantaneous enforcement and can tolerate temporary violations, the proposed algorithm remains effective under reasonable communication delays.
\subsection{Impact of Damping Coefficients Uncertainties}

\begin{figure}[h] \centering \begin{minipage}{0.48\linewidth} \centering \includegraphics[width=\linewidth]{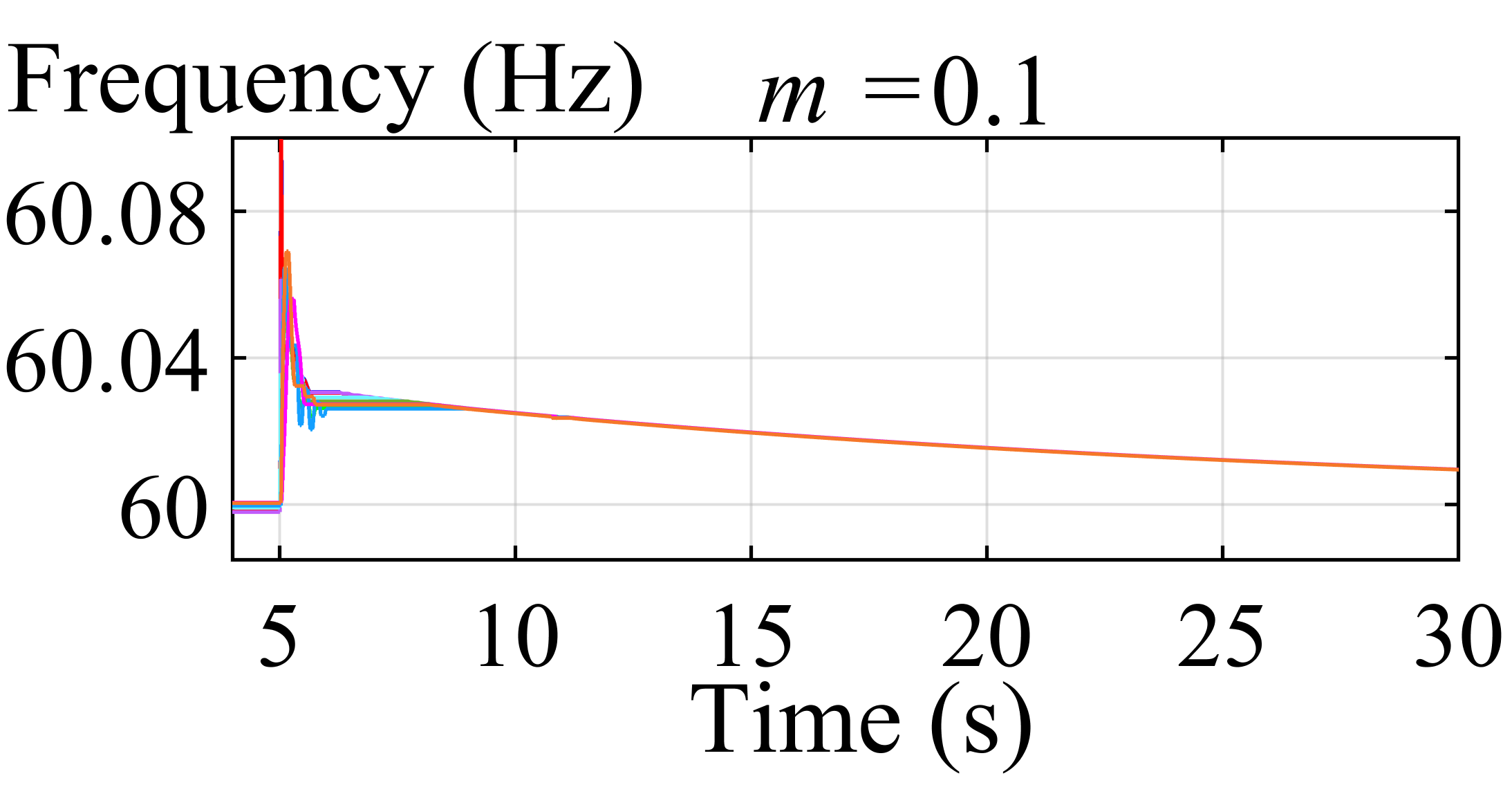} \end{minipage} \hfill \begin{minipage}{0.48\linewidth} \centering \includegraphics[width=\linewidth]{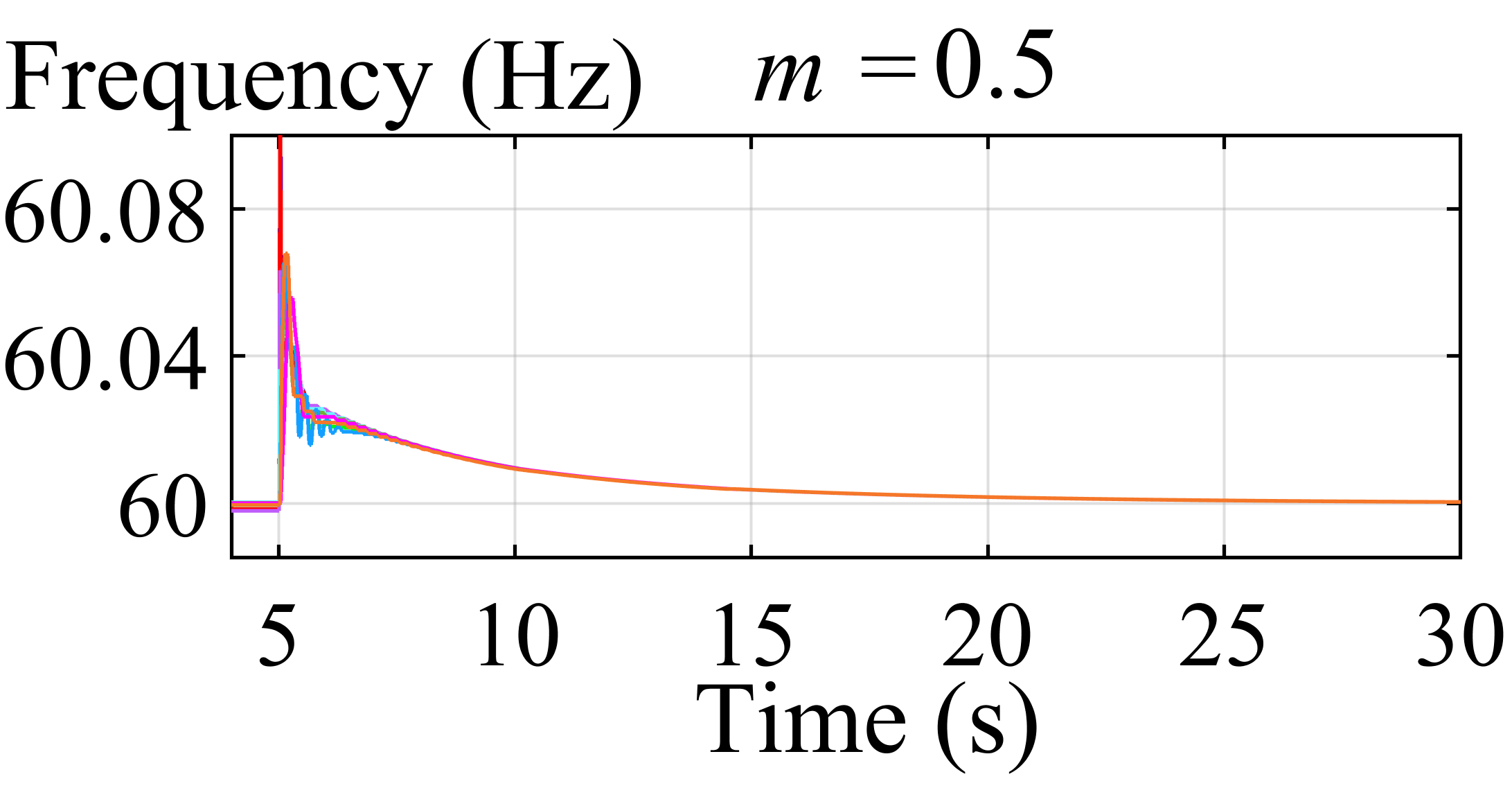} \end{minipage} 
\vspace{15pt}
\begin{minipage}{0.48\linewidth} \centering \includegraphics[width=\linewidth]{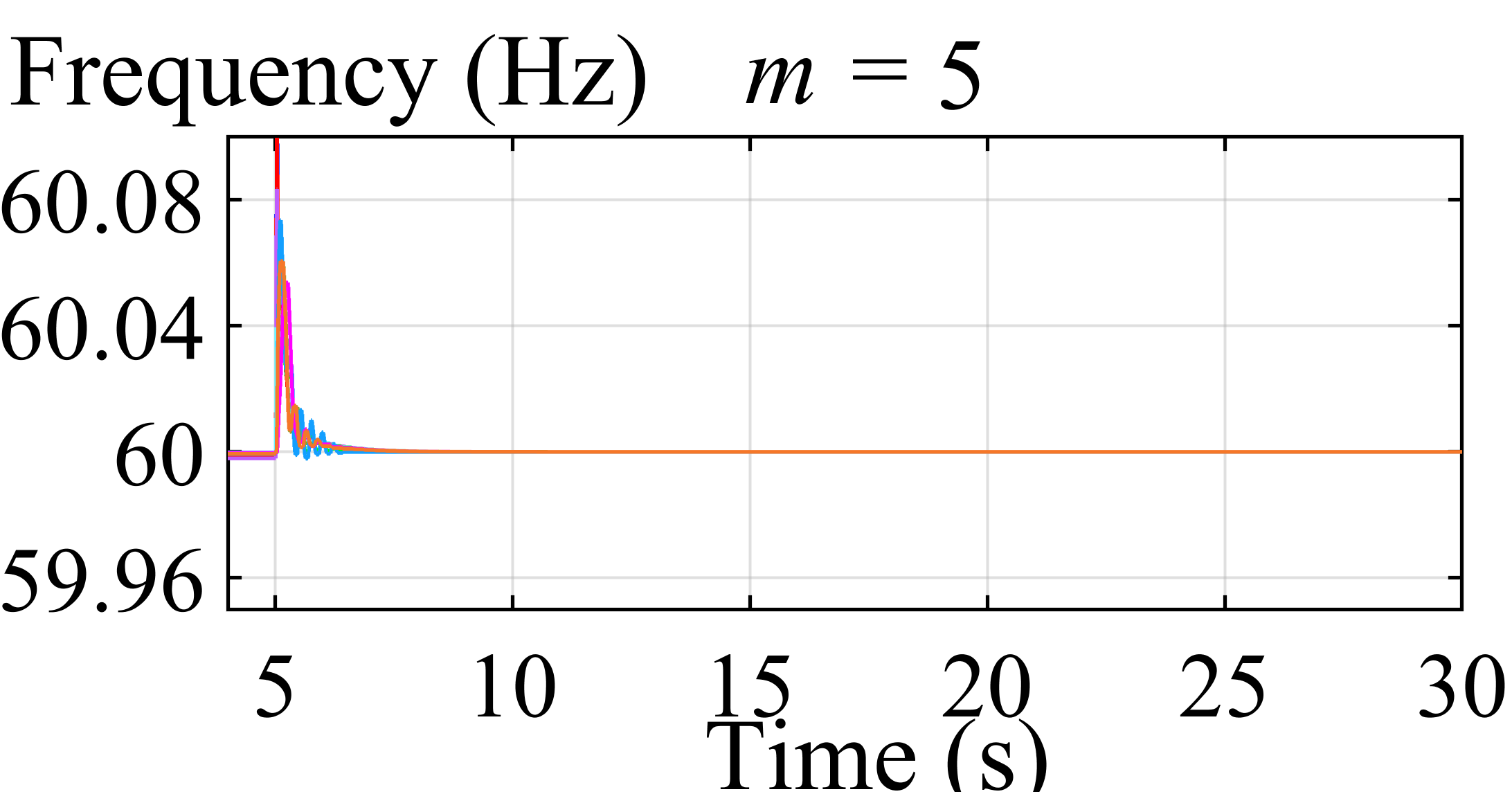} \end{minipage} \hfill \begin{minipage}{0.48\linewidth} \centering \includegraphics[width=\linewidth]{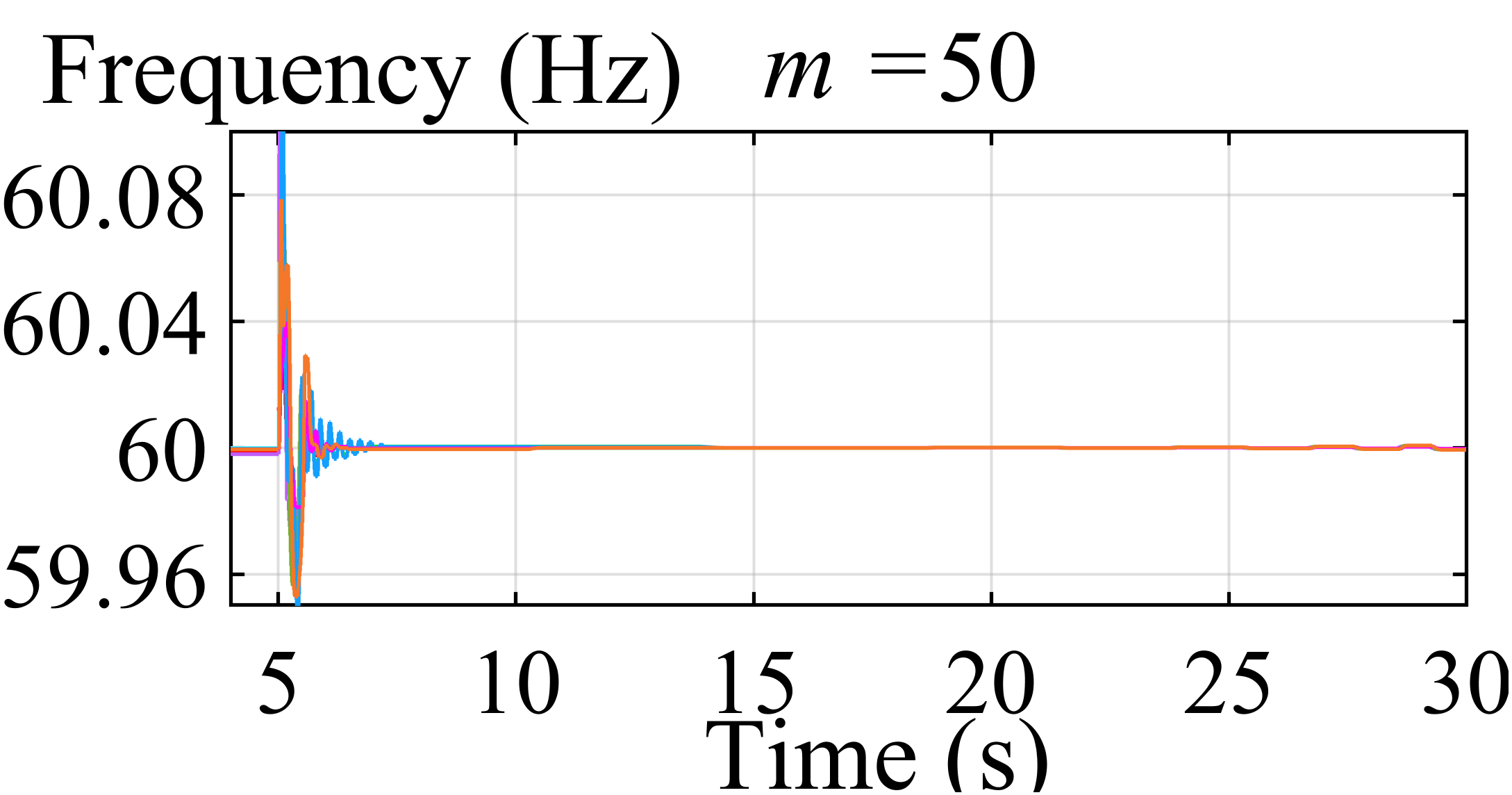} \end{minipage} 

\caption{Frequency dynamics under the impact of damping coefficients $k_i$.}\label{fig:derror_comparison} \end{figure}

Consider the inaccurate damping coefficient $k_i$ in the proposed fully local OFC method. Let the inaccurate damping coefficient used by the controller ${\tilde k_i}$ be $m$ times the real value. Use the step power change case of Section \ref{section:fulltloct_sim}. We tune $m=$ 0.1, 0.5, 5, and 50 to see the effect of inaccurate parameters. The frequency dynamics are shown in Figure \ref{fig:derror_comparison}. When the damping coefficient is estimated to be smaller, the frequency restoration becomes slower. This can be explained from the control block diagram in Figure \ref{fig:localcontroller}, where the damping coefficient determines the penalty applied to the frequency deviation. When the damping coefficient $k_i$ is zero, the proposed algorithm ceases to operate, and secondary frequency control is disabled.
Conversely, a larger damping coefficient enables faster frequency recovery; however, an excessively large damping value leads to frequency oscillations, as it imposes an overly strong penalty on the frequency deviation. When the damping coefficient lies within a reasonable range, e.g., $k_i = 0.5$, the proposed algorithm remains stable and effective. Moreover, the impact of inaccurate damping estimation can be mitigated, and the convergence speed can be regulated, by tuning the control time constant $\epsilon_{\mu_i}$.

\subsection{Time-Varying Power Capacity Limits of IBRs} \label{section:Timevaryinghupper}
\begin{figure}[ht]
    \centering
    \includegraphics[width=1\linewidth]{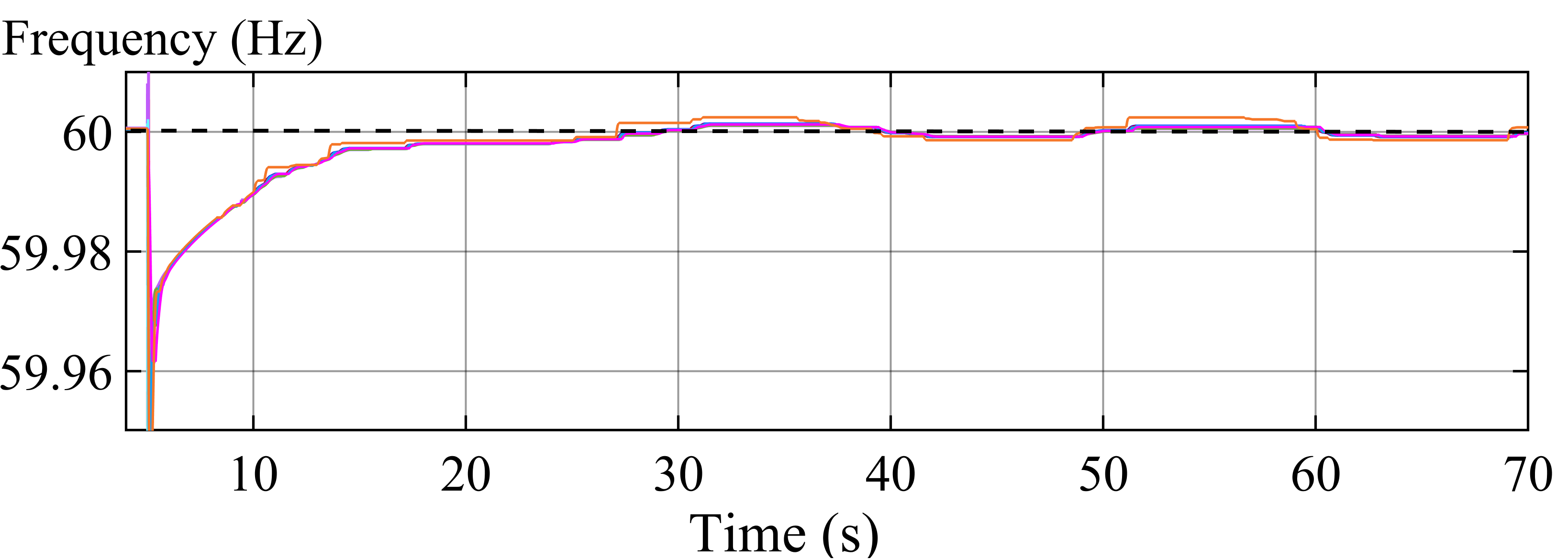}
    \caption{Frequency dynamics under step-increasing power disturbance and time-varying power capacity limits of IBR 10.}
    \label{fig:fre_dynamicupper}
\end{figure}
\begin{figure}[ht]
    \centering
    \includegraphics[width=1\linewidth]{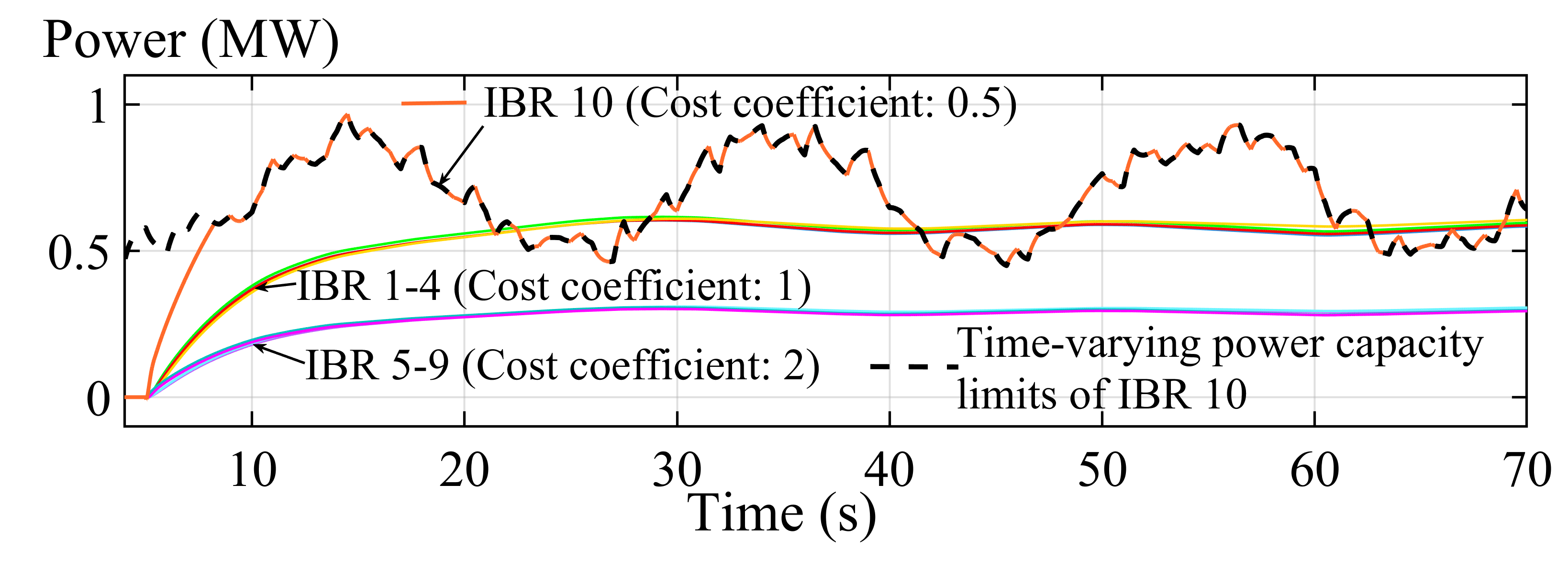}
    \caption{Power adjustments of all IBRs under step-increasing power disturbance and time-varying power capacity limits of IBR 10.} 
    \label{fig:power_dynamicupper}
\end{figure}

We evaluate the applicability of the proposed algorithm under continuously time-varying IBR power capacity limits. Specifically, a continuously time-varying upper-bound power capacity constraint is imposed on IBR 10, as illustrated by the dotted black curve in Figure \ref{fig:power_dynamicupper}.
Under a 5 MW step decrease in PV power, the system frequency response regulated by the proposed fully local OFC algorithm is shown in Figure \ref{fig:fre_dynamicupper}. It is observed that the frequency deviation caused by the power reduction is rapidly restored. The power adjustments of IBRs 1–10 are depicted in Figure \ref{fig:power_dynamicupper}, where the power output of IBR 10 is strictly constrained within its time-varying capacity limit throughout the entire transient. For IBRs 1–9, whose capacity limits are not reached, the power imbalance is allocated according to their respective cost coefficients. As the upper-bound capacity of the lowest-cost IBR 10 varies over time, the power outputs of the remaining IBRs adjust correspondingly to maintain system frequency.
These results demonstrate that the proposed algorithm based on projected primal-dual gradient dynamics can strictly and instantaneously enforce time-varying power constraints, owing to the real-time projection applied to the secondary frequency control power commands.

%% file: Conclusion.tex
\section{Conclusion}
This paper presents a fully distributed control algorithm for coordinating large-scale GFM and GFL IBRs to achieve grid-level optimal frequency control. By leveraging the structure of the primary control dynamics of IBRs, we interpret the projected primal-dual gradient dynamics for solving an optimal frequency control problem as a combination of the physical grid dynamics and the proposed control mechanism. This design outsources substantial computations to the physical system and 
enables the distributed implementation that only needs local measurement and local communication with neighbors. Furthermore, when line thermal constraints are not considered, the algorithm supports the fully local implementation without any communication while still ensuring control optimality. Using Lyapunov stability analysis, we prove that the proposed algorithm globally asymptotically converges to the optimal solution. High-fidelity EMT simulations validate the algorithm’s effectiveness in maintaining nominal frequency under step and continuous power disturbances.

%% file: Appendix.tex
\appendix
\section{Appendix: Lemmas and Proofs} \label{Appendix A}
\subsection{Proof of Proposition \ref{proposition:eq=saddle}} \label{proof:proposition:eq=saddle}

According to \cite[Theorem 3.25]{Ruszczynski2011}, the KKT conditions of the saddle point problem \eqref{saddle point problem} are given by:
     \begin{subequations} \label{KKT}
         \begin{align}
         & \bm{0} \in \nabla c(\Vx^*) \!+\! {\Vlambda^*}^\top \nabla \bm{h}(\Vx^*) \!+\!{\bm{\sigma}^*}^\top \nabla \bm{g}(\Vx^*)\!+\! N_{\sX}(\Vx^*),\\
         &\Vx^*\in {\sX}, {\Vlambda^*} \!\in\! \R^l, {\bm{\sigma}^*} \!\in\! \R^m_{\geq 0}, \bm{g}(\Vx^*)\le \boldsymbol{0}, \bm{h}(\Vx^*)=0,\\
         & {\bm{\sigma}^*}^\top \bm{g}(\Vx^*)=0,
         \end{align}
     \end{subequations}
     where $N_{\sX}(\Vx^*)$ is the normal cone of ${\sX}$ at $\Vx^*$.
     By definition, the equilibrium points of dynamics \eqref{dynamic} satisfy \eqref{equilibrium}:
     \begin{subequations}\label{equilibrium}
   \begin{align}
   0&=\epsilon_x[{\text{Proj}}_{\sX}(\boldsymbol{x}-\alpha_{x} \nabla_{\boldsymbol{x}}\sL)-\boldsymbol{x}],\\
      0&=\epsilon_{\lambda}\alpha_{\lambda} \nabla_{\boldsymbol{\lambda}}\sL,\\  
   0&=\epsilon_{\sigma}[\text{Proj}_{\R_+^m}(\boldsymbol{\sigma}+\alpha_{\sigma} \nabla_{\boldsymbol{\sigma}}\sL)-\boldsymbol{\sigma}].
\end{align}
\end{subequations}
 The solution of \eqref{equilibrium} is the same as \eqref{KKT}, and thus Proposition \ref{proposition:eq=saddle} is proved.

\subsection{Proof of Lemma \ref{asconvergence_final}} \label{Proofdvdt0eqequil}
We have the following property for the projected primal-dual gradient dynamics \eqref{dynamicz} as shown in Proposition \ref{Proposition:dynamic_equilibriumpoint}.
\begin{proposition} \label{Proposition:dynamic_equilibriumpoint} \cite{Gao2003Mar}
$\bz^* \in \sZ$ is an equilibrium point of dynamics \eqref{dynamicz} if and only if $\bz^*$ is a solution to the variational inequality problem that finds $\bz^* \in { \sZ}$ such that
 $    (\bz-\bz^*)^\top \boldsymbol{\Omega}(\bz^*) \ge 0, \forall \bz \in \sZ  $.
\end{proposition}

From \eqref{eq:dvdtle0}, when $\frac{dV}{dt}=0$,  we must have $ (\bz-\bz^*)^\top{\bm{\Omega}}(\bz)=0 $. Since ${\bm{\Omega}}(\bz)$ is a  monotone gradient mapping, by \cite[Definition 4.1]{Bianchi2000Jan}, there exists a positive function $k(\bm{x},\bm{y})$ on $\sZ \times \sZ$ such that, for all $\bx, \by \in \sZ$:
\begin{align}
     (\by-\bx)^\top\bm{\Omega}(\bx)=0 \Longrightarrow \bm{ \Omega}(\bm{x})=k(\bm{x,y})\bm{\Omega}(\bm{y}).
\end{align}
Thus, $ (\bz-\bz^*)^\top{\bm{\Omega}}(\bz)=0 $ implies that there exists a positive function $k(\bz,\bz^*)$ such that ${\bm{\Omega}}(\bz)=k(\bz,\bz^*){\bm{\Omega}}(\bz^*)$. Hence,  we can obtain that 
 for any $\hat{\bz} \in \sZ$:
\begin{align}
  \hspace{-15pt}    (\hat{\bz} -\bz)^\top\bm{\Omega}(\bz)&=   (\hat{\bz} -\bz^*)^\top\bm{\Omega}(\bz) +  (\bz^* -\bz)^\top\bm{\Omega}(\bz) \nonumber \\
  &= k(\bz,\bz^*) (\hat{\bz} -\bz^*)^\top {\bm{\Omega}}(\bz^*) + 0\ge 0 .
\end{align}

According to Proposition \ref{Proposition:dynamic_equilibriumpoint}, when $\frac{dV(\bz)}{dt}=0$, $\bz$ is also an equilibrium point of dynamics \ref{dynamicz}.
Thus, combining with Proposition \ref{proposition:eq=saddle}, any point in the largest invariant set $M$   is an optimal solution of the saddle point problem \eqref{saddle point problem}. Thus, $\Bar{\bz} \in M \subseteq B \subseteq {\sZ}^* $.

Due to the boundedness of $\sS(\bz(t_0))$ in \eqref{eq:bound}, $\{\bz(t)|t\ge t_0\}$ is also bounded. Thus, there exists a limit point $\Bar{\bz}$ and a time sequence $\{t_n\}$ with $t_0<t_1<t_2<...<t_n<...$ and $t_n \to +\infty$ as $n \to + \infty$ \cite[Theorem~4.4, Proof]{khalil2002nonlinear} such that:
\begin{align} \label{sequence conve}
    \lim_{n \to + \infty} \bz(t_n)= \Bar{\bz},
\end{align}
where $ \Bar{\bz} \in {\sZ} $ is an ${{\omega}}$-limit point of $\{ \bz(t)| t \ge t_0 \}$. 
Next, we prove that dynamics \eqref{dynamicz} converges to one fixed point in the optimal solution set of \eqref{saddle point problem}.

Similar to the definition of $V(\bz)$, we define 
  \begin{align}
   \hspace{-5pt}   \Bar{  V}(\bz)=&\frac{1}{2}||\bz\!-\!\Bar{\bz}||^2 \!+\!  \alpha(\bz\!-\!\bP_{{\sZ}}(\by)){\bm{\Omega}}(\bz)\!-\!\frac{1}{2}||\bz\!-\!\bP_{{\sZ}}(\by)||^2.
    \end{align}
    Then, similar to the proof of { Lemma \ref{Property_Ly}}, one can conclude that $\Bar{V}(\bz) \ge \frac{1}{2}||\bz-\Bar{\bz}||^2$ and $\Bar{V}(\bz)$ is monotonically non-increasing on $t \in[t_0, + \infty)$. By the continuity of function $\Bar{V}(\bz)$, it follows that $\forall \epsilon >0, \exists \delta>0$, such that
       \begin{align}  \label{Vcontinous}
           \Bar{V}(\bz)\le \frac{ \epsilon^2}{2},\quad  \text{if }  ||\bz-\Bar{\bz}|| \le \delta.
       \end{align}
       From \eqref{sequence conve} and \eqref{Vcontinous}, there exists a natural number $N$ such that:
\begin{align}
    ||\bz-\Bar{\bz}||^2 \le \epsilon^2, \text{when } t \ge t_N,
\end{align}
which leads to 
$\lim_{ t \to + \infty}\bz(t) =\Bar{\bz}$.  
This implies that the solution $\bz(t)$ of \eqref{dynamicz} converges to a fixed point in ${\sZ}^*$. Hence, for any $\bz(t_0)\in {\sZ}$,  there exists a unique continuous solution trajectory $\bz(t) \in {\sZ}$ and it will converge to a fixed saddle point of problem \eqref{saddle point problem}. Hence, Lemma  \ref{asconvergence_final} is proved.

%% file: IEEEabrv.bib
@STRING{IEEE_J_AC         = "{IEEE} Trans. Autom. Control"}

@STRING{IEEE_J_PWRS       = "{IEEE} Trans. Power Syst."}


%% file: mybibfile.bib
@article{Rebours2007Jan,
	author = {Rebours, Yann G. and Kirschen, Daniel S. and Trotignon, Marc and Rossignol, Sbastien},
	title = {{A Survey of Frequency and Voltage Control Ancillary Services{\ifmmode---\else\textemdash\fi}Part I: Technical Features}},
	journal = {IEEE Trans. Power Syst.},
	volume = {22},
	number = {1},
	pages = {350--357},
	year = {2007},
	month = jan,
	publisher = {IEEE},
	doi = {10.1109/TPWRS.2006.888963}
}

@article{Tang2019Mar,
  title={Semi-global exponential stability of augmented primal--dual gradient dynamics for constrained convex optimization},
  author={Tang, Yujie and Qu, Guannan and Li, Na},
  journal={Systems \& Control Letters},
  volume={144},
  pages={104754},
  year={2020},
  publisher={Elsevier}
}

@article{Du2020Aug,
	author = {Du, Wei and Tuffner, Francis K. and Schneider, Kevin P. and Lasseter, Robert H. and Xie, Jing and Chen, Zhe},
	title = {{Modeling of Grid-Forming and Grid-Following Inverters for Dynamic Simulation of Large-Scale Distribution Systems}},
	journal = {IEEE Trans. Power Delivery},
	volume = {36},
	number = {4},
	pages = {2035--2045},
	year = {2020},
	month = aug,
	publisher = {IEEE},
	doi = {10.1109/TPWRD.2020.3018647}
}

@article{Li2015Jul,
	author = {Li, Na and Zhao, Changhong and Chen, Lijun},
	title = {{Connecting Automatic Generation Control and Economic Dispatch From an Optimization View}},
	journal = {IEEE Trans. Control Network Syst.},
	volume = {3},
	number = {3},
	pages = {254--264},
	year = {2015},
	month = jul,
	publisher = {IEEE},
	doi = {10.1109/TCNS.2015.2459451}
}

@article{Zhao2014Jan,
	author = {Zhao, Changhong and Topcu, Ufuk and Li, Na and Low, Steven},
	title = {{Design and Stability of Load-Side Primary Frequency Control in Power Systems}},
	journal = {IEEE Trans. Autom. Control},
	volume = {59},
	number = {5},
	pages = {1177--1189},
	year = {2014},
	month = jan,
	publisher = {IEEE},
	doi = {10.1109/TAC.2014.2298140}
}

@article{Chen2020Sep,
	author = {Chen, Xin and Zhao, Changhong and Li, Na},
	title = {{Distributed Automatic Load Frequency Control With Optimality in Power Systems}},
	journal = {IEEE Trans. Control Network Syst.},
	volume = {8},
	number = {1},
	pages = {307--318},
	year = {2020},
	month = sep,
	publisher = {IEEE},
	doi = {10.1109/TCNS.2020.3024489}
}

@article{Mallada2017Jun,
	author = {Mallada, Enrique and Zhao, Changhong and Low, Steven},
	title = {{Optimal Load-Side Control for Frequency Regulation in Smart Grids}},
	journal = {IEEE Trans. Autom. Control},
	volume = {62},
	number = {12},
	pages = {6294--6309},
	year = {2017},
	month = jun,
	publisher = {IEEE},
	doi = {10.1109/TAC.2017.2713529}
}

@ARTICLE{10488734,
  author={Wang, Yifan and Liu, Shuai and Cao, Xianghui and Chow, Mo-Yuen},
  journal={IEEE Trans. Autom. Control}, 
  title={An Operator Splitting Scheme for Distributed Optimal Load-Side Frequency Control With Nonsmooth Cost Functions}, 
  year={2024},
  volume={69},
  number={9},
  pages={6442-6449},
  doi={10.1109/TAC.2024.3384341}}

@article{WANG2020104607,
title = {Distributed optimal load frequency control considering nonsmooth cost functions},
journal = {Syst. Control Lett.},
volume = {136},
pages = {104607},
year = {2020},
issn = {0167-6911},
doi = {https://doi.org/10.1016/j.sysconle.2019.104607},
author = {Zhaojian Wang and Feng Liu and Changhong Zhao and Zhiyuan Ma and Wei Wei}
}

@article{chen2017robust,
  title={Robust capacity assessment of distributed generation in unbalanced distribution networks incorporating ANM techniques},
  author={Chen, Xin and Wu, Wenchuan and Zhang, Boming},
  journal={IEEE Trans. Sustain. Energy},
  volume={9},
  number={2},
  pages={651--663},
  year={2017},
  publisher={IEEE}
}

@article{chen2019aggregate,
  title={Aggregate power flexibility in unbalanced distribution systems},
  author={Chen, Xin and Dall’Anese, Emiliano and Zhao, Changhong and Li, Na},
  journal={IEEE Trans. Smart Grid},
  volume={11},
  number={1},
  pages={258--269},
  year={2019},
  publisher={IEEE}
}

@inproceedings{Johnson2022Jan,
  title={A Generic Primary-control Model for Grid-forming Inverters: Towards Interoperable Operation \& Control.},
  author={Johnson, Brian B and Roberts, TG and Ajala, Olaoluwapo and Dom{\'\i}nguez-Garc{\'\i}a, Alejandro D and Dhople, Sairaj V and Ramasubramanian, Deepak and Tuohy, Aidan and Divan, Deepak and Kroposki, Benjamin},
  booktitle={HICSS},
  pages={1--10},
  year={2022}
}

@article{Li2022Nov,
	author = {Li, Zhongwen and Cheng, Zhiping and Liang, Jing and Si, Jikai},
	title = {{Distributed Cooperative AGC Method for New Power System With Heterogeneous Frequency Regulation Resources}},
	journal = {IEEE Trans. Power Syst.},
	volume = {38},
	number = {5},
	pages = {4928--4939},
	year = {2022},
	month = nov,
	publisher = {IEEE},
	doi = {10.1109/TPWRS.2022.3218583}
}

@article{Markovic2021Feb,
	author = {Markovic, Uros and Stanojev, Ognjen and Aristidou, Petros and Vrettos, Evangelos and Callaway, Duncan and Hug, Gabriela},
	title = {{Understanding Small-Signal Stability of Low-Inertia Systems}},
	journal = {IEEE Trans. Power Syst.},
	volume = {36},
	number = {5},
	pages = {3997--4017},
	year = {2021},
	month = feb,
	publisher = {IEEE},
	doi = {10.1109/TPWRS.2021.3061434}
}

@article{Pogaku2007Mar,
	author = {Pogaku, Nagaraju and Prodanovic, Milan and Green, Timothy C.},
	title = {{Modeling, Analysis and Testing of Autonomous Operation of an Inverter-Based Microgrid}},
	journal = {IEEE Trans. Power Electron.},
	volume = {22},
	number = {2},
	pages = {613--625},
	year = {2007},
	month = mar,
	publisher = {IEEE},
	doi = {10.1109/TPEL.2006.890003}
}

@article{Gao2003Mar,
	author = {Gao, Xing-Bao},
	title = {{Exponential stability of globally projected dynamic systems}},
	journal = {IEEE Trans. Neural Networks},
	volume = {14},
	number = {2},
	pages = {426--431},
	year = {2003},
	month = mar,
	publisher = {IEEE},
	doi = {10.1109/TNN.2003.809409}
}

@ARTICLE{9779512,
  author={Cui, Wenqi and Jiang, Yan and Zhang, Baosen},
  journal={IEEE Trans. Power Syst.}, 
  title={Reinforcement Learning for Optimal Primary Frequency Control: A Lyapunov Approach}, 
  year={2023},
  volume={38},
  number={2},
  pages={1676-1688},
  doi={10.1109/TPWRS.2022.3176525}}

@book{Ruszczynski2011,
  author    = {A. Ruszczynski},
  title     = {Nonlinear Optimization},
  publisher = {Princeton University Press},
  year      = {2011}
}

@misc{Slotine1991AppliedNC,
	author = {Slotine, J. and Li, Weiping},
	title = {{Applied Nonlinear Control}},
	year = {1991},
	note = {[Online; accessed 15. Apr. 2025]},
	url = {https://www.semanticscholar.org/paper/Applied-Nonlinear-Control-Slotine-Li/1ae0d9625f9f580a3b8d8e92a0edbc2087a1cc0e}
}

@article{doi:10.1137/1023111,
author = {Evans, Lawrence C.},
title = {An Introduction to Variational Inequalities and Their Applications (D. Kinderlehrer and G. Stampacchia)},
journal = {SIAM Review},
volume = {23},
number = {4},
pages = {539-543},
year = {1981},
doi = {10.1137/1023111}
}

@article{Bianchi2000Jan,
	author = {Bianchi, M. and Schaible, S.},
	title = {{An Extension of Pseudolinear Functions and Variational Inequality Problems}},
	journal = {J. Optim. Theory Appl.},
	volume = {104},
	number = {1},
	pages = {59--71},
	year = {2000},
	month = jan,
	issn = {1573-2878},
	publisher = {Kluwer Academic Publishers-Plenum Publishers},
	doi = {10.1023/A:1004672604998}
}

@article{Lyu2024Dec,
	author = {Lyu, Chenghao and Wang, Weiquan and Wang, Junyue and Bai, Yilin and Song, Zhengxiang and Wang, Wei and Meng, Jinhao},
	title = {{The role of co-optimization in trading off cost and frequency regulation service for industrial microgrids}},
	journal = {Appl. Energy},
	volume = {375},
	pages = {124131},
	year = {2024},
	month = dec,
	issn = {0306-2619},
	publisher = {Elsevier},
	doi = {10.1016/j.apenergy.2024.124131}
}

@article{Xia2000Jul,
	author = {Xia, Y. S. and Wang, J.},
	title = {{On the Stability of Globally Projected Dynamical Systems}},
	journal = {J. Optim. Theory Appl.},
	volume = {106},
	number = {1},
	pages = {129--150},
	year = {2000},
	month = jul,
	issn = {1573-2878},
	publisher = {Kluwer Academic Publishers-Plenum Publishers},
	doi = {10.1023/A:1004611224835}
}

@article{Wu2024Sep,
	author = {Wu, Chenyu and Wu, Zhi and Gu, Wei and Yi, Zhongkai and Xi, Chen and Shi, Zhengkun},
	title = {{Source-Load Collaborative Frequency Control With Real-Time Optimality in Power Networks}},
	journal = {IEEE Trans. Smart Grid},
	volume = {16},
	number = {1},
	pages = {164--172},
	year = {2024},
	month = sep,
	publisher = {IEEE},
	doi = {10.1109/TSG.2024.3446792}
}

@article{Xu2021Jun,
	author = {Xu, Yiqiao and Dong, Zhen and Li, Zhongguo and Liu, Yixing and Ding, Zhengtao},
	title = {{Distributed Optimization for Integrated Frequency Regulation and Economic Dispatch in Microgrids}},
	journal = {IEEE Trans. Smart Grid},
	volume = {12},
	number = {6},
	pages = {4595--4606},
	year = {2021},
	month = jun,
	publisher = {IEEE},
	doi = {10.1109/TSG.2021.3089421}
}

@article{Yi2016Dec,
	author = {Yi, Peng and Hong, Yiguang and Liu, Feng},
	title = {{Initialization-free distributed algorithms for optimal resource allocation with feasibility constraints and application to economic dispatch of power systems}},
	journal = {Automatica},
	volume = {74},
	pages = {259--269},
	year = {2016},
	month = dec,
	issn = {0005-1098},
	publisher = {Pergamon},
	doi = {10.1016/j.automatica.2016.08.007}
}

@article{Cherukuri2016Dec,
	author = {Cherukuri, Ashish and Cort{\ifmmode\acute{e}\else\'{e}\fi}s, Jorge},
	title = {{Initialization-free distributed coordination for economic dispatch under varying loads and generator commitment}},
	journal = {Automatica},
	volume = {74},
	pages = {183--193},
	year = {2016},
	month = dec,
	issn = {0005-1098},
	publisher = {Pergamon},
	doi = {10.1016/j.automatica.2016.07.003}
}

@article{Wang2018Aug,
	author = {Wang, Zhaojian and Liu, Feng and Pang, John Z. F. and Low, Steven H. and Mei, Shengwei},
	title = {{Distributed Optimal Frequency Control Considering a Nonlinear Network-Preserving Model}},
	journal = {IEEE Trans. Power Syst.},
	volume = {34},
	number = {1},
	pages = {76--86},
	year = {2018},
	month = aug,
	publisher = {IEEE},
	doi = {10.1109/TPWRS.2018.2861941}
}

@ARTICLE{9869334,
  author={Jiang, Yan and Cui, Wenqi and Zhang, Baosen and Cortés, Jorge},
  journal={IEEE Open J. Control Syst.}, 
  title={Stable Reinforcement Learning for Optimal Frequency Control: A Distributed Averaging-Based Integral Approach}, 
  year={2022},
  volume={1},
  number={},
  pages={194-209},
  doi={10.1109/OJCSYS.2022.3202202}}

@ARTICLE{8399490,
  author={Qu, Guannan and Li, Na},
  journal={IEEE Control Systems Letters}, 
  title={On the Exponential Stability of Primal-Dual Gradient Dynamics}, 
  year={2019},
  volume={3},
  number={1},
  pages={43-48},
  doi={10.1109/LCSYS.2018.2851375}}

@article{Feijer2010Dec,
	author = {Feijer, Diego and Paganini, Fernando},
	title = {{Stability of primal{\textendash}dual gradient dynamics and applications to network optimization}},
	journal = {Automatica},
	volume = {46},
	number = {12},
	pages = {1974--1981},
	year = {2010},
	month = dec,
	issn = {0005-1098},
	publisher = {Pergamon},
	doi = {10.1016/j.automatica.2010.08.011}
}

@article{Cherukuri2016Jan,
	author = {Cherukuri, Ashish and Mallada, Enrique and Cort{\ifmmode\acute{e}\else\'{e}\fi}s, Jorge},
	title = {{Asymptotic convergence of constrained primal{\textendash}dual dynamics}},
	journal = {Systems Control Lett.},
	volume = {87},
	pages = {10--15},
	year = {2016},
	month = jan,
	issn = {0167-6911},
	publisher = {North-Holland},
	doi = {10.1016/j.sysconle.2015.10.006}
}

@article{Chen2022Jun,
  author={Chen, Xin and Poveda, Jorge I. and Li, Na},
  journal=IEEE_J_AC, 
  title={Continuous-Time Zeroth-Order Dynamics With Projection Maps: Model-Free Feedback Optimization With Safety Guarantees}, 
  year={2025},
  volume={70},
  number={8},
  pages={5005-5020}
}

@book{clarke1990optimization,
  author    = {Frank H. Clarke},
  title     = {Optimization and Nonsmooth Analysis},
  year      = {1990},
  publisher = {Society for Industrial and Applied Mathematics},
  address   = {Philadelphia},
  series    = {Wiley-Interscience Series in Discrete Mathematics and Optimization},
  note      = {Originally published by Wiley, reprinted by SIAM},
}

@article{Cortes2008May,
	author = {Cortes, Jorge},
	title = {{Discontinuous dynamical systems}},
	journal = {IEEE Control Syst. Mag.},
	volume = {28},
	number = {3},
	pages = {36--73},
	year = {2008},
	month = may,
	publisher = {IEEE},
	doi = {10.1109/MCS.2008.919306}
}

@article{Shen2023Oct,
	author = {Shen, Yukang and Wu, Wenchuan and Sun, Shumin},
	title = {{Stochastic Model Predictive Control Based Fast-Slow Coordination Automatic Generation Control}},
	journal = {IEEE Trans. Power Syst.},
	volume = {39},
	number = {3},
	pages = {5259--5271},
	year = {2023},
	month = oct,
	publisher = {IEEE},
	doi = {10.1109/TPWRS.2023.3321847}
}

@book{khalil2002nonlinear,
  author    = {H. K. Khalil and J. W. Grizzle},
  title     = {Nonlinear Systems},
  edition   = {3rd},
  publisher = {Prentice Hall},
  address   = {Upper Saddle River, NJ},
  year      = {2002}
}

@misc{tisp_projection,
  author       = {Shailesh Kumar},
  title        = {Topics in Signal Processing},
  howpublished = {\url{https://tisp.indigits.com/cvxopt/projection}},
  note         = {Accessed: 2025-05-09},
  year         = {2025}
}

@inproceedings{Wang2025PESGM,
  author    = {Wang, Xiaoyang and Chen, Xin},
  title     = {Distributed Coordination of Grid-Forming and Grid-Following Inverter-Based Resources for Optimal Frequency Control in Power Systems},
  booktitle = {Proc. IEEE Power \& Energy Society General Meeting (PESGM)},
  address   = {Austin, TX, USA},
  year      = {2025},
  pages     = {1--5},
  doi       = {10.1109/PESGM52009.2025.11225381}
}

@ARTICLE{10696981,
  author={Zhao, Yunzheng and Liu, Tao and Hill, David J.},
  journal=IEEE_J_PWRS, 
  title={Distributed Attention-Enabled Multi-Agent Reinforcement Learning Based Frequency Regulation of Power Systems}, 
  year={2025},
  volume={40},
  number={3},
  pages={2427-2437},
  doi={10.1109/TPWRS.2024.3469132}}

@ARTICLE{8588380,
  author={Khayat, Yousef and Naderi, Mobin and Shafiee, Qobad and Batmani, Yazdan and Fathi, Mohammad and Guerrero, Josep M. and Bevrani, Hassan},
  journal=IEEE_J_PWRS, 
  title={Decentralized Optimal Frequency Control in Autonomous Microgrids}, 
  year={2019},
  volume={34},
  number={3},
  pages={2345-2353},
  doi={10.1109/TPWRS.2018.2889671}}

@ARTICLE{10955180,
  author={Dos Santos, Lucas Elias and Flores Huaman, Jesus and Carlos Ugaz Peña, José and Duarte Queiroz, Eliabe and Dotta, Daniel},
  journal={IEEE Trans. Smart Grid},
  title={A Data-Driven Centralized Secondary Control Design Methodology for Microgrids}, 
  year={2025},
  volume={16},
  number={4},
  pages={2738-2751},
  doi={10.1109/TSG.2025.3558108}}

@ARTICLE{10124785,
  author={Liu, Jiayi and Song, Huihui and Chen, Chenyue and Guerrero, Josep M. and Liu, Meng and Qu, Yanbin},
  journal={IEEE Transactions on Smart Grid}, 
  title={Decentralized Secondary Frequency Control of Autonomous Microgrids via Adaptive Robust {L2}-Gain Performance}, 
  year={2024},
  volume={15},
  number={1},
  pages={67-76},
  doi={10.1109/TSG.2023.3274235}}

@ARTICLE{7112129,
  author={Simpson-Porco, John W. and Shafiee, Qobad and Dörfler, Florian and Vasquez, Juan C. and Guerrero, Josep M. and Bullo, Francesco},
  journal={IEEE Transactions on Industrial Electronics}, 
  title={Secondary Frequency and Voltage Control of Islanded Microgrids via Distributed Averaging}, 
  year={2015},
  volume={62},
  number={11},
  pages={7025-7038},
  doi={10.1109/TIE.2015.2436879}}

@ARTICLE{8556089,
  author={Weitenberg, Erieke and Jiang, Yan and Zhao, Changhong and Mallada, Enrique and De Persis, Claudio and Dörfler, Florian},
  journal={IEEE Transactions on Automatic Control}, 
  title={Robust Decentralized Secondary Frequency Control in Power Systems: Merits and Tradeoffs}, 
  year={2019},
  volume={64},
  number={10},
  pages={3967-3982},
  doi={10.1109/TAC.2018.2884650}}

@ARTICLE{9442926,
  author={Zholbaryssov, Madi and Domínguez-García, Alejandro D.},
  journal={IEEE Transactions on Power Systems}, 
  title={Safe Data-Driven Secondary Control of Distributed Energy Resources}, 
  year={2021},
  volume={36},
  number={6},
  pages={5933-5943},
  doi={10.1109/TPWRS.2021.3084440}}

@ARTICLE{8892668,
  author={Khayat, Yousef and Shafiee, Qobad and Heydari, Rasool and Naderi, Mobin and Dragičević, Tomislav and Simpson-Porco, John W. and Dörfler, Florian and Fathi, Mohammad and Blaabjerg, Frede and Guerrero, Josep M. and Bevrani, Hassan},
  journal={IEEE Transactions on Power Electronics}, 
  title={On the Secondary Control Architectures of {AC} Microgrids: An Overview}, 
  year={2020},
  volume={35},
  number={6},
  pages={6482-6500},
  doi={10.1109/TPEL.2019.2951694}}

@standard{institute2003ieee,
  title = {{IEEE} guide for synchronous generator modeling practices and applications in power system stability analyses},
  organization = {IEEE},
  year         = {2003},
  note         = {{IEEE} Std 1110-2002},
}

@misc{wang2025ieee39ibr,
  author       = { Wang, Xiaoyang and Chen, Xin},
  title        = { {IEEE} 39-Bus Power System with 100\% Inverter-Based Resources},
  year         = {2025},
  howpublished = {\url{https://github.com/Xiaoyang-Wang-TAMU/IEEE39BusSystemwithIBR}},
}
